\documentclass[11pt,a4paper]{article}
\pdfoutput=1

\usepackage{jheppub}
\usepackage{latexsym}
\usepackage{color}
\usepackage[usenames,dvipsnames,svgnames,table]{xcolor}
\usepackage{graphicx}
\usepackage{epsfig}  
\usepackage{epsf}   
\usepackage{dcolumn}
\usepackage{bm}
\usepackage[toc,page]{appendix}
\usepackage{dcolumn}
\usepackage{textcomp}
\usepackage[force]{feynmp-auto}
\usepackage{float}
\usepackage{hypcap}
\usepackage[]{hyperref}
\usepackage{booktabs}
\usepackage{multirow}
\usepackage{diagbox}
\usepackage{alphalph}
\usepackage{textcomp}
\usepackage{epstopdf}
\usepackage{amssymb}
\usepackage{wasysym}
\usepackage{subcaption}
\usepackage{orcidlink}
\usepackage{lineno, blindtext}
\usepackage{caption}
\usepackage{lineno}
\usepackage{soul} 
\usepackage[normalem]{ulem}

\newcommand{\ie}{{\it i.e.}}

\newcommand{\eg}{{\it e.g.}}

\newcommand{\eq}{Eq.}

\newcommand{\fig}{Fig.}

\newcommand{\Refe}{Ref.}
\newcommand{\Refes}{Refs.}

\newcommand{\equ}[1]{\eq~(\ref{equ:#1})}
\newcommand{\figu}[1]{\fig~\ref{fig:#1}}

\newcommand{\bi}{\begin{itemize}}
\newcommand{\ei}{\end{itemize}}

\newcommand{\dcp}{\delta_{\rm CP}}

\newcommand{\beq}{\begin{equation}}
\newcommand{\eeq}{\end{equation}}
\newcommand{\beqa}{\begin{eqnarray}}
\newcommand{\eeqa}{\end{eqnarray}}
\definecolor{orcidlogocol}{HTML}{A6CE39}

\preprint{IP/BBSR/2023-05}

\title{Present and future constraints on flavor-dependent long-range interactions of high-energy astrophysical neutrinos}

\author[a,b,c]{\orcidlink{0000-0002-9714-8866}Sanjib Kumar Agarwalla,}
\author[d]{\orcidlink{0000-0001-6923-0865}Mauricio Bustamante,}
\author[a,b]{\orcidlink{0000-0002-5508-7751}Sudipta Das,}
\author[a]{\orcidlink{0000-0002-4191-7806}Ashish Narang}

\affiliation[a]{Institute of Physics, Sachivalaya Marg, Sainik School Post, Bhubaneswar 751005, India}
\affiliation[b]{Homi Bhabha National Institute, Training School Complex, Anushakti Nagar, Mumbai 400094}
\affiliation[c]{Department of Physics \& Wisconsin IceCube Particle Astrophysics Center, University of Wisconsin,
	Madison, WI 53706, U.S.A.}
\affiliation[d]{Niels Bohr International Academy, Niels Bohr Institute,
University of Copenhagen, DK-2100 Copenhagen, Denmark}

\emailAdd{sanjib@iopb.res.in}
\emailAdd{mbustamante@nbi.ku.dk}
\emailAdd{sudipta.d@iopb.res.in}
\emailAdd{ashish.narang@iopb.res.in}

\abstract{The discovery of new, flavor-dependent neutrino interactions would provide compelling evidence of physics beyond the Standard Model. We focus on interactions generated by the anomaly-free, gauged, abelian lepton-number symmetries, specifically $L_e-L_\mu$, $L_e-L_\tau$, and $L_\mu-L_\tau$, that introduce a new matter potential sourced by electrons and neutrons, potentially impacting neutrino flavor oscillations.  We revisit, revamp, and improve the constraints on these interactions that can be placed via the flavor composition of the diffuse flux of high-energy astrophysical neutrinos, with TeV--PeV energies, \ie, the proportion of $\nu_e$, $\nu_\mu$, and $\nu_\tau$ in the flux.  Because we consider mediators of these new interactions to be ultra-light, lighter than $10^{-10}$~eV, the interaction range is ultra-long, from km to Gpc, allowing vast numbers of electrons and neutrons in celestial bodies and the cosmological matter distribution to contribute to this new potential. We leverage the present-day and future sensitivity of high-energy neutrino telescopes and of oscillation experiments to estimate the constraints that could be placed on the coupling strength of these interactions. We find that, already today, the IceCube neutrino telescope demonstrates potential to constrain flavor-dependent long-range interactions significantly better than existing constraints, motivating further analysis. We also estimate
the improvement in the sensitivity due to the next-generation neutrino telescopes such as IceCube-Gen2, Baikal-GVD, KM3NeT, P-ONE, 
and TAMBO.}


\arxivnumber{2305.03675}

\begin{document}
\maketitle



\section{Introduction}
\label{sec:introduction}

The high-energy astrophysical neutrinos discovered by the IceCube neutrino telescope~\cite{Aartsen:2013bka, Aartsen:2013jdh, Aartsen:2014gkd, Aartsen:2015rwa, Aartsen:2016xlq, Ahlers:2018fkn, IceCube:2020wum, Ackermann:2022rqc} provide unprecedented potential to explore high-energy fundamental physics~\cite{Gaisser:1994yf, Ahlers:2018mkf, Ackermann:2019cxh, Arguelles:2019rbn, AlvesBatista:2021eeu, Ackermann:2022rqc}.  Because they reach energies in the TeV--PeV range, they are sensitive to possible new physics at energy scales beyond the reach of terrestrial colliders.  Because they travel Gpc-scale distances --- essentially the size of the observable Universe --- minute new-physics effects, ordinarily undetectable, may accumulate en route to Earth and become detectable.   (They also provide unique insight into the most energetic astrophysical sources~\cite{Ackermann:2019ows, AlvesBatista:2021eeu, Ackermann:2022rqc}.)  We gear this potential to far-reaching studies of new, long-range neutrino interactions.

\begin{figure}[t!]
 \centering
 \includegraphics[width=.7\textwidth]{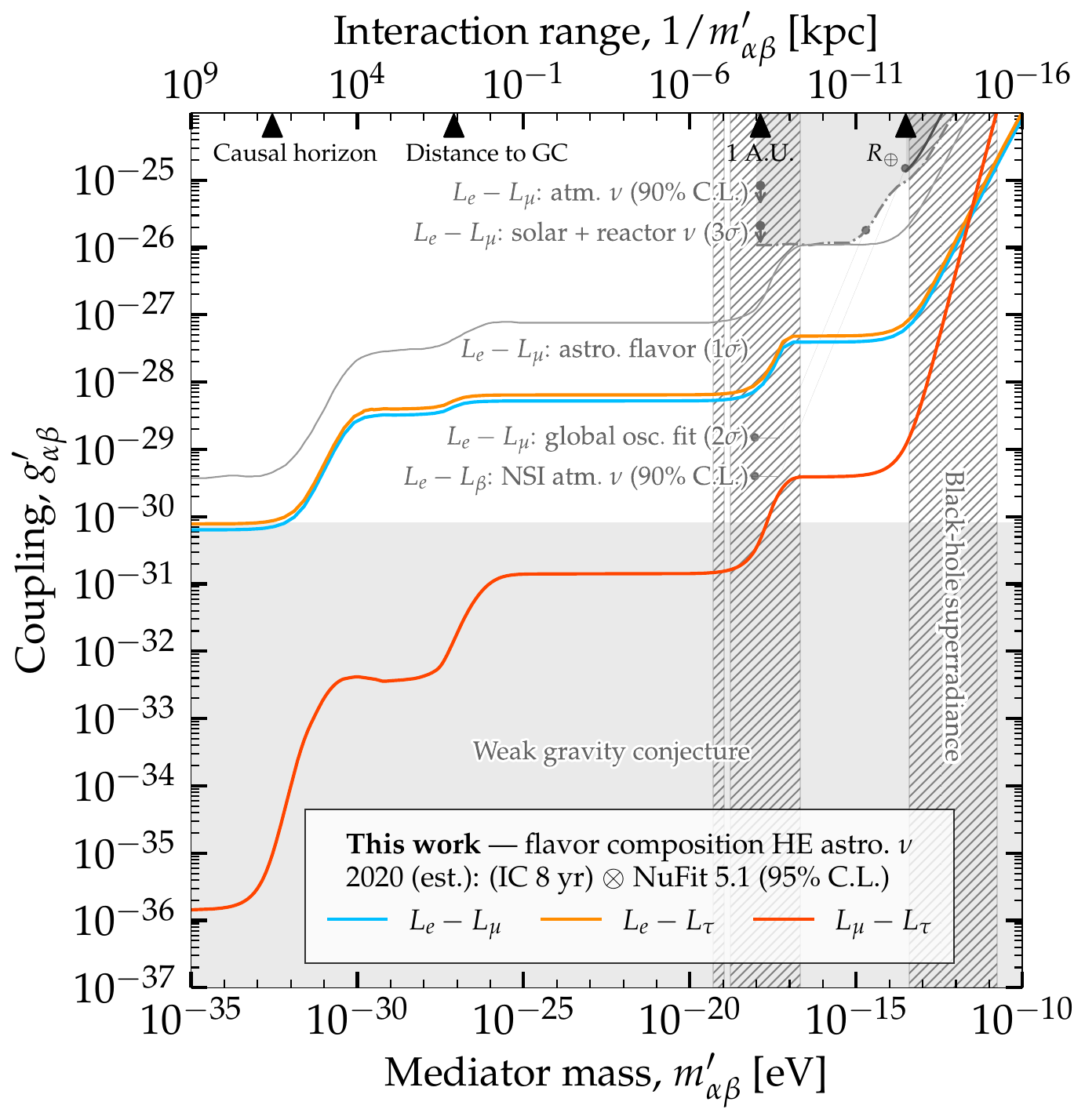}
 \caption{\label{fig:limits_3models}\textbf{\textit{Estimated present-day upper limits on the coupling strength, $g_{\alpha\beta}^\prime$, of the new boson, $Z_{\alpha\beta}^\prime$, with mass $m_{\alpha\beta}^\prime$, that mediates flavor-dependent long-range neutrino interactions.}}  Our results are based on estimates of the measurement of flavor composition of high-energy astrophysical neutrinos in IceCube using 8~years of data~\cite{IceCube-Gen2:2020qha} (through-going muons plus HESE, see Section~\ref{sec:he_nu}) and on present-day uncertainties in the neutrino mixing parameters~\cite{Esteban:2020cvm, NuFIT}, assuming normal neutrino mass ordering.  Limits are on neutrino-electron interactions, under the $L_e-L_\mu$ and $L_e-L_\tau$ symmetries, and on neutrino-neutron interactions, under the $L_\mu-L_\tau$ symmetry.  For the latter, we assume a $Z_{\mu\tau}^\prime$--$Z$ mixing strength of $(\xi - \sin\theta_W\chi) = 5 \times 10^{-24}$~\cite{Heeck:2010pg}.  Existing limits are from a recent global oscillation fit~\cite{Coloma:2020gfv}, atmospheric neutrinos~\cite{Joshipura:2003jh}, solar and reactor neutrinos~\cite{Bandyopadhyay:2006uh}, and non-standard interactions (NSI)~\cite{Super-Kamiokande:2011dam, Ohlsson:2012kf, Gonzalez-Garcia:2013usa}.  For comparison, we show the proof-of-principle sensitivity ($1\sigma$) based on 2015 IceCube flavor-composition measurements~\cite{IceCube:2015gsk} from \Refe~\cite{Bustamante:2018mzu}.  Indirect limits~\cite{Wise:2018rnb} are from 
 black-hole superradiance (90\% C.L.)~\cite{Baryakhtar:2017ngi} and the weak gravity conjecture~\cite{Arkani-Hamed:2006emk}, assuming a lightest neutrino mass of $0.01$~eV.  Figures~\ref{fig:g_vs_m_120} and \ref{fig:plots_for_et} show projections for the year 2040; \figu{posterior_IMO} shows results assuming inverted mass ordering.  See Section~\ref{sec:results} for details.  {\it Our results show that limits obtained from present-day flavor-composition measurements may improve significantly on existing ones.}}
\end{figure}

Presently, theory and experiment leave room for the possibility that neutrinos undergo interactions beyond the Standard Model weak interactions that would have gone undetected so far.  In particular, new interactions that affect different neutrino flavors differently may modify neutrino oscillations, and so may be tested in oscillation experiments that use neutrinos from a variety of sources, natural and man-made.  References~\cite{Joshipura:2003jh, Bandyopadhyay:2006uh} pioneered the idea of introducing such flavor-dependent interactions by gauging native global lepton-number symmetries of the Standard Model, generated by $L_\alpha-L_\beta$, where $L_\alpha$ ($\alpha = e, \mu, \tau$) is the $\alpha$-flavor lepton number.  Doing so introduces new neutrino-matter interactions mediated by a boson that, if light, makes it possible for distant matter to source a long-range potential that affects neutrino oscillation probabilities.  Previous works have explored this possibility and placed constraints on the strength of the new interaction using atmospheric neutrinos~\cite{Joshipura:2003jh}, solar and reactor neutrinos~\cite{Bandyopadhyay:2006uh}, accelerator neutrinos~\cite{Heeck:2010pg,Chatterjee:2015gta}, and a global fit to oscillation data~\cite{Coloma:2020gfv}.  

Recently, \Refe~\cite{Bustamante:2018mzu} proposed a new way to probe these interactions using high-energy astrophysical neutrinos.  Because the relative contribution of the standard oscillation mechanism, driven by the neutrino masses, falls with neutrino energy, at high energies the contribution from new interactions may become prominent and more easily detectable.  By modifying oscillations --- in an extreme case, by suppressing flavor transitions --- the new interactions would also modify the flavor composition of high-energy astrophysical neutrinos, \ie, the proportion of $\nu_e$, $\nu_\mu$, and $\nu_\tau$ in their flux.   IceCube measures the flavor composition of the diffuse flux of high-energy astrophysical neutrinos~\cite{IceCube:2015rro, IceCube:2015gsk, Aartsen:2018vez, IceCube:2020fpi} (see also \Refes~\cite{Mena:2014sja, Palomares-Ruiz:2015mka, Vincent:2016nut}) and, though the measurement is challenging, it is a versatile probe of high-energy neutrino physics~\cite{Rasmussen:2017ert, Ahlers:2018mkf, Ackermann:2019cxh, AlvesBatista:2021eeu, MammenAbraham:2022xoc, Berryman:2022hds, Ackermann:2022rqc, Arguelles:2022tki, Soto:2021vdc} and astrophysics~\cite{Ackermann:2019ows, AlvesBatista:2021eeu, Ackermann:2022rqc}.  In particular, the sensitivity to new neutrino interactions stems from looking for characteristic deviations in the flavor composition relative to its expectation under standard oscillations. 

Based on this, \Refe~\cite{Bustamante:2018mzu} provided a proof of principle for using the flavor composition measured by IceCube and forecast for its planned upgrade, IceCube-Gen2~\cite{IceCube-Gen2:2020qha}, to constrain flavor-dependent long-range neutrino-electron interactions generated by gauging the $L_e-L_\mu$ and $L_e-L_\tau$ symmetries.  It focused on ultra-light mediators, lighter than $10^{-10}$~eV, which makes the interaction range ultra-long, from km to Gpc, depending on the mass.  As a result, vast numbers of electrons in the Earth, the Moon, the Sun, the Milky Way, and in the distant Universe could source a sizable long-range potential.  Yet, while the proof-of-principle prospects in \Refe~\cite{Bustamante:2018mzu} were promising, the statistical methods used therein are not amenable to being generalized to compute sensitivity beyond the $1\sigma$ statistical significance.  

To address this, we revisit, revamp, and expand the tests of flavor-dependent long-range neutrino interactions based on flavor composition.  There are four main improvements in our work.  First, we adopt a Bayesian approach to compute constraints, based on the methods from \Refe~\cite{Song:2020nfh}.  It allows us to combine consistently the uncertainties in the measurement of the flavor composition in IceCube and other neutrino telescopes, and the uncertainties in the neutrino mixing parameters that drive standard oscillations, and to compute constraints at arbitrary statistical significance.  Second, not only do we consider neutrino-electron interactions induced by the $L_e-L_\mu$ and $L_e-L_\tau$ symmetries, as in \Refe~\cite{Bustamante:2018mzu}, but also neutrino-neutron interactions induced by the $L_\mu-L_\tau$ symmetry.  Third, we study not only the sensitivity of IceCube and IceCube-Gen2, but also the sensitivity achieved by combining them with other neutrino telescopes expected by 2040, Baikal-GVD~\cite{Avrorin:2019vfc}, KM3NeT~\cite{Adrian-Martinez:2016fdl}, P-ONE~\cite{P-ONE:2020ljt}, and TAMBO~\cite{Romero-Wolf:2020pzh}.  Fourth, we use projected measurements of the mixing parameters in upcoming neutrino oscillation experiments, the Deep Underground Neutrino Experiment (DUNE)~\cite{Abi:2020wmh}, Hyper-Kamiokande (HK)~\cite{Abe:2018uyc}, and the Jiangmen Underground Neutrino Observatory (JUNO)~\cite{An:2015jdp}, that will shrink the uncertainty on their values.

Figure~\ref{fig:limits_3models} shows our main results, in the form of upper limits on the coupling strength of the new interactions, conservatively based on approximate present-day flavor-measurement capabilities in IceCube and of uncertainties in the mixing parameters.  For mediators lighter than $10^{-18}$~eV, our estimated upper limits confirm~\cite{Bustamante:2018mzu} the reach of flavor-composition measurements to test ultra-long-range interactions, now at higher statistical significance.  We find that, \textbf{\textit{already today, high-energy astrophysical neutrinos may be able to constrain flavor-dependent long-range neutrino interactions more strongly than existing limits}}.  By 2040, our conservative forecast limits, not shown in \figu{limits_3models}, improve marginally (Section~\ref{sec:results}), though more power may be reaped from improvements to our analysis that could be afforded by significantly higher detection rates of high-energy neutrinos and precision in the mixing parameters.  We point these out later (Section~\ref{sec:future}).

We provide our constraints as approximate, rather than as definitive, since they rely on specific, albeit realistic assumptions on present-day and upcoming experimental capabilities.  For instance, they are obtained assuming that measurements of the flavor composition are performed on an astrophysical neutrino spectrum that is a power law $\propto E^{-2.5}$, where $E$ is the neutrino energy.  While this is compatible with IceCube observations~\cite{IceCube:2015gsk}, our results do not account for deviations from this assumption that lie within experimental error, and which could make the spectrum softer (see, \eg, \Refe~\cite{IceCube:2020wum}) or harder (see, \eg, \Refe~\cite{IceCube:2021uhz}).  Accounting for this would require deriving the IceCube flavor sensitivity for different choices of the neutrino spectrum, as in \Refe~\cite{IceCube:2015gsk}, which involves combining different data sets that are not available outside the Collaboration.  Doing so lies beyond the scope of this paper, but we comment on possible future developments (Section~\ref{sec:future}).

This paper is organized as follows. Section~\ref{sec:modelandLRI} introduces long-range interactions and their effect on neutrino oscillations.   Section~\ref{sec:he_nu} provides an overview of high-energy astrophysical neutrinos and shows the effect of long-range interactions on their flavor composition.  Section~\ref{sec:analysis} presents out statistical methods and results.  Section~\ref{sec:future} points out future improvements of our analysis.  Section~\ref{sec:conc} summarizes and concludes.


\section{New long-range, lepton-number neutrino interactions}
\label{sec:modelandLRI}


\subsection{New lepton-number symmetries: $L_e-L_\mu$, $L_e-L_\tau$, and $L_\mu-L_\tau$}
\label{subsec:models}

The Standard Model (SM), in addition to the gauge group $SU(3)_{\rm C}\times SU(2)_{\rm L}\times U(1)_{\rm Y}$, contains global  $U(1)$ symmetries associated to the baryon number and the three lepton numbers, $L_e$, $L_\mu$, and $L_\tau$.  While they cannot be gauged individually without introducing anomalies, certain combinations of them can be; for an extensive list, see \Refe~\cite{Coloma:2020gfv}.  We focus on three well-motivated, anomaly-free symmetries that gauge lepton number differences~\cite{Foot:1990mn, He:1990pn, He:1991qd}: $U(1)_{L_e-L_\mu}$, $U(1)_{L_e-L_\tau}$, and $U(1)_{L_\mu-L_\tau}$.  Each one introduces a new neutral gauge vector boson, $Z_{e \mu}^\prime$, $Z_{e \tau}^\prime$, and $Z_{\mu\tau}^\prime$, that mediates new neutrino interactions~\cite{He:1990pn,He:1991qd,Foot:1994vd} with electrons and neutrons, as we show below.  
(The Higgs sector differentiates between different lepton flavors~\cite{Foot:2005uc}; however, here we focus only on the gauge interactions through the new boson.)  
The $L_{\alpha}-L_{\beta}$ gauge symmetries and their extensions have been explored in numerous scenarios, including as possible solutions of the Hubble tension~\cite{Araki:2021xdk}, of the electron and muon $(g-2)$ anomalies~\cite{Chen:2020jvl, Bodas:2021fsy, Panda:2022kbn, Borah:2021khc}, considering the new boson to be dark matter~\cite{Baek:2015fea, Asai:2020qlp, Alonso-Alvarez:2023tii} and as radiation from compact binary systems~\cite{KumarPoddar:2019ceq}, leptogenesis~\cite{Asai:2020qax}, their influence on muon-beam dumps at the TeV-scale~\cite{Cesarotti:2022ttv}, possible production of dark photons at the MUonE experiment~\cite{GrillidiCortona:2022kbq}, and explaining the electron and positron excess in the cosmic-ray flux~\cite{Duan:2017qwj,He:2009ra}.  However, these studies consider mediators that are heavier than the ones we consider here, and couplings (see below) that are stronger.

The effective neutrino-matter interaction receives three contributions, mediated by the SM $Z$ boson, by the new $Z^\prime_{\alpha \beta}$ boson, and via the mixing between $Z^\prime_{\alpha \beta}$ and $Z$, \ie,
\begin{equation}
 \label{equ:Lagrangian_1} 
 \mathcal{L}_{\text{eff}} = \mathcal{L}_{\text{SM}} +\mathcal{L}_{Z'}+\mathcal{L}_{\text{mix}} \;.
\end{equation}
The first term in \equ{Lagrangian_1}, $\mathcal{L}_{\text{SM}}$, is the SM contribution, \ie,
\begin{equation}
 \label{equ:lag_sm}
 \mathcal{L}_{\text{SM}}
 =
 \frac{e}{\sin \theta_W \cos \theta_W}Z_\mu \left[-\frac{1}{2}\bar{l}_{\alpha}\gamma^\mu P_L l_{\alpha}+\frac{1}{2}\bar{\nu}_{\alpha}\gamma^\mu P_L \nu_{\alpha}+\frac{1}{2}\bar{u}\gamma^\mu P_L u-\frac{1}{2}\bar{d}\gamma^\mu P_L d\right] \;,
\end{equation}
where $e$ is the unit electric charge, $\theta_W$ is the Weinberg angle, $l_\alpha$ and $\nu_\alpha$ are, respectively, the charged lepton and neutrino of flavor $\alpha = e$, $\mu$, or $\tau$, $u$ and $d$ are the up and down quarks, and $P_L$ is the left-handed projection operator.  (Equation~(\ref{equ:lag_sm}), and also \equ{lag_mix} below, assumes that matter is electrically neutral, \ie, that it contains equal numbers of electrons and protons~\cite{Heeck:2010pg}; this is also what we assume later when computing the new matter potential; see Section~\ref{subsec:LRI}.)

The second term in \equ{Lagrangian_1}, $\mathcal{L}_{Z'}$, describes neutrino-matter interactions via the new mediator~\cite{He:1990pn,He:1991qd,Heeck:2010pg,Coloma:2020gfv}, \ie,
\begin{equation}
 \label{equ:lag_zprime}
 \mathcal{L}_{Z'}
 =
 g_{\alpha \beta}' Z'_{\sigma}(\bar{l}_{\alpha}\gamma^{\sigma}l_{\alpha}-\bar{l}_{\beta}\gamma^{\sigma}l_{\beta}+ \bar{\nu}_{\alpha}\gamma^{\sigma}P_{L}\nu_{\alpha}-\bar{\nu}_{\beta}\gamma^{\sigma}P_{L}\nu_{\beta}) \;,
\end{equation}
where $g^{\prime}_{\alpha\beta}$ ($\alpha,\beta$ = $e,\mu,\tau$, $\alpha \neq \beta$) is the adimensional coupling constant associated to the $L_\alpha-L_\beta$ gauge symmetry.  Because naturally occurring muons and tauons are scarce, we consider this contribution only under $L_e-L_\mu$ and $L_e-L_\tau$, and the corresponding interaction to be sourced only by abundant electrons.

The third term in \equ{Lagrangian_1}, $\mathcal{L}_{\text{mix}}$, contains terms that mix $Z^\prime_{\alpha\beta}$ with $Z$~\cite{Babu:1997st,Heeck:2010pg,Joshipura:2019qxz}, \ie,
$\mathcal{L}_{ZZ^\prime} = -\frac{1}{2} \sin \chi \hat{Z}'_{\mu \nu}\hat{B}^{\mu \nu}+\delta \hat{M}^2 \hat{Z}'_{\mu}\hat{Z}^{\mu}$, where $\hat{Z}'_{\mu \nu}$ and $\hat{B}^{\mu \nu}$ are the field strength tensors for $U(1)^{\prime}$ and $U(1)_Y$, respectively, $\hat{Z}$ and $\hat{Z}^\prime$ are the gauge eigenstates corresponding to the neutral massive gauge bosons of the SM and the new $U(1)^{\prime}$ gauge symmetry, $\chi$ is the kinetic mixing angle, and $\delta \hat{M}^2$ is the squared-mass difference between $\hat{Z}$ and $\hat{Z}^\prime$. Diagonalizing this Lagrangian redefines the fields in terms of physical states: the photon and two massive bosons, $Z$ and $Z'$, that are related to $\hat{Z}$ and $\hat{Z'}$ through a new mixing angle $\xi$~\cite{Babu:1997st}, \ie, $\mathcal{L}_{ZZ^\prime} \supset (\xi-\sin\theta_W\chi) Z'_{\mu}Z^{\mu}$.  This introduces a four-fermion neutrino-matter interaction term via $Z$--$Z_{\alpha\beta}^\prime$ mixing, \ie,
\begin{equation}
 \label{equ:lag_mix}
 \mathcal{L}_{\rm mix}
 =
 -g_{\alpha\beta}^\prime
 (\xi-\sin\theta_W\chi)\frac{e}{\sin\theta_W \cos\theta_W}
 J'_\rho J_3^\rho \;,
\end{equation}
where $J^\prime_\rho = \bar{\nu}_\alpha \gamma_\rho P_L\nu_\alpha-\bar{\nu}_\beta \gamma_\rho P_L\nu_\beta$ and $J_3^\rho = -\frac{1}{2}\bar{e}\gamma^\rho P_L e+\frac{1}{2}\bar{u}\gamma^\rho P_L u-\frac{1}{2}\bar{d}\gamma^\rho P_L d$.
To illustrate the effect of mixing, below we fix the value of the mixing strength to $(\xi-\sin \theta_W \chi) = 5 \times 10^{-24}$~\cite{Heeck:2010pg}, the upper limit for a new interaction whose range is of the order of the distance between the Earth and the Sun~\cite{Schlamminger:2007ht,Adelberger:2009zz,Heeck:2010pg}.  (The upper limit on the mixing strength for an interaction range of the order of the size of the Earth is slightly weaker and allows for larger mixing~\cite{Schlamminger:2007ht,Adelberger:2009zz,Heeck:2010pg}.)  Because of our analysis choices, $Z$--$Z_{\alpha\beta}^{\prime}$ mixing affects the results only under the $L_\mu-L_\tau$ symmetry; we explain this below.

\begin{figure}[t!]
\centering
\includegraphics[width=\textwidth]{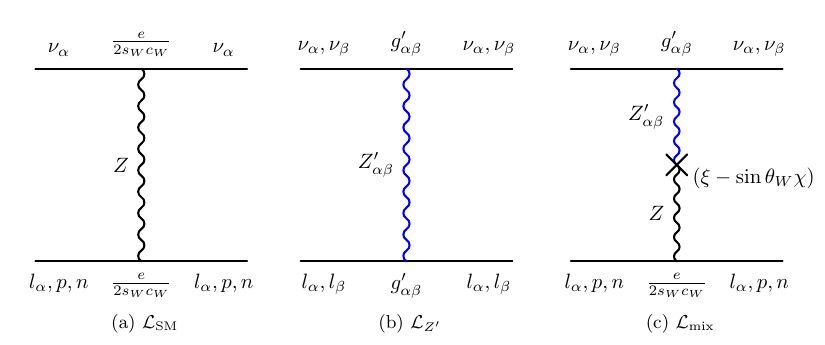}
 \caption{\textbf{\textit{Feynman diagrams for the contributions to the neutrino-matter interaction Lagrangian, \equ{Lagrangian_1}.}}  (a) Neutrino-matter interaction in the Standard Model, mediated by the neutral gauge boson $Z$.  In our analysis, this is important only for neutrinos inside the Earth, where matter densities are high.  (b) Neutrino-matter interaction in the $U(1)_{L_\alpha - L_\beta}$ model, mediated by the new $Z_{\alpha \beta}^\prime$ gauge boson. (c) Neutrino-matter interaction via $Z_{\alpha \beta}^\prime$--$Z$ mixing.  In our analysis, for the $L_e-L_\mu$ and $L_e-L_\tau$ symmetries, only diagram (a) (and only inside the Earth) and diagram (b), sourced by electrons, contribute, and we do not consider diagram (c).  For the $L_\mu-L_\tau$ symmetry, only diagram (a) (and only inside the Earth) and diagram (c), sourced by neutrons, contribute.  See Section~\ref{subsec:models} for details.}\label{fig:feyn}
\end{figure}

Figure~\ref{fig:feyn} shows the Feynman diagrams corresponding to the above contributions.  Regarding $\mathcal{L}_{\rm SM}$, because the $Z$ boson is heavy, the SM neutral-current interaction is short-range and significant only in relatively high-density regions such as inside the Earth (which we account for later).  Regarding $\mathcal{L}_{Z^\prime}$, it induces new interactions between neutrinos and charged leptons.  However, in practice, the scarcity of muons and tauons in Nature means that interactions are sourced only by electrons.  As a result, this term in the Lagrangian is active only for the $L_e-L_\mu$ and $L_e-L_\tau$ symmetries.  Regarding $\mathcal{L}_{\rm mix}$, it induces interactions between neutrinos and electrons, protons, and neutrons.  Under the $L_\mu-L_\tau$ symmetry, this is the dominant contribution; its size depends on the value of the mixing strength (see above).  Further, for a macroscopic collection of ordinary matter, like the ones we consider below, the contribution of electrons is nullified by the contribution of protons, leaving only neutrons to source the new interaction; see Appendix~\ref{app:potential_lmu_ltau} for details.  Under the $L_e-L_\mu$ and $L_e-L_\tau$ symmetries, the relative importance of $\mathcal{L}_{\rm mix}$ vs.~$\mathcal{L}_{Z^\prime}$ depends on the value of the mixing strength; to be conservative, we assume no mixing for these two symmetries. 

In summary, under the $L_e-L_\mu$ and $L_e-L_\tau$ symmetries, the new neutrino interaction is via $\mathcal{L}_{Z^\prime}$ and is sourced by electrons only.  Under the $L_\mu-L_\tau$ symmetry, it is via $\mathcal{L}_{\rm mix}$ and is sourced by neutrons only.  Under all symmetries, $\mathcal{L}_{\rm SM}$ acts only inside the Earth.

\subsection{Long-range neutrino interactions}
\label{subsec:LRI}

A neutrino separated by a distance $d$ from a collection of $N_e$ electrons, under $L_e-L_\mu$ and $L_e-L_\tau$, or of $N_n$ neutrons, under $L_\mu-L_\tau$, experiences a Yukawa potential of
\begin{equation}
\label{equ:pot_general}
 V_{\rm \alpha \beta}
 =
 \mathcal{G}_{\alpha \beta}\frac{e^{-m_{\alpha \beta}^{\prime} d}}{4 \pi d}
 \times
 \left\{
  \begin{array}{lll}
   N_e & , & {\rm for}~\alpha, \beta = e, \mu ~{\rm or}~ e, \tau \\
   N_n & , & {\rm for}~\alpha, \beta = \mu, \tau \\
  \end{array}
 \right. \;,
\end{equation}
where $m_{\alpha \beta}^{\prime}$ is the mass of the $Z'_{\alpha \beta}$ boson and the coupling strength is
\begin{equation}
 \label{equ:Gab}
 \mathcal{G}_{\alpha \beta}
 =
 \left\{
  \begin{array}{lll}
   g^{\prime 2}_{e \mu} & , & ~{\rm for}~\alpha, \beta = e, \mu \\
   g^{\prime 2}_{e \tau} & , & ~{\rm for}~\alpha, \beta = e, \tau \\
   g^{\prime}_{\mu \tau} (\xi-\sin \theta_W \chi) \frac{e}{4 \sin \theta_W \cos \theta_W} & , & ~{\rm for}~\alpha, \beta = \mu, \tau \\
  \end{array}
 \right. \;.
\end{equation}
The range of the interaction is $\sim$$1/m_{\alpha\beta}^\prime$; beyond this distance from the source of the potential, it is suppressed due to the mediator mass.  Depending on the interaction range, the potential receives contributions from nearby bodies --- if the mediator is heavy --- or also from distant bodies --- if the mediator is light.  We consider masses of $10^{-35}$--$10^{-10}$~eV, for which the interaction range is km--Gpc, which encompasses the Earth ($\oplus$), Moon ($\leftmoon$), Sun ($\astrosun$), Milky Way (MW) and the cosmological matter distribution (cos).  Thus, the potential experienced by neutrinos is the sum of the contributions sourced by each of them, \ie,
\begin{eqnarray}
 \label{equ:pot_total}
 V_{\alpha \beta} = V_{\alpha \beta}^\oplus + V_{\alpha \beta}^{\leftmoon} + V_{\alpha \beta}^{\astrosun} + V_{\alpha \beta}^{\rm MW} + \langle V_{\alpha \beta}^{\rm cos} \rangle \;,
\end{eqnarray}
and the relative contribution of each term depends on the value of $m_{\alpha\beta}^\prime$.  

We do not compute the changing potential along the underground trajectories of the neutrinos inside the Earth or inside the Sun; see \Refe~\cite{Coloma:2020gfv} for such treatment. Instead, like \Refe~\cite{Bustamante:2018mzu}, we compute the average potential experienced by the neutrinos at their point of detection in IceCube.  This approximation is especially valid for mediators lighter than about $10^{-18}$~eV, for which the interaction range is longer than the Earth-Sun distance (see \figu{limits_3models}), and so all of the electrons and neutrons on the Earth or the Sun contribute to the potential experienced by a neutrino regardless of its position inside them.  Below $10^{-18}$~eV is also where we place limits in an unexplored range of mediator mass.

In what follows, we use \equ{pot_general} as a basis to calculate the potential induced by these sources. We adopt the methods introduced in \Refe~\cite{Bustamante:2018mzu} for the $L_e-L_\mu$ and $L_e-L_\tau$ cases.  Below, we revisit them, extend them to the $L_\mu-L_\tau$ case, and introduce refinements in the computation of the potential due to solar and cosmological electrons and neutrons. Throughout, we assume that matter is electrically neutral, so that the number of electrons and protons is the same, and that matter is isoscalar, so that the number of electrons and neutrons is the same, except for the Sun and for the cosmological distribution of matter.

\paragraph{The Earth.}
Astrophysical neutrinos travel inside the Earth from its surface to IceCube, located $d_{\rm IC} = 1.5$~km underground at the South Pole.  Along the way, they undergo long-range interactions with underground electrons or neutrons.  Under the $L_\alpha - L_\beta$ symmetry, the potential sourced by them is
\begin{equation}
 \label{equ:pot_earth}
 V_{\alpha \beta}^\oplus
 =
 \frac{\mathcal{G}_{\alpha\beta}}{2} 
 \int_0^\pi d\theta_z \int_0^{r_{\max}(\theta_z)} 
 dr ~r 
 \sin \theta ~e^{-m_{\alpha \beta}^{\prime} r}
 \times
 \left\{
  \begin{array}{lll}
   \langle n_{e,\oplus} \rangle_{\theta_z} & , & {\rm for}~\alpha, \beta = e, \mu ~{\rm or}~ e, \tau \\
   \langle n_{n,\oplus} \rangle_{\theta_z} & , & {\rm for}~\alpha, \beta = \mu, \tau \\
  \end{array}
 \right. \;,
\end{equation}
where $\theta_z$ is the zenith angle along which the neutrinos travel, measured from the South Pole, the chord length traveled along this direction is $r_{\max}(\theta_z) = (R_\oplus-d_{\rm IC}) \cos \theta_z + \left[ (R_\oplus-d_{\rm IC})^2 \cos^2 \theta_z + (2R_\oplus-d_{\rm IC}) d_{\rm IC} \right]^{1/2}$, and the radius of the Earth is $R_\oplus = 6371$~km.  The average densities of electrons and neutrons along direction $\theta_z$ are $\langle n_{e,\oplus} \rangle_{\theta_z} = \langle n_{n,\oplus} \rangle_{\theta_z}$, assuming matter is electrically neutral and isoscalar.  To compute them, we adopt the matter density profile of the Preliminary Reference Earth Model~\cite{Dziewonski:1981xy}, and we assume an electron fraction of $Y_e \equiv N_e / (N_e + N_p) = 0.5$.  The total number of electrons and neutrons in the Earth is $N_{e, \oplus} \approx N_{n, \oplus} \sim 4 \times 10^{51}$.  Because we do not track the propagation of neutrinos inside the Earth and the changing long-range potential that they experience at different points along their trajectories, our treatment is approximate; yet, it allows us explore efficiently the parameter space.  For a detailed treatment, see \Refe~\cite{Coloma:2020gfv}.

\paragraph{The Moon and the Sun.}
We treat the Moon and the Sun as point sources of electrons and neutrons.  Under the $L_\alpha-L_\beta$ symmetry, the potential sourced by the Moon is
\begin{equation}
 \label{equ:pot_moon}
 V^{\leftmoon}_{\alpha \beta}
 =
 -\mathcal{G}_{\alpha\beta}
 \frac{e^{-m_{\alpha \beta}^\prime d_{\leftmoon}}} { 4\pi d_{\leftmoon}}
 \times
 \left\{
  \begin{array}{lll}
   N_{e,\leftmoon} & , & {\rm for}~\alpha, \beta = e, \mu ~{\rm or}~ e, \tau \\
   N_{n,\leftmoon} & , & {\rm for}~\alpha, \beta = \mu, \tau \\
  \end{array}
 \right. \;,
\end{equation}
where the distance between the Earth and the Moon is $d_{\leftmoon} \approx 4 \cdot 10^5$~km, and the number of electrons and neutrons in the Moon is $N_{e,\leftmoon} = N_{n,\leftmoon}\sim 5 \cdot 10^{49}$, assuming the lunar matter is electrically neutral and isoscalar.  Similarly, the potential sourced by the Sun is
\begin{equation}
 V^{\astrosun}_{\alpha \beta} 
 =
 -\mathcal{G}_{\alpha\beta}
 \frac{e^{-m_{\alpha \beta}^\prime d_{\astrosun}}} { 4\pi d_{\astrosun}}
 \times
 \left\{
  \begin{array}{lll}
   N_{e,\astrosun} & , & {\rm for}~\alpha, \beta = e, \mu ~{\rm or}~ e, \tau \\
   N_{n,\astrosun} & , & {\rm for}~\alpha, \beta = \mu, \tau \\
  \end{array}
 \right. \;,
\end{equation}
where the distance between the Earth and the Sun is $d_{\astrosun} =1$~A.U., the number of electrons in the Sun is $N_{e,\astrosun} \sim 10^{57}$, and the number of neutrons in it is $N_{n,\astrosun} = N_{e,\astrosun}/4$.  Like for Earth, we do not compute the changing long-range matter potential inside the Moon and Sun as neutrinos propagate inside them; for the latter, see \Refe~\cite{Coloma:2020gfv}.

\paragraph{The Milky Way.}
A neutrino of extragalactic origin traverses the Milky Way before reaching the Earth and may be affected by the long-range potential sourced by Galactic electrons and neutrons.  We do not track the propagation of neutrinos inside the Milky Way.  Instead, we estimate the effect of long-range interactions by computing the potential experienced by neutrinos at the location of the Earth by integrating the Galactic electron and neutron column densities across all possible neutrino trajectories that have the Earth as the endpoint.  Under the $L_\alpha - L_\beta$ symmetry, this is
\begin{equation}
 \label{equ:pot_mw}
 V_{\alpha \beta}^{\rm MW}
 =
 \frac{\mathcal{G}_{\alpha\beta}}{4\pi} 
 \int_0^\infty
 dr 
 \int_0^\pi d\theta \int_0^{2\pi} 
 d\phi 
 ~r 
 \sin\theta 
 ~e^{-m_{\alpha \beta}^{\prime} r}
 \times
 \left\{
  \begin{array}{lll}
   n_{e, {\rm MW}}(r, \theta, \phi) & , & {\rm for}~\alpha, \beta = e, \mu ~{\rm or}~ e, \tau \\
   n_{n, {\rm MW}}(r, \theta, \phi) & , & {\rm for}~\alpha, \beta = \mu, \tau \\
  \end{array}
 \right. \;,
\end{equation}
where the coordinate system is centered at the position of the Earth, 8.33~kpc away from the Galactic Center, and the densities of electrons and neutrons are $n_{e, {\rm MW}} = n_{n, {\rm MW}}$, assuming matter is electrically neutral and isoscalar.  In the Milky Way, $N_{e, {\rm MW}} \approx N_{n, {\rm MW}} \sim 10^{67}$ electrons and neutrons are contained in stars and cold gas --- distributed in a central bulge, a thick disc, and a thin disc --- and in hot gas --- distributed in a diffuse halo.  Following \Refe~\cite{Bustamante:2018mzu}, we adopt the ``conventional model'' from \Refe~\cite{McMillan:2011wd} for the matter density of the central bulge, thick disc, and thin disc, and the spherical saturated matter density from \Refe~\cite{Miller:2013nza} for the diffuse halo; Fig.~A1 in \Refe~\cite{Bustamante:2018mzu} shows the total matter distribution.

\paragraph{Cosmological electrons and neutrons.}
For $m_{\alpha\beta}^\prime \lesssim 10^{-25}$~eV, the dominant contribution to the long-range potential is from the large-scale cosmological distribution of electrons and neutrons.  We follow \Refe~\cite{Bustamante:2018mzu} to compute the potential, including the effect of the adiabatic cosmological expansion on the densities of electrons and neutrons; we defer to it for a derivation of the potential.  While \Refe~\cite{Bustamante:2018mzu} assumed that the cosmological distribution is isoscalar, we instead account for the fact that, after the recombination epoch, the neutron-to-proton ratio saturates to about $1/7$~\cite{Steigman:2007xt}.  Using this and assuming that cosmological matter is electrically neutral, the number of cosmological electrons is seven times higher than that of neutrons.  Under the $L_\alpha - L_\beta$ symmetry, the potential at redshift $z$ is
\begin{eqnarray}
 \label{equ:pot_cos_z}
 V_{\alpha \beta}^{\rm cos}(z)
 &=&
 \frac{3}{4\pi} 
 \frac{\mathcal{G}_{\alpha\beta}}{m_{\alpha \beta}^{\prime~2} d_{\rm H}^3(z)} 
 \left\{ 
 1 - e^{-m_{\alpha \beta}^\prime d_{\rm H}(z)} [1+m_{\alpha \beta}^\prime d_{\rm H}(z)] 
 \right\} 
 \nonumber \\
 &&
 \qquad
 \times
 \left\{
  \begin{array}{lll}
   N_{e, {\rm cos}}(z) & , & {\rm for}~\alpha, \beta = e, \mu ~{\rm or}~ e, \tau \\
   N_{n, {\rm cos}}(z) & , & {\rm for}~\alpha, \beta = \mu, \tau \\
  \end{array}
 \right. \;,
\end{eqnarray}
where $N_{e, {\rm cos}} \left( z \right) \simeq 7 M_{\rm H}(z) / (8 m_p + 7 m_e)$ is the number of electrons, $m_p$ and $m_e$ are the proton and electron mass, respectively, $N_{n, {\rm cos}}(z) = N_{e, {\rm cos}}(z)/7$ is the number of neutrons, $M_{\rm H}$ is the baryonic mass inside casual horizon (see Eq.~(16.105) in \Refe~\cite{Giunti:2007ry}), and $d_{\rm H}$ is the size of the causal horizon, i.e., the radius of the largest sphere centered on the Earth within which events can be causally connected~\cite{Weinberg:2008zzc}.  We adopt a $\Lambda$CDM cosmology with Hubble constant $H_0 = 100 h$ km s$^{-1}$ Mpc$^{-1}$, where $h = 0.673$~\cite{ParticleDataGroup:2014cgo}, the vacuum energy density $\Omega_\Lambda = 0.692$ and the matter density $\Omega_{\rm M} = 0.308$~\cite{Planck:2015fie}.  Because we use the diffuse flux of high-energy astrophysical neutrinos, we account for the evolution with redshift of the number density of neutrino sources, $\rho_{\rm src}$, by averaging the  potential over $z$, \ie,
\begin{equation}
 \langle V_{\alpha \beta}^{\rm cos} \rangle \propto
 \int dz~ 
 \rho_{\rm src}(z) 
 \frac{dV_{\rm c}}{dz} 
 V_{\alpha \beta}^{\rm cos}(z) \;,
\end{equation}
where $V_{\rm c}$ is the comoving volume~\cite{Hogg:1999ad} and  $\rho_{\rm src}$ follows the star formation rate~\cite{Hopkins:2006bw, Yuksel:2008cu, Kistler:2009mv}.

\medskip

\begin{figure}[t!]
 \centering
 \includegraphics[width=.49\textwidth]{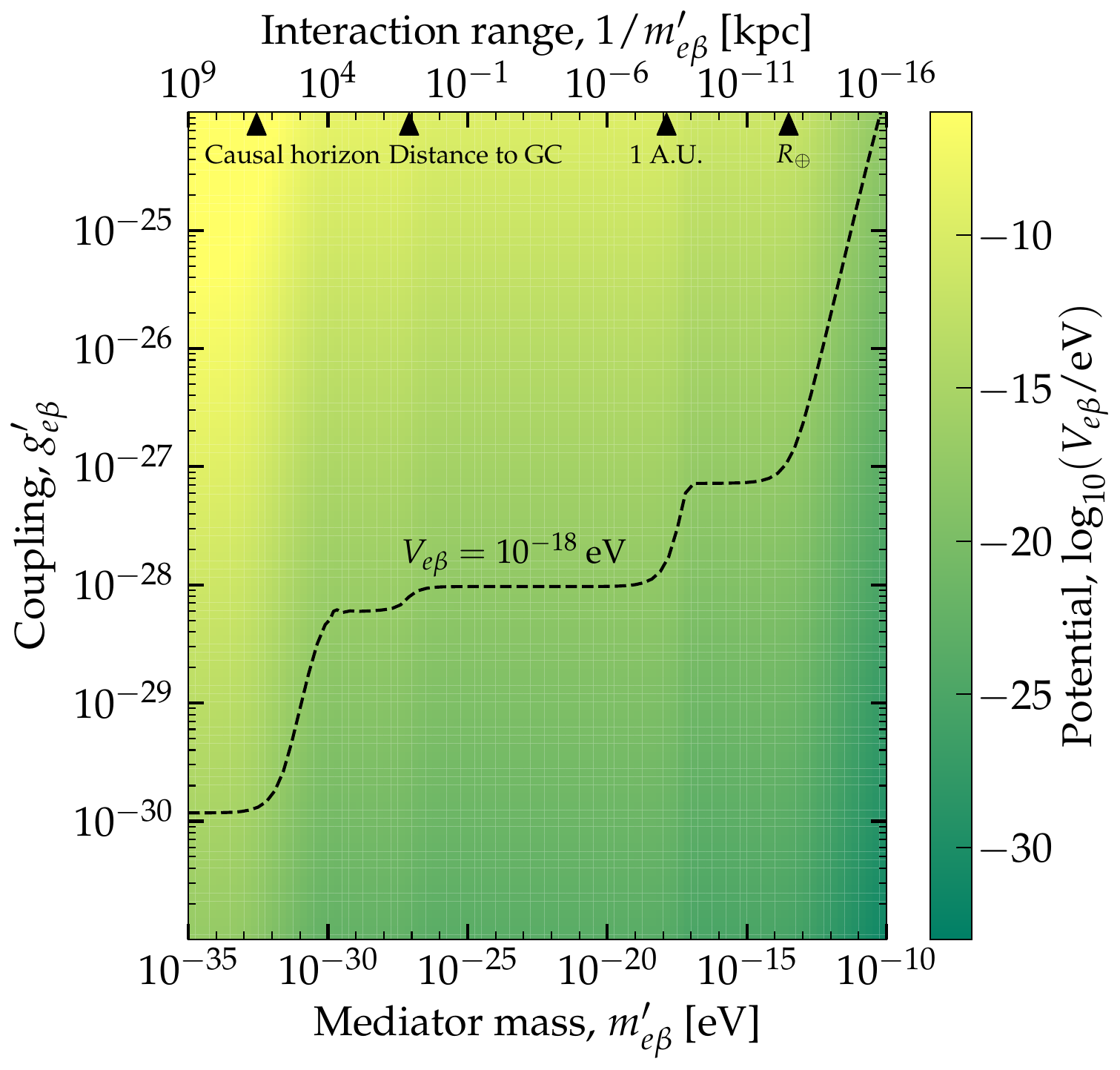}
 \includegraphics[width=.49\textwidth]{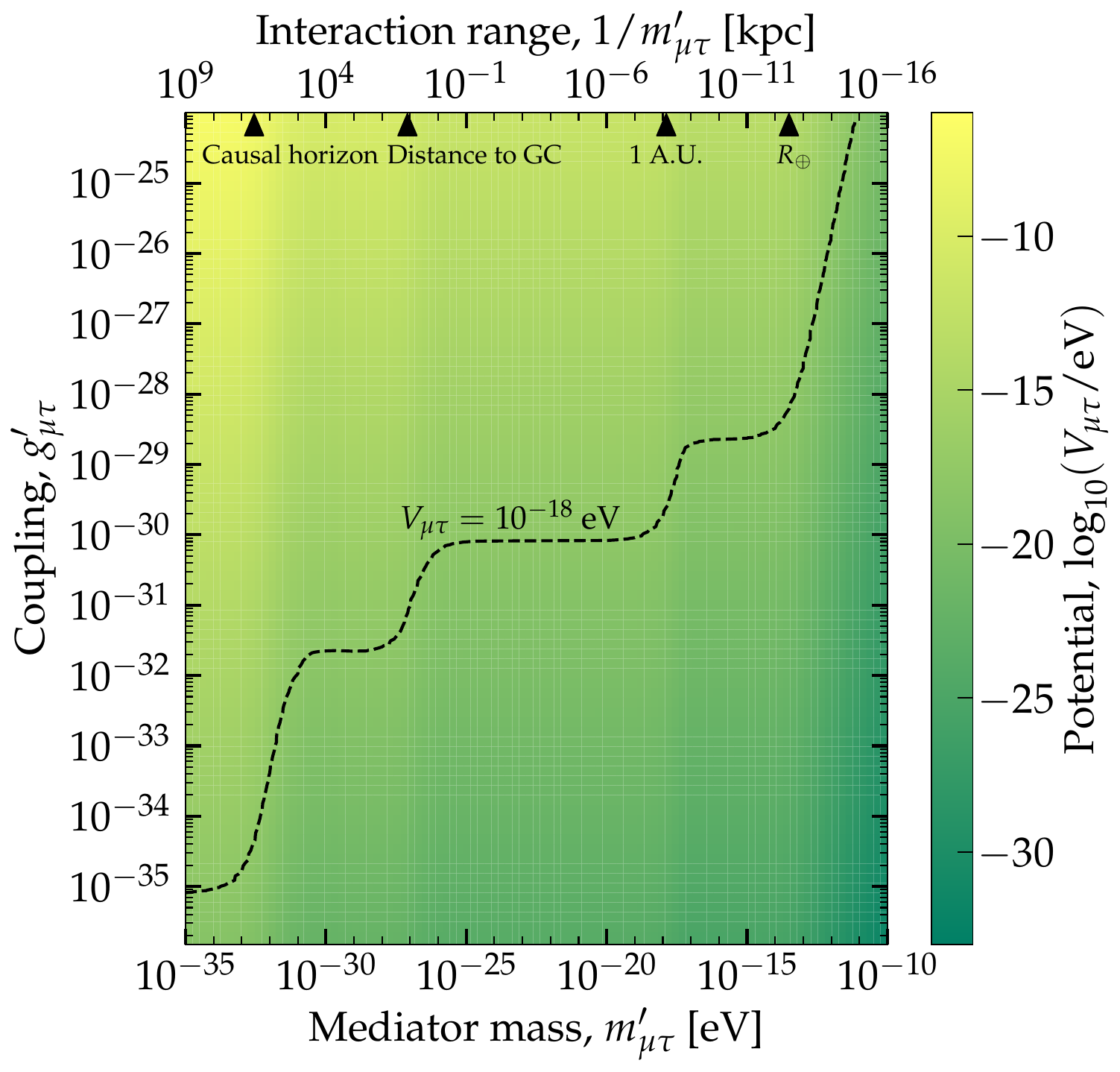}
 \caption{\textbf{\textit{Long-range matter potential, $V_{\alpha \beta}$, experienced by a neutrino at Earth, \equ{pot_total}.}}  The potential is sourced by electrons and neutrons in the Earth, Moon, Sun, Milky Way, and the distant Universe as a function of the mass and adimensional coupling of the new $U(1)$ gauge symmetry.  \textit{Left:} For the $L_e-L_{\beta}$ symmetries ($\beta = \mu, \tau$), sourced by electrons. \textit{Right:} For the $L_{\mu}-L_{\tau}$ symmetry, sourced by neutrons. Isocontours are drawn at the illustrative values of $V_{e\beta} = V_{\mu\tau} =  10^{-18}$~eV.  (For the $L_{\mu}-L_{\tau}$ symmetry, values of the coupling run lower due to the mixing between the new mediator and the SM $Z$ boson.)  The step-like transitions in the potential signal the onset of contributions from sources of electrons and neutrons located at different distances relative to the interaction range.  See Section~\ref{subsec:LRI} for details.}
 \label{fig:potential}
\end{figure}

Figure~\ref{fig:potential} shows the total potential, \equ{pot_total}, as a function of the mediator mass and coupling, for the $L_e-L_\beta$ ($\beta = \mu, \tau$) and $L_\mu-L_\tau$ symmetries.  To illustrate its behavior, we show the isocontour for the illustrative value of $V_{\alpha \beta} = 10^{-18}$~eV (which is close to the final constraint that we obtain in Section~\ref{sec:results}). As the mass shrinks, the interaction range grows.  As a result, the potential isocontour undergoes several step-like transitions, each of which represents the inclusion of the contribution of a more distant collection of electrons and neutrons into the total potential. From $m_{\alpha\beta}^\prime \sim 10^{-10}$~eV to $10^{-18}$~eV, the potential is dominated by the Earth, with a minor contribution from the Moon. The step at $m_{\alpha\beta}^\prime \sim 10^{-18}$~eV represents the inclusion of the potential sourced by the Sun, which becomes dominant. Because the Sun contains far more electrons and neutrons than the Earth and the Moon, a smaller coupling is enough to achieve the same potential.  The step at $m_{\alpha\beta}^\prime \sim 10^{-27}$~eV represents the onset of dominance of Milky-Way electrons and neutrons, which far outnumber those in the Sun, and are concentrated in the Galactic Center. The last step, at $m_{\alpha\beta}^\prime \sim 10^{-33}$~eV, represents the onset of the dominance of cosmological distribution of electrons and neutrons, which far outnumber those in the Milky Way.  

For the $L_\mu-L_\tau$ case, \figu{potential} shows that smaller couplings are needed to achieve the same illustrative value of the potential because it scales $\propto g_{\mu\tau}^\prime$, rather than $\propto g_{e\beta}^{\prime~2}$, as in the $L_e-L_\mu$ and $L_e-L_\tau$ cases; see \equ{Gab}.  In our work, including in \figu{potential}, for the $L_\mu-L_\tau$ case we fixed the $Z_{\mu\tau}^\prime$--$Z$ mixing strength, $(\xi-\sin \theta_W \chi)$, to its maximum allowed value (see Section~\ref{subsec:models}).  Decreasing or increasing the mixing strength would shift the potential isocontour in \figu{potential} up or down, respectively.

\subsection{Neutrino flavor-transition probabilities}
\label{subsec:transition-prob}

In the presence of the new $L_\alpha - L_\beta$ gauge symmetry, the Hamiltonian that drives neutrino propagation, written in the flavor basis, is
\begin{equation}
 \label{equ:hamiltonian_tot}
 \mathbf{H}
 =
 \mathbf{H}_{\rm vac}
 +
 \mathbf{V}_{\rm mat}
 +
 \mathbf{V}_{\alpha\beta} \;.
\end{equation}
The contributions on the right-hand side describe, respectively, oscillations in vacuum, standard neutrino-matter interactions, and the new neutrino-matter interactions.

\begin{figure}[t!]
 \centering
 \includegraphics[width=\textwidth]{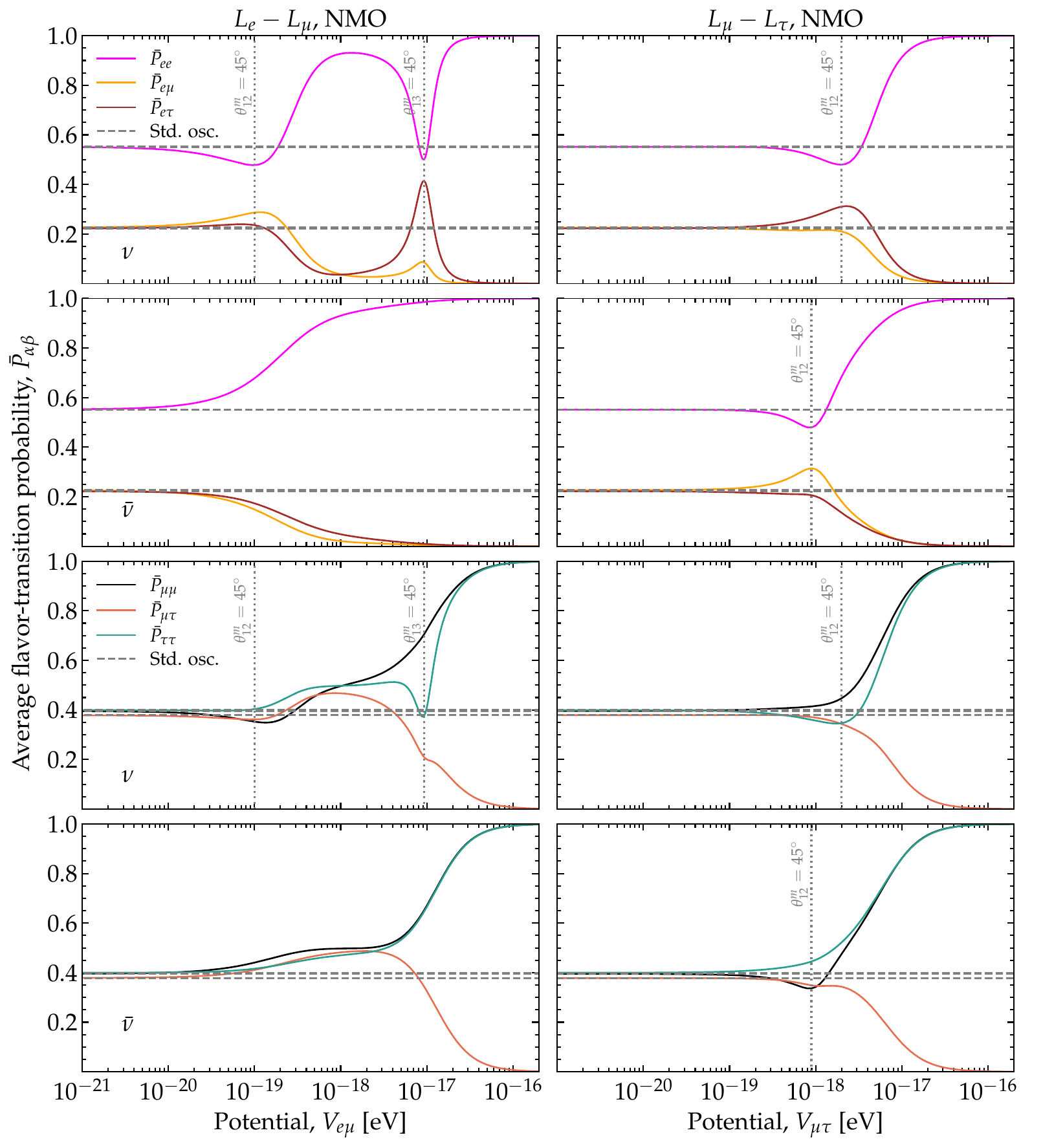}
 \caption{\textit{\textbf{Average flavor-transition probabilities, \equ{prob_avg}, as functions of the new matter potential induced by the $U(1)$ gauge symmetries $L_e-L_\mu$ (left column) and $L_\mu-L_\tau$ (right column).}}  Probabilities are for neutrinos and antineutrinos, at a fixed energy of 100~TeV, assuming normal mass ordering (NMO).  The features in the probabilities are due to resonances induced by the new potential. Vertical lines mark the values of the potential for which $\theta_{12}^m$ and $\theta_{13}^m$ become resonant.  Probabilities under standard oscillations (``Std.~osc.''), \ie, for $V_{\alpha\beta} = 0$, are shown for comparison. The mixing parameters are fixed at their present-day best-fit values from NuFit 5.1~\cite{Esteban:2020cvm, NuFIT}.  Appendix~\ref{app:results_le_ltau} contains results for $L_e-L_\tau$; they are similar to $L_e-L_\mu$. Appendix~\ref{appendix_e} contains results assuming inverted mass ordering.  See Section~\ref{subsec:transition-prob} for details.}
 \label{fig:prob_lemmt_nmo}
\end{figure}

Oscillations in vacuum are driven by
\begin{equation}
 \mathbf{H}_{\rm vac}
 =
 \frac{1}{2 E}
 \mathbf{U}~
 {\rm diag}(0, \Delta m^2_{21}, \Delta m^2_{31})
 ~\mathbf{U}^{\dagger} \;,
\end{equation}
where $E$ is the neutrino energy, $\Delta m^2_{31} \equiv m^2_3-m^2_1$, and $\Delta m^2_{21} \equiv m^2_2-m^2_1$ are the atmospheric and solar mass-squared splittings, respectively, $m_i$ the mass of the $\nu_i$ mass eigenstate, and $\mathbf{U}$ is the Pontecorvo-Maki-Nakagawa-Sakata (PMNS) mixing matrix, parameterized as a product of three rotation matrices, \ie, $R(\theta_{23}) R(\theta_{13}, \delta_{\text{CP}}) R(\theta_{12})$, where $\theta_{ij}$ ($i,j = 1,2,3$) are three mixing angles and $\delta_{\text{CP}}$ is the Dirac CP-violating phase.  Under normal mass ordering (NMO), where $m_1<m_2<m_3$, the present-day best-fit values of the mixing parameters from the global oscillation analysis of \Refes~\cite{Esteban:2020cvm, NuFIT} are: $\theta_{23} = 42.1^\circ$, $\theta_{13} = 8.62^\circ$, $\theta_{12} = 33.45^\circ$, $\delta_{\text{CP}}=230^\circ$, $\Delta m^2_{31} = 2.51\times10^{-3}$~eV$^2$, and $\Delta m^2_{21} = 7.42\times10^{-5}$~eV$^2$.  Below, in Figs.~\ref{fig:prob_lemmt_nmo} and \ref{fig:prob_let_nmo}, we adopt these values to illustrate the effect of the new interaction on the neutrino flavor-transition probabilities.  Later, in our statistical analysis, we let their values float within their present-day and future predicted allowed ranges.

In \equ{hamiltonian_tot}, the contribution of the SM potential from coherent forward neutrino scattering on electrons is
\begin{equation}
 \mathbf{V}_{\rm mat}
 =
 {\rm diag}(V_{\rm CC}, 0, 0) \;,
\end{equation}
where $V_{\rm CC} = \sqrt{2}G_{F} \langle n_{e, \oplus} \rangle_{\theta_z}$ is the charged-current neutrino-electron interaction potential and $G_F$ is the Fermi constant.  In our calculations, this contribution is relevant only inside Earth, where matter densities are high.  The potential above is for neutrinos; for antineutrinos, it flips sign, \ie, $\mathbf{V}_{\rm mat} \to -\mathbf{V}_{\rm mat}$.

Finally, in \equ{hamiltonian_tot} the contribution from the new matter interaction is
\begin{equation}
 \label{equ:pot_lri_matrix}
 \mathbf{V}_{\alpha\beta}
 =
 \left\{
  \begin{array}{ll}
   {\rm diag}(V_{e\mu}, -V_{e\mu}, 0), & {\rm for}~ \alpha, \beta = e, \mu \\
   {\rm diag}(V_{e\tau}, 0, -V_{e\tau}), & {\rm for}~ \alpha, \beta = e, \tau \\
   {\rm diag}(0, V_{\mu\tau}, -V_{\mu\tau}), & {\rm for}~ \alpha, \beta = \mu, \tau \\   
  \end{array}
 \right. \;,
\end{equation}
where $V_{\alpha\beta}$ is the potential, calculated using \equ{pot_total}.  Depending on the mediator mass, it is sourced by nearby and faraway sources of electrons and neutrons.  The potential above is for neutrinos; for antineutrinos, it flips sign, \ie, $\mathbf{V}_{\alpha\beta} \to -\mathbf{V}_{\alpha\beta}$.

The $\nu_\alpha \to \nu_\beta$ flavor-transition probability associated to the Hamiltonian, \equ{hamiltonian_tot}, is
\begin{equation}
 \label{eq:osc_prob_1}
 P_{\alpha\beta}
 =
 \left\vert
 \sum^3_{i=1} U^m_{\alpha i}
 \exp\left[-\frac{\Delta \Tilde{m}^2_{i1}L}{2E}\right]U^{m^{\ast}}_{\beta i}
 \right\vert^2 \;,
\end{equation}
where $L$ is the distance traveled by the neutrino from its point of production to the Earth, $\Delta \tilde{m}^2_{ij} \equiv \tilde{m}_i^2 - \tilde{m}_j^2$, with $\tilde{m}^2_i/2E$ the eigenvalues of the Hamiltonian, and $\mathbf{U}^m$ is the matrix that diagonalizes the Hamiltonian, parameterized like the PMNS matrix, but in terms of new mixing parameters $\theta_{23}^m$, $\theta_{13}^m$, $\theta_{12}^m$,  $\delta_{\rm CP}^m$.  In Appendix~\ref{app:evol_mix_angles}, we show the evolution of the three modified mixing angles with the long-range potential.  The flavor-transition probability is oscillatory, but for high-energy neutrinos that travel cosmological-scale distances, like the ones we consider, the oscillations are rapid, \ie, $\Delta \Tilde{m}^2_{ij} L / (2E) \gg 1$. Given the limited energy resolution of neutrino telescopes, they are only sensitive to the average probability,
\begin{equation}
 \label{equ:prob_avg}
 \bar{P}_{\alpha\beta}
 =
 \sum^3_{i=1}|U^m_{\alpha i}|^2|U^m_{\beta i}|^2 \;.
\end{equation}
We use this expression to produce all our results below.  Under standard oscillations, \ie, when $V_{\alpha\beta} = 0$, the average oscillation probability loses its dependence on neutrino energy. In contrast, in the presence of the long-range interactions, it retains the energy dependence via the interplay of the vacuum and potential contributions to the Hamiltonian, \equ{hamiltonian_tot}.

Figure~\ref{fig:prob_lemmt_nmo} shows average oscillation probabilities as functions of the potential $V_{\alpha\beta}$, computed for a fixed neutrino energy of $100$~TeV, representative of high-energy astrophysical neutrinos.  In what follows, we focus on the probabilities under the $L_e-L_\mu$ symmetry. Figure~\ref{fig:prob_lemmt_nmo} shows that for low values of the potential, \ie, $V_{\alpha\beta} \ll \Delta m_{ij}^2 / (2 E)$, oscillations are standard.  Deviations start to appear when the potential becomes comparable to the Hamiltonian in vacuum, \ie, $V_{\alpha\beta} \approx 10^{-20}$~eV. For neutrinos, sharp features --- dips or peaks --- appear at $V_{e\mu} \approx 10^{-17}$~eV in the $\nu_e\to\nu_\beta$ ($\beta = e,\mu,\tau$) probabilities, reflecting that $\theta_{13}^m \approx 45^\circ$ (see Appendix~\ref{app:evol_mix_angles}), which maximizes $U_{e3}^m$ and makes flavor transitions resonant. At $V_{e\mu} \approx 10^{-19}$~eV, less prominent features reflect that $\theta_{12}^m = 45^\circ$; they are less prominent because in vacuum $\theta_{12} \approx 45^\circ$ already. 
Similar features appear in the $\nu_\mu \to \nu_\tau$ and $\nu_\tau \to \nu_\tau$ probabilities, for analogous reasons.  For large values of the potential, \ie, $V_{\alpha\beta} \gg 10^{-17}$~eV, the term $\mathbf{V}_{\alpha\beta}$ dominates the Hamiltonian and, because it is diagonal, it suppresses flavor transitions. 
For $L_e-L_\tau$, shown in Fig.~\ref{fig:prob_let_nmo}, the flavor-transition probabilities are similar to $L_e-L_\mu$.  Figure~\ref{fig:prob_lemmt_nmo} shows that, for $L_\mu-L_\tau$, no resonance due to $\theta_{13}^m$ occurs neither for neutrinos nor antineutrinos because $\theta_{13}^m$ never reaches $45^\circ$, but there is a small dip or jump in the probabilities around $V_{\mu\tau}\approx 10^{-18}$~eV, due to resonant $\theta_{12}^m$.  

In the main text, we produce results assuming normal neutrino mass ordering, motivated by recent hints from global oscillation fits~\cite{deSalas:2020pgw, Capozzi:2021fjo, Esteban:2020cvm, NuFIT, Gonzalez-Garcia:2021dve}.  For antineutrinos, the sharp features above do not appear because $\theta_{13}^m$ and $\theta_{12}^m$ never become resonant under normal neutrino ordering.  However, they do appear under inverted mass ordering, because $\Delta m_{31}^2$ is negative; see Appendix~\ref{appendix_e}.


\section{High-energy astrophysical neutrinos}
\label{sec:he_nu}


\subsection{Overview}
\label{sec:he_nu_overview}

High-energy astrophysical neutrinos, discovered by IceCube in 2013~\cite{Aartsen:2013bka, Aartsen:2013jdh}, have the highest neutrino energies detected so far, between tens of TeV to a few PeV~\cite{IceCube:2020wum, IceCube:2021uhz}.  The origin of the bulk of them --- which makes up their diffuse flux --- remains unknown~\cite{Ahlers:2018fkn, Ackermann:2019ows, Ackermann:2022rqc}, although a handful of promising candidate sources have been identified~\cite{IceCube:2018dnn, Stein:2020xhk, IceCube:2021xar, IceCube:2022der}.  In our analysis, we consider exclusively the diffuse neutrino flux, since only for it are measurements of the flavor composition available~\cite{IceCube:2015rro, IceCube:2015gsk, Aartsen:2018vez, IceCube:2020fpi}.  These neutrinos are likely made in extragalactic cosmic-ray accelerators~\cite{Anchordoqui:2013dnh, Halzen:2019qkf, AlvesBatista:2019tlv, Ackermann:2022rqc}.  Possible candidate source populations~\cite{Anchordoqui:2013dnh, AlvesBatista:2019tlv} include active galactic nuclei~\cite{Murase:2022feu}, galaxy clusters~\cite{Berezinsky:1996wx, Murase:2008yt, Kotera:2009ms, Murase:2013rfa, Fang:2017zjf, Hussain:2021dqp}, gamma-ray bursts~\cite{Paczynski:1994uv, Waxman:1997ti, Murase:2006mm, Bustamante:2014oka, Senno:2015tsn, Pitik:2021xhb, Guarini:2021gwh},  starburst galaxies~\cite{Loeb:2006tw, Thompson:2006qd, Stecker:2006vz, Tamborra:2014xia, Palladino:2018bqf, Peretti:2018tmo, Peretti:2019vsj, Ambrosone:2020evo}, supernovae~\cite{Senno:2017vtd, Esmaili:2018wnv, Sarmah:2022vra}, and tidal disruption events~\cite{Wang:2011ip, Dai:2016gtz, Senno:2016bso, Lunardini:2016xwi, Zhang:2017hom, Winter:2020ptf}, among others.  In them, protons and other charged nuclei might be accelerated to energies of at least tens of PeV --- and possibly much higher --- within magnetized environments by means of collisionless shocks or other mechanisms~\cite{AlvesBatista:2019tlv}.

Inside the sources, the high-energy protons interact with surrounding matter, in proton-proton ($pp$) collisions~\cite{Margolis:1977wt, Stecker:1978ah, Kelner:2006tc}, or radiation, in proton-photon ($p\gamma$) collisions~\cite{Stecker:1978ah, Mucke:1999yb, Murase:2005hy, Hummer:2010vx}.  These generate high-energy pions that, upon decaying, make neutrinos, \ie, $\pi^+ \to \mu^+ + \nu_\mu$, followed by $\mu^+ \to \bar{\nu}_\mu + e^+ + \nu_e$, and their charge-conjugated processes.  (At higher energies, the decay of kaons and other production channels become important~\cite{Mucke:1999yb, Murase:2005hy, Hummer:2010vx, Hummer:2011ms, Morejon:2019pfu}.)  Each neutrino carries, on average, 5\% of the energy of the parent proton~\cite{Mucke:1999yb, Kelner:2006tc}.  In production via $pp$ collisions, the neutrino spectrum is a power law $\propto E^{-\gamma}$ that follows that of the parent protons, with the value of $\gamma$ loosely expected to be in the interval [2,3].  In production via $p\gamma$ collisions, its shape depends on those of the parent protons and photons, and it is bump-like, on account of the photon spectrum peaking at a characteristic energy.  Because the diffuse neutrino flux results from the addition of the emissions of all neutrino sources, the above features in the neutrino spectrum are softened.  So far, IceCube observations are described well by an unbroken power with the value of $\gamma$ dependent on the data set used and with little to no preference for alternative spectral shapes~\cite{IceCube:2020wum, IceCube:2021uhz}, though this may change with further observations~\cite{Fiorillo:2022rft}.


\subsection{Neutrino detection and flavor composition}
\label{sec:he_nu_detection}

Presently, IceCube is the largest neutrino telescope~\cite{IceCube:2016zyt, Ahlers:2018fkn}: it consists of 1~km$^3$ of deep Antarctic ice instrumented with photomultipliers.  In it, high-energy neutrinos interact with nucleons in the ice, mainly via deep inelastic scattering~\cite{IceCube:2017roe, Bustamante:2017xuy, Aartsen:2018vez, IceCube:2020rnc}, either neutral-current (NC), \ie, $\nu_l + N \to \nu_l + X$, where $l = e, \mu, \tau$ and $X$ are final-state hadrons, or charged-current (CC), \ie, $\nu_l + N \to l + X$.  Final-state charged particles shower and emit Cherenkov radiation, which propagates through the ice and is detected by the photomultipliers.  From the number of photons detected and from their spatial and temporal distributions, IceCube reconstructs the energy, direction, and flavor of the interacting neutrino~\cite{IceCube:2013dkx, IceCube:2016zyt}.

Broadly stated, IceCube detects three types of neutrino-induced events: tracks, cascades, and double cascades.  Tracks are made mainly by CC interactions of $\nu_\mu$, where the final-state hadrons initiate a shower around the interaction vertex, and the final-state muon leaves a track of Cherenkov light in its wake, several kilometers in length and easily identifiable.  Tracks are also made in 17\% of $\nu_\tau$ CC interactions where the final-state tauon decays into a muon~\cite{ParticleDataGroup:2022pth}.  Cascades are made by CC interactions of $\nu_e$ and $\nu_\tau$, where both the final-state lepton and hadrons shower around the interaction vertex, and by NC interactions of neutrinos of all flavors, where only the final-state hadrons shower.  Double cascades are made by CC interactions of $\nu_\tau$ in which the final-state hadrons make a first shower, centered around the interaction vertex. The final-state tauon is energetic enough to decay some distance away, generating a second shower~\cite{Learned:1994wg, Athar:2000rx, IceCube:2020fpi}.

{\it Starting events} are those where the neutrino interacts within the detector volume.  Because a large fraction of the neutrino energy is deposited in the ice, starting events trace neutrino energy closely~\cite{IceCube:2013dkx}.  A subset of them above 60~TeV are known as High-Energy Starting Events (HESE)~\cite{IceCube:2020wum}: they are subjected to a self-veto~\cite{Schonert:2008is, Gaisser:2014bja} that reduces the contamination of atmospheric neutrinos and muons and, as a result, they have the highest content of neutrinos of astrophysical origin.  However, its detection rate is relatively low, of approximately 10 events per year~\cite{IceCube:2020wum}.  {\it Through-going tracks} are those where $\nu_\mu$ (and $\nu_\tau$) interact outside the detector and produce muon tracks that cross part of it~\cite{IceCube:2021uhz}.  Their detection rate is orders-of-magnitude higher than for HESE, but, because they have lower energies, they are dominated by atmospheric neutrinos and muons, made in cosmic-ray interactions in the atmosphere.  Further, for tracks, because the location of the neutrino interaction is unknown, the neutrino energy can only be inferred uncertainly~\cite{IceCube:2013dkx}.  In our analysis below, we exploit the combined capabilities of HESE and through-going events for flavor measurements, motivated by \Refe~\cite{IceCube:2015gsk}.

Because different event types can be made by more than one neutrino flavor, a straightforward, event-by-event correspondence between detected event type and neutrino flavor is typically unfeasible, except for double cascades, which are only made by $\nu_\tau$.  (In addition, showers made by the Glashow resonance~\cite{Glashow:1960zz}, recently discovered by IceCube~\cite{IceCube:2021rpz}, are triggered exclusively by the interaction of 6.3-PeV $\bar{\nu}_e$ with electrons~\cite{Bhattacharya:2011qu, Bhattacharya:2012fh, Biehl:2016psj, Huang:2019hgs}.)  Still, from the relative number of detected events of different types, it is possible to infer statistically the flavor composition of the neutrino flux, \ie, the proportion of $\nu_e$, $\nu_\mu$, and $\nu_\tau$ in it, even if, because of the above limitations, the measurement uncertainties are significant.  Our analysis below rests on the capabilities of IceCube and upcoming neutrino telescopes to infer the flavor composition in such a way, following the same spirit as \Refes~\cite{Song:2020nfh, Schumacher:2021hhm, Fiorillo:2022rft}

The flavor composition can be inferred either using only HESE showers, tracks, and double cascades, or by combining them with through-going muons.  IceCube has reported the flavor composition in several analyses: using the first 3 years of HESE data~\cite{IceCube:2015rro}, a combination of 4 years of contained events plus 2 years of through-going tracks~\cite{IceCube:2015gsk},
5 years of contained events starting at lower energies~\cite{Aartsen:2018vez}, and, most recently, 7.5 years of HESE~data, including a dedicated search for double cascades~\cite{IceCube:2020fpi}.  Independent analyses have reported complementary results; see, \eg, \Refes~\cite{Mena:2014sja, Palomares-Ruiz:2015mka, Vincent:2016nut, Brdar:2018tce, Palladino:2019pid}.  The precision of these analyses is limited by the relative scarcity of HESE events.  Presently, the most precise measurements come from the combined analysis of \Refe~\cite{IceCube:2015gsk}, where the large number of through-going tracks pins down the flavor content of $\nu_\mu$.  Such a combined analysis has not been updated since 2015~\cite{IceCube:2015gsk} in spite of the availability of larger event samples, on account of its complexity.  Our analysis below is based on realistic estimates for what such analysis could look like with present and future data; see Section~\ref{sec:he_nu_analysis_choices}.


\subsection{Flavor composition of high-energy astrophysical neutrinos}
\label{sec:he_nu_flavor_ratios}

Astrophysical sources emit high-energy $\nu_e$, $\nu_\mu$, and $\nu_\tau$ in the proportions $(f_{e, {\rm S}}, f_{\mu, {\rm S}}, f_{\tau, {\rm S}})$, where $f_{\alpha, {\rm S}} \leq 1$ ($\alpha = e, \mu, \tau$) is the ratio of $\nu_\alpha + \bar{\nu}_\alpha$ to the total emitted flux.  We consider three benchmark scenarios of the flavor composition emitted by the sources often studied in the literature (see, \eg, \Refes~\cite{Barenboim:2003jm, Lipari:2007su, Hummer:2010ai, Mehta:2011qb, Bustamante:2015waa, Arguelles:2015dca, Palladino:2015zua, Rasmussen:2017ert, Bustamante:2019sdb, Song:2020nfh}): $(1/3,2/3,0)_{\rm S}$, the canonical expectation from the full pion decay chain (see Section~\ref{sec:he_nu_overview}); $(0,1,0)_{\rm S}$, the muon-damped scenario where the intermediate muons created in pion decay cool by synchrotron radiation before decaying; and $(1,0,0)_{\rm S}$, from the beta-decay of neutrons and isotopes.  Below, we show the effect of the new matter potential in these three scenarios.  Later, in Section~\ref{sec:analysis}, we perform a full statistical analysis only for the full pion decay chain scenario.

For a given choice of flavor composition at the sources, en route to Earth neutrino oscillations modify it into
\begin{equation}
 \label{eq:flavatearth}
 f_{\alpha,\oplus}
 =
 \sum_{\beta=e,\mu,\tau} \bar{P}_{\beta\alpha} f_{\beta, {\rm S}} \;,
\end{equation}
where the average oscillation probability, $\bar{P}_{\beta\alpha}$, is computed using \equ{prob_avg}.  During propagation, the total number of neutrinos is conserved; flavor transitions simply redistribute them into different flavors.  Figure~\ref{fig:flav_ratio} shows the allowed regions of flavor composition at Earth under standard oscillations~\cite{Bustamante:2015waa, Song:2020nfh}, \ie, when the potential $V_{\alpha \beta} = 0$, for the three scenarios, and varying the values of the standard mixing parameters within their present-day $1\sigma$ allowed ranges~\cite{Esteban:2020cvm, NuFIT}.  In the presence of long-range interactions, \ie, when $V_{\alpha\beta}$ is nonzero, because the flavor-transition probabilities are modified, so is the flavor composition at Earth.  In this case, the flavor composition becomes energy-dependent, via the probability, in contrast to the flavor composition computed under standard oscillations.

Figure~\ref{fig:flav_ratio} illustrates the expected flavor composition at Earth in the presence of the new matter potential, under the $L_e-L_\mu$ and $L_\mu-L_\tau$ symmetries, for a representative neutrino energy of $100$~TeV, and for the three benchmark scenarios of flavor composition at the sources.  Results under the $L_e-L_\tau$ symmetry are similar to those under $L_e-L_\mu$; see \figu{flav_ratio_let}.   In \figu{flav_ratio}, we compute $f_{\alpha, \oplus}$ for varying values of the long-range matter potential, $V_{\alpha\beta}$, and of the standard mixing parameters, within their $1\sigma$ allowed ranges~\cite{Esteban:2020cvm, NuFIT}.  For small values of the potential, \ie, $V_{\alpha\beta} \lesssim \Delta m^2_{21}/(2E) \sim 10^{-20}$~eV, the flavor composition is close to the expectation from standard oscillations.  As the potential grows, the behavior of the flavor composition traces that of the flavor-transition probabilities (Figs.~\ref{fig:prob_lemmt_nmo} and \ref{fig:prob_let_nmo}): the wiggles in $f_{\alpha, \oplus}$ reflect the resonant features in the probabilities.  For large values of the potential, \ie, $V_{\alpha\beta} \gtrsim 10^{-16}$~eV, the new matter potential becomes the dominant contribution to the Hamiltonian, \equ{hamiltonian_tot}.  In that case, because the contribution of the new potential is flavor-diagonal, flavor transitions are suppressed (Section~\ref{subsec:transition-prob}).  Accordingly, \figu{flav_ratio} shows that the flavor composition at Earth is the same as at the sources, \ie, $f_{\alpha, \oplus} \approx f_{\alpha, {\rm S}}$.


\subsection{Analysis choices}
\label{sec:he_nu_analysis_choices}

\paragraph{Flavor composition at the sources independent of energy.}  Because different neutrino production channels become available at different energies~\cite{Mucke:1999yb, Kelner:2006tc, Hummer:2010vx, Morejon:2019pfu}, the flavor composition emitted by the sources, $f_{\alpha, {\rm S}}$, may vary with energy~\cite{Kashti:2005qa, Kachelriess:2006fi, Lipari:2007su, Kachelriess:2007tr, Hummer:2010ai, Bustamante:2015waa}, contingent on key source properties like the magnetic field intensity~\cite{Winter:2013cla, Bustamante:2020bxp}.  Nevertheless, adopting the same practice as IceCube flavor measurements~\cite{IceCube:2015rro, IceCube:2015gsk, Aartsen:2018vez, IceCube:2020fpi}, we take $f_{\alpha, \rm S}$ to be constant within the energy range of our analysis.  This is justified by the limited number of HESE events, which would be further diluted by attempting to measure the flavor composition in multiple energy bins.  Under this assumption, any energy dependence that the flavor composition at the Earth, $f_{\alpha, \oplus}$, might have is due solely to the presence of the new neutrino-matter interaction.  Reference~\cite{Mehta:2011qb} explores the interplay between energy dependence that stems from neutrino production and from new physics.

\paragraph{Averaging $f_{\alpha, \oplus}$ between $\nu$ and $\bar{\nu}$.}  Because, in high-energy neutrino telescopes, events triggered by neutrinos and antineutrinos are so far indistinguishable from one another, we average the flavor composition at Earth between neutrinos and antineutrinos, assuming they are present in equal proportions in the flux that reaches Earth.  In actuality, their proportions are practically unknown (see, however, \Refes~\cite{Bustamante:2020niz, IceCube:2021rpz}), but the above assumption aligns with theoretical expectations, especially from multi-pion neutrino production at high energies; see, \eg, \Refe~\cite{Hummer:2010vx}.  As a result of averaging, the resonant features in the oscillation probabilities, which are more prominent for neutrinos than for antineutrinos (\figu{prob_lemmt_nmo}), appear softened in the  flavor composition at Earth that we use.

\begin{figure}[t!]
 \centering
 \includegraphics[width=.49\textwidth]{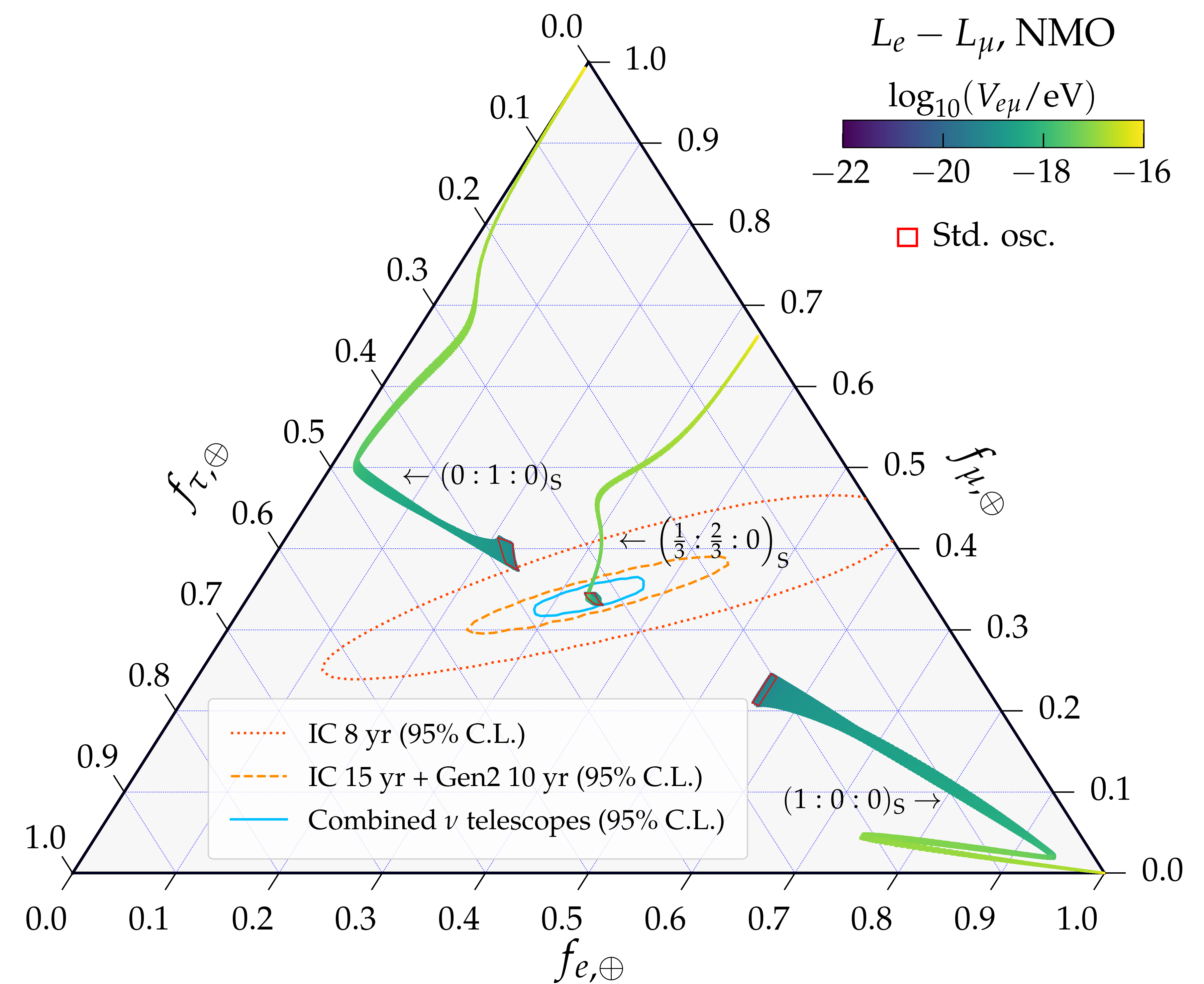}
 \includegraphics[width=.49\textwidth]{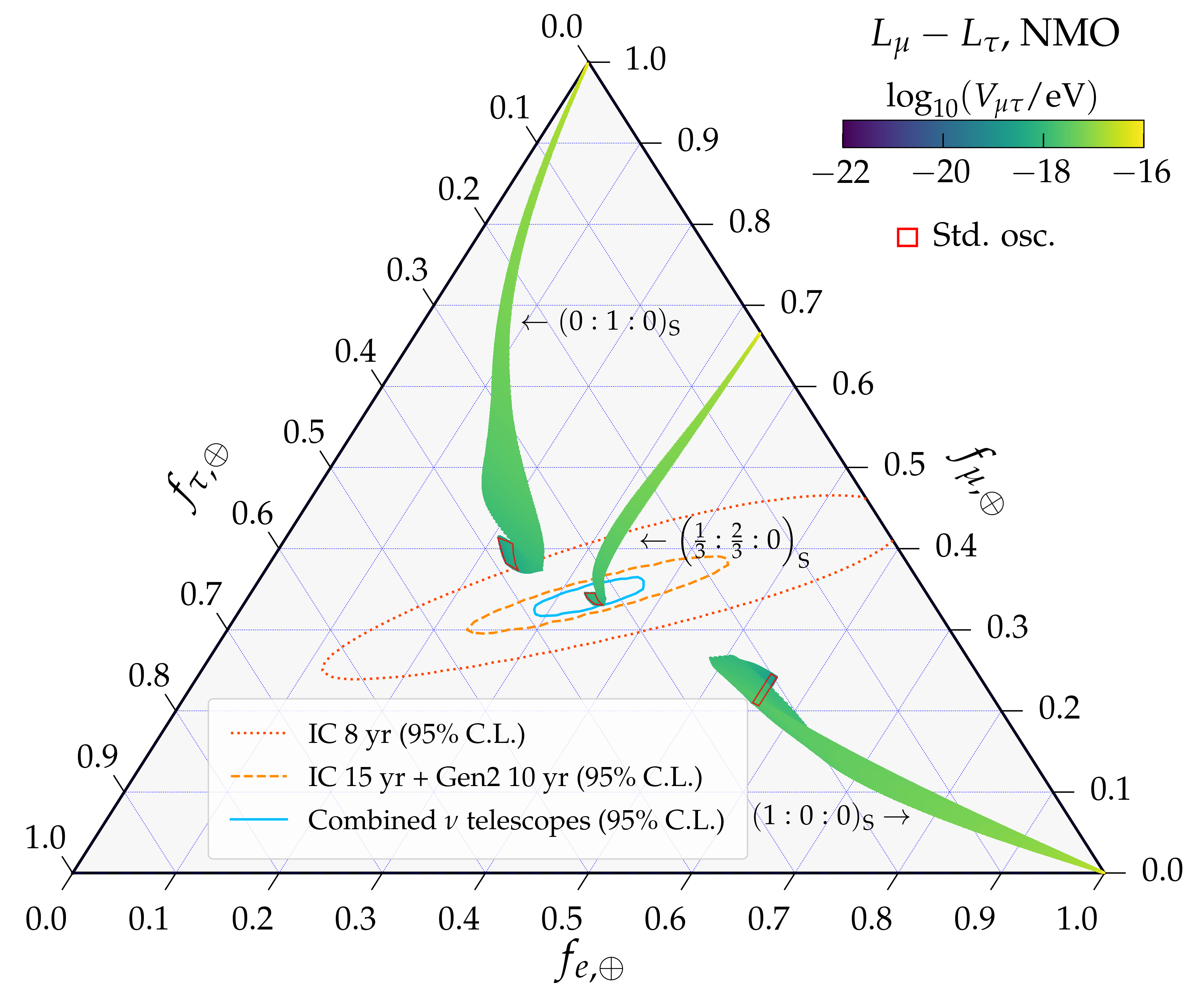}
 \caption{\label{fig:flav_ratio}\textbf{\textit{Flavor composition of high-energy astrophysical neutrinos at Earth, $f_{\alpha, \oplus}$, as a function of the long-range matter potential $V_{e\mu}$, under $L_e-L_\mu$ (left), or $V_{\mu\tau}$, under $L_\mu-L_\tau$ (right).}} We show results for three benchmark choices of the flavor composition at the sources, $(f_{e, {\rm S}}, f_{\mu, {\rm S}}, f_{\tau, {\rm S}}) = (1/3, 2/3, 0)$, the canonical expectation, $(0,1,0)$, and $(1,0,0)$, compared to the expectation from standard oscillations (``Std.~osc.''). The flavor composition is averaged between neutrinos and antineutrinos, assuming they exist in equal proportion in the flux. For this plot only, we fix the neutrino energy to 100~TeV, assume normal neutrino mass ordering, and vary the standard mixing parameters within their $1\sigma$ allowed ranges; our statistical methods systematically vary these choices. Flavor-composition predictions are compared against three estimates of the flavor sensitivity of neutrino telescopes using 8~years of IceCube data (``IC 8~yr''), 15~years of IceCube data plus 10~years of IceCube-Gen2 (``IC 15~yr + Gen2 10~yr''), and the combined exposure of all neutrino telescopes available by 2040 (``Combined $\nu$ telescopes''). For $L_e-L_\tau$, results are similar as for $L_e-L_\mu$; see Appendix~\ref{app:results_le_ltau}.  See Section~\ref{sec:he_nu_flavor_ratios} for details.}
\end{figure}

\paragraph{Averaging $f_{\alpha, \oplus}$ over energy.}  In Figs.~\ref{fig:prob_lemmt_nmo} and \ref{fig:flav_ratio}, we illustrated the effect of the new matter potential on the flavor-transition probabilities and flavor ratios for a fixed neutrino energy of 100~TeV.  However, IceCube detects neutrinos with different energies in the range TeV--PeV.  Within it, the spectrum of $\nu_\alpha$ is described well by a power law $\Phi_{\alpha}(E)\propto f_{\alpha,\oplus}(E)E^{-\gamma}$, where the value of the spectral index is $\gamma \in [2,3]$.  To produce our results, we fix $\gamma = 2.5$~\cite{IceCube:2015gsk}.  As in \Refe~\cite{Bustamante:2018mzu}, we account for the neutrino energy distribution by averaging the flux within the energy range of 25~TeV--2.8~PeV~\cite{IceCube:2015gsk}, \ie, $\langle \phi_\alpha \rangle \approx (2.8\,\text{PeV})^{-1}\int dE \phi_\alpha(E)$.  Below, to produce our results, we use the energy-averaged flavor ratios computed from the average fluxes, \ie, $\langle f_{\alpha,\oplus} \rangle \approx \langle \phi_\alpha \rangle /\sum_{\beta} \langle \phi_\beta \rangle$.

\paragraph{Estimates of present and future flavor sensitivity.}  In our analysis, we use estimates of the present-day flavor sensitivity of IceCube and of future detectors; see \figu{flav_ratio}.  These are based on estimated combined analyses of HESE events plus through-going tracks (see Section~\ref{sec:he_nu_detection}), motivated by \Refe~\cite{IceCube:2015gsk}, which offer the greatest sensitivity.  We adopt the flavor measurement projections for IceCube-Gen2 from \Refe~\cite{IceCube-Gen2:2020qha}, which were generated assuming a plausible neutrino spectrum $\propto E^{-2.5}$.  Because these projections were only generated assuming a flavor composition at the sources of $(1/3,2/3,0)_{\rm S}$, later we only derive constraints on long-range neutrino interactions for that one benchmark scenario.  As in \Refe~\cite{Song:2020nfh}, we isolate the contributions of IceCube and IceCube-Gen2 to the flavor sensitivity from \Refe~\cite{IceCube-Gen2:2020qha}.  This allows us to generate results based on estimates of the present-day flavor sensitivity, for the year 2020, using 8 years of IceCube, and forecasts for the year 2040, using 15 years of IceCube plus 10 years of IceCube-Gen2, and using the additional contribution of future neutrino telescopes Baikal-GVD~\cite{Avrorin:2019vfc}, KM3NeT~\cite{Adrian-Martinez:2016fdl}, P-ONE~\cite{P-ONE:2020ljt}, and TAMBO~\cite{Romero-Wolf:2020pzh}.  Following \Refe~\cite{Song:2020nfh}, we assume that future detectors will have the same detection efficiency of HESE and through-going tracks as IceCube (see Section~\ref{sec:analysis}), though with different rates, and rescale the IceCube flavor sensitivity to their respective expected exposures.  This is admittedly a simplification, made necessary by the current absence of details on the capabilities of future detectors.
   
\section{Statistical analysis and results}
\label{sec:analysis}

Our goal is to compute the bounds on the long-range-interaction parameters --- the mediator mass, $m_{\alpha\beta}^\prime$, and the coupling, $g_{\alpha\beta}^\prime$ --- that can be placed by measuring the flavor composition.  To do this, we contrast our predictions of the flavor composition obtained under long-range interactions (Section~\ref{sec:he_nu_flavor_ratios}) against the estimated present and future flavor-measuring capabilities of IceCube and other upcoming detectors (Section~\ref{sec:he_nu_analysis_choices}).  First, we compute bounds on the long-range matter potential, $V_{\alpha\beta}$; then, we translate them into bounds on the mediator mass and coupling.


\subsection{Statistical analysis}
\label{sec:stat_analysis}

\begin{table}[t!]
 \centering
 \resizebox{\linewidth}{!}{%
 \begin{tabular}{|c c c|}
  \hline 
  Observation epoch & Neutrino telescopes & Neutrino mixing parameters \\ 
  \hline
  2020 (estimated) & IC 8 yr & NuFit~5.1 (2021) \\ 
  2040 (projected) & IC 15 yr + IC-Gen2 10 yr & NuFit~5.1 + JUNO + DUNE + HK \\ 
  2040 (projected) & Combined $\nu$ telescopes     & NuFit~5.1 + JUNO + DUNE + HK\\ 
  \hline 
 \end{tabular}}
 \caption{\textbf{\textit{Observation epochs of our analysis: years 2020 and 2040.}}  For each epoch, we show the neutrino telescopes that  measure the flavor composition of high-energy astrophysical neutrinos and the oscillation experiments that constrain the value of the neutrino mixing parameters.  We assume that upcoming neutrinos have flavor-measuring capabilities similar to those of IceCube.  In 2040, the combined telescopes include IceCube, IceCube-Gen2~\cite{IceCube-Gen2:2020qha}, Baikal-GVD~\cite{Avrorin:2019vfc}, KM3NeT~\cite{Adrian-Martinez:2016fdl}, P-ONE~\cite{P-ONE:2020ljt}, and TAMBO~\cite{Romero-Wolf:2020pzh}.  See Section~\ref{sec:he_nu_detection} for details.  For our 2040 projections of the measurement of mixing parameters, we assume that their real values are the present-day best-fit values from the NuFit~5.1 global oscillation fit~\cite{Esteban:2020cvm, NuFIT}.  See Section~\ref{subsec:transition-prob} for details.}
 \label{tab:proj}
\end{table}

We adopt the Bayesian statistical methods introduced in \Refe~\cite{Song:2020nfh}.  For given flavor ratios at the source, $f_{\alpha, \rm S}$, we assume one of the three benchmark scenarios outlined in Section~\ref{sec:he_nu_flavor_ratios}.  Then, for test values of the long-range potential, $V_{\alpha\beta}$, and of the mixing parameters, $\pmb{\vartheta} \equiv (s_{12}^2, s_{23}^2, s_{13}^2, \dcp)$, with $s_{ij} \equiv \sin \theta_{ij}$, we compute the associated energy-averaged flavor composition at Earth, $\langle \pmb{f}_\oplus \rangle \equiv ( \langle f_{e, \oplus} \rangle,\allowbreak \langle f_{\mu, \oplus} \rangle,\allowbreak \langle f_{\tau, \oplus} \rangle )$, using the procedure introduced in Section~\ref{sec:he_nu_analysis_choices}.  We assess the compatibility of these predictions with measurements of the flavor composition in neutrino telescopes through three factors: $\pi(V_{\alpha\beta})$, the prior associated to the value of $V_{\alpha\beta}$; $\pi(\pmb{\vartheta})$, the prior associated to the value of $\pmb{\vartheta}$; and $\mathcal{L}(\langle \pmb{f}_{\oplus} \rangle)$, the likelihood of having measured the flavor composition $\langle \pmb{f}_{\oplus} \rangle$.  We expand on these factors below.  Using them, we compute the posterior probability density of $V_{\alpha \beta}$, marginalized over all possible values of the mixing parameters, \ie,
\begin{equation}
 \label{equ:posterior}
 \mathcal{P}\left(V_{\alpha \beta}\right)
 =
 \int d\pmb{\vartheta}
 \mathcal{L}
 \left(\langle\pmb{f}_\oplus\left(V_{\alpha \beta}, \pmb{\vartheta}\right)\rangle\right)
 \pi(\pmb{\vartheta})
 \pi\left(V_{\alpha \beta}\right) \;.
\end{equation}
This method represents an improvement over the one used in the original study of long-range interactions using the flavor composition, \Refe~\cite{Bustamante:2018mzu}.  There, the sensitivity to long-range interactions was derived from a straightforward comparison of the predicted flavor composition vs.~the flavor sensitivity of IceCube, and the method was not amenable to being generalized to compute sensitivity beyond the $1\sigma$ statistical significance.  Here, in contrast, the Bayesian approach allows us to derive limits at higher statistical significance and to properly account for the prior on mixing parameters.

Table~\ref{tab:proj} summarizes the two epochs for which we perform the above analysis: the present, represented by the year 2020, and the future, represented by the year 2040.  Each epoch has an associated precision on mixing parameters and flavor measurements associated to it, encoded in the prior $\pi(\pmb{\vartheta})$ and the likelihood $\mathcal{L}\left(\pmb{f}_\oplus \right)$.  We expand on them below.

\paragraph{Prior on the long-range matter potential, $\pi(V_{\alpha\beta})$.}  To avoid introducing unnecessary bias, we use a uniform prior on $V_{\alpha\beta}$ in the interval $10^{-24}$--$10^{-16}$~eV.  This interval covers the full range of long-range-interaction effects: towards the lower end, standard oscillations dominate, and towards the upper end, long-range interactions dominate.

\paragraph{Prior on the mixing parameters, $\pi(\pmb{\vartheta})$.}  We take the prior of each mixing parameter, today and in the future, to be centered at its present-day best-fit value from the NuFit~5.1~\cite{Esteban:2020cvm, NuFIT} global oscillation fit.  The distributions for $s_{12}^2$ and $s_{13}^2$ are each normal and uncorrelated with other parameters, while those of $s_{23}^2$ and $\delta_{\rm CP}$ are correlated.  For our 2020 estimates, we use the NuFit~5.1 $\chi^2$ distributions to compute the joint likelihood, $-2 \ln\mathcal{L}_{2020} \equiv \chi^2(s_{12}^2) + \chi^2(s_{13}^2) + \chi^2(s_{23}^2, \dcp)$.  For our 2040 projections, we combine this with future measurements of $\theta_{12}$ by JUNO, as computed in \Refe~\cite{Song:2020nfh}, and of $\theta_{23}$ and $\dcp$ by DUNE~\cite{Abi:2020wmh} and HK~\cite{Abe:2018uyc}, which we generate ourselves.  We do so in dedicated simulations using {\sc GLoBES}~\cite{Huber:2007ji}, and adopting the detector descriptions for DUNE~\cite{DUNE:2021cuw,DUNE:2016ymp}, using 3.5 years of runtime in neutrino and antineutrino modes each (using a revised plan of 5 years for each would not change results significantly~\cite{DUNE:2020jqi}), and for HK~\cite{Abe:2018uyc}, using 2.5 years of runtime in neutrino mode and 7.5 years in antineutrino mode.  (For $\theta_{23}$ and $\dcp$, our projections differ from those of \Refe~\cite{Song:2020nfh} because we generate them using the best-fit values from NuFit~5.1~\cite{Esteban:2020cvm, NuFIT} instead of NuFit~5.0, as a result of which the value of $\theta_{23}$ has shifted from the upper octant to the lower octant~\cite{FernandezMenendez:2021jfk}.)  For $s_{13}^2$, we keep the width of its prior fixed to its present-day size, which is dominated by measurements in Daya Bay~\cite{DayaBay:2016ssb}, as they are not expected to be improved upon significantly by 2040.  In summary, for 2040, we use the likelihood $-2\ln \mathcal{L}_{2040} \equiv -2\ln \mathcal{L}_{2020} + \sum_\mathcal{E} \chi_{\mathcal{E}}^2$, where the contribution of DUNE, HK, and JUNO is each $\chi_{\mathcal{E}}^2 =\sum_{i,j}(\vartheta_i-\bar \vartheta_i)\Sigma^{-1}_{\mathcal{E},ij}(\vartheta_j-\bar \vartheta_j)$, $\Sigma_{ij}$ is the covariance matrix for the mixing parameters $\vartheta_i$, $\vartheta_j$, and $\bar{\vartheta}_i$, $\bar{\vartheta}_j$ are their assumed real values. 

\paragraph{Likelihood of flavor measurement, $\mathcal{L}\left(\langle \pmb{f}_\oplus \rangle \right)$.}  The likelihood of flavor measurement represents the precision with which neutrino telescopes can measure the flavor composition.  For the 2020 and 2040 epochs, we adopt the same flavor likelihoods as \Refe~\cite{Song:2020nfh}.  By construction, they are centered on the canonical flavor composition at Earth, near $(1/3,1/3,1/3)_\oplus$, that is expected from neutrino production via the full pion decay chain (Section~\ref{sec:he_nu_overview}).  Hence, the bounds on long-range interactions that we derive later are obtained under the assumption that the true value of the measured flavor is such.  For our 2020 estimates, we adopt an estimate of the IceCube flavor sensitivity that would be obtained using 8 years of HESE events and through-going tracks~\cite{IceCube-Gen2:2020qha}, assuming a flux with spectral index $\gamma = 2.5$.  For our 2040 projections, we adopt either the expected flavor sensitivity from 15 years of IceCube plus 10 years of IceCube-Gen2 measurements, extracted from \Refe~\cite{IceCube-Gen2:2020qha}, or the combined sensitivity from that plus the expected 2040 sensitivity of Baikal-GVD~\cite{Avrorin:2019vfc}, KM3NeT~\cite{Adrian-Martinez:2016fdl}, P-ONE~\cite{P-ONE:2020ljt}, and TAMBO~\cite{Romero-Wolf:2020pzh}, extracted from \Refe~\cite{Song:2020nfh}, which we defer to for details.  As illustration, \figu{flav_ratio} shows 95\% C.L.~contours for the 2020 and 2040 flavor-measurement likelihoods.


\subsection{Results}
\label{sec:results}

\begin{figure}[t!]
 \centering
 \includegraphics[width=.49\textwidth]{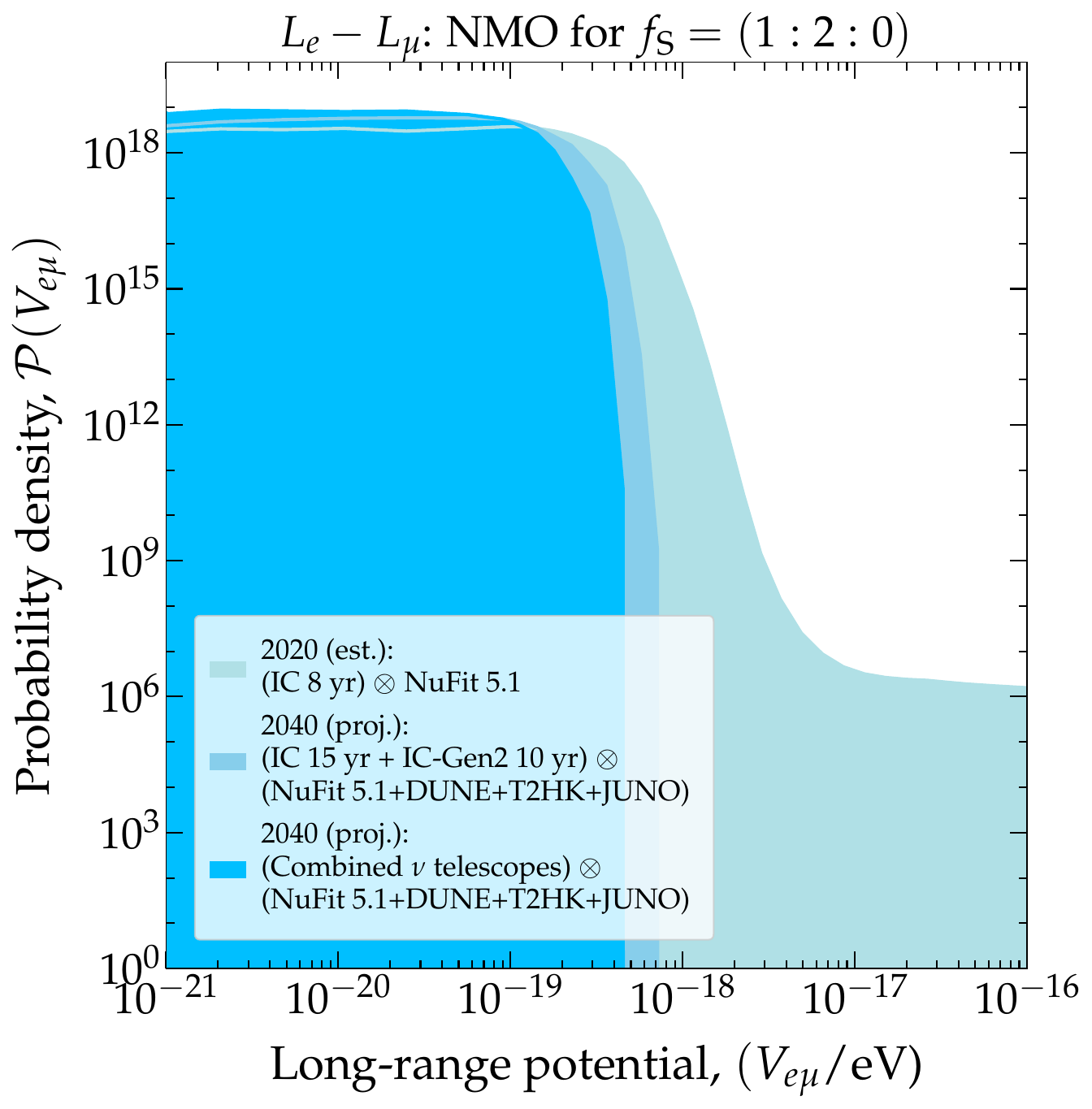}
 \includegraphics[width=.49\textwidth]{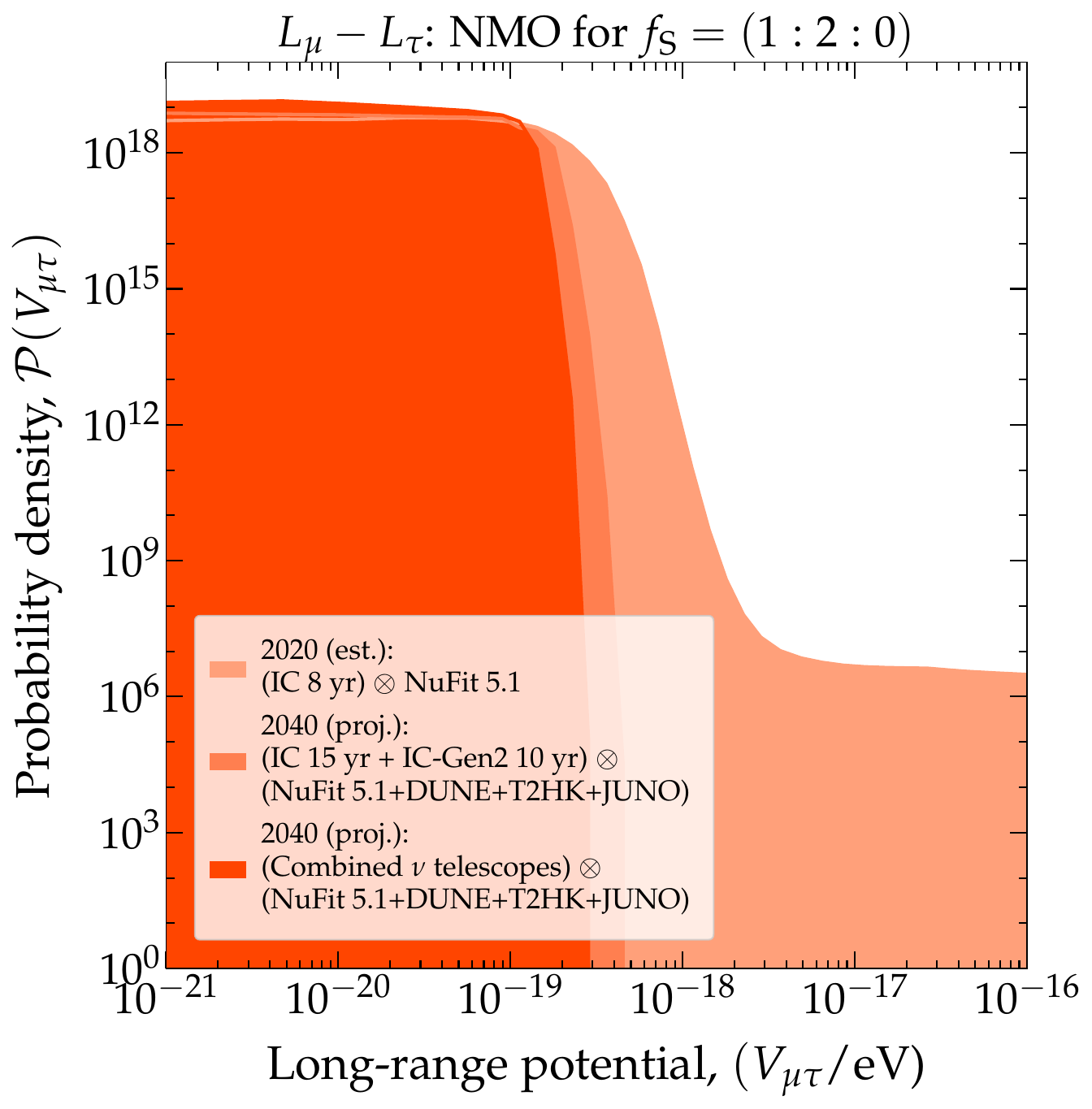}
 \caption{\textbf{\textit{Posterior probability density of the long-range matter potentials $V_{e \mu}$, under the $L_e-L_\mu$ symmetry ({left}) and $V_{\mu \tau}$, under the $L_\mu-L_\tau$ symmetry ({right}).}}  We assume a measurement of the flavor composition at Earth centered around the canonical expectation of about $(1/3:1/3:1/3)_{\oplus}$, corresponding to $f_{\rm S} = (1/3:2/3:0)$ at the sources; see \figu{flav_ratio}.  See \figu{potential_limits} for the limits on the potential derived from the posterior.  Results for $V_{e\tau}$, under the $L_e-L_\tau$ symmetry, are similar to $V_{e\mu}$; see Appendix~\ref{app:results_le_ltau}.  We assume that all neutrino telescopes available by 2040 have detection efficiencies similar to that of IceCube (Section~\ref{sec:he_nu_analysis_choices}).  Results here are obtained assuming normal neutrino mass ordering (NMO); results for inverted ordering are in Appendix~\ref{appendix_e}.  See Section~\ref{sec:stat_analysis} for details. }
 \label{fig:posterior}
\end{figure}

Figure~\ref{fig:posterior} shows the resulting posteriors of $V_{e\mu}$, under the $L_e-L_\mu$ symmetry, and of $V_{\mu\tau}$, under the $L_\mu-L_\tau$ symmetry, computed using \equ{posterior}, for the years 2020 and 2040.  Results for $V_{e\tau}$, under the $L_e-L_\tau$ symmetry, are similar to $V_{e\mu}$; see Appendix~\ref{app:results_le_ltau}.  The posteriors are maximum at a potential of roughly $10^{-19}$~eV.  Because the posteriors plateau at lower values, we can place upper limits on the values of the potentials.  Values above roughly $10^{-18}$~eV are noticeably less favored, and they become more so when moving from 2020 to 2040 due to improvements in the precision of the mixing parameters and flavor composition, \ie, to narrower $\pi(\pmb{\vartheta})$ and $\mathcal{L}\left(\pmb{f}_\oplus\right)$, respectively.  

\begin{figure}[t!]
 \centering
 \includegraphics[width=\textwidth]{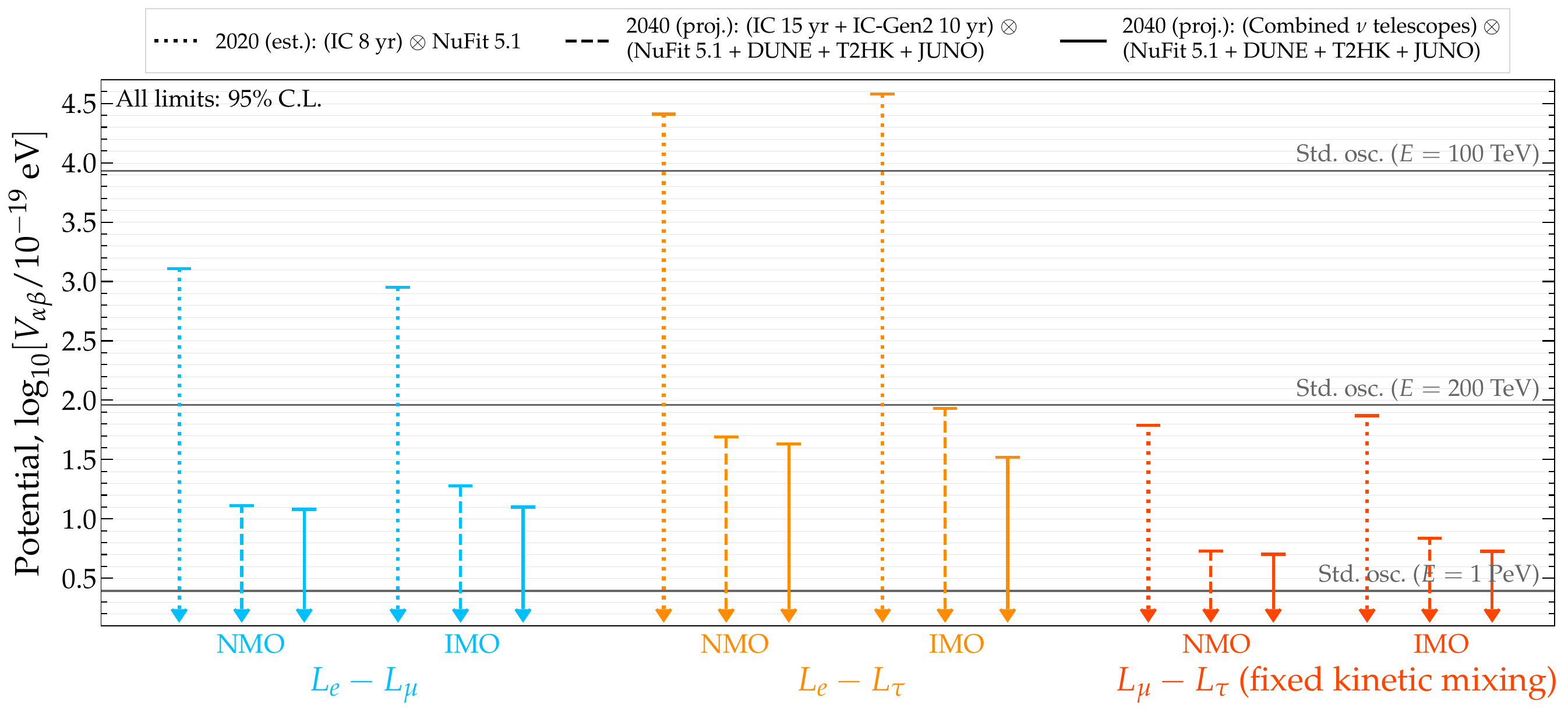}
 \caption{\textbf{\textit{Upper limits (95\% C.L.) on the long-range potential induced by the $L_e-L_\mu$, $L_e-L_{\tau}$, and $L_\mu-L_\tau$ symmetries, estimated for 2020 and projected for 2040.}}  For comparison, we show the standard-oscillation contribution (``Std.~osc.'') to the Hamiltonian, $\Delta m_{21}^2/(2E)$, evaluated at the present-day best-fit value of $\Delta m_{21}^2$~\cite{Esteban:2020cvm, NuFIT} and at different values of the neutrino energy, $E$. We assume that all neutrino telescopes available by 2040 have detection efficiencies similar to that of IceCube (Section~\ref{sec:he_nu_analysis_choices}). Figure~\ref{fig:g_vs_m_120} shows these limits translated into limits on the coupling, $g_{\alpha\beta}^\prime$.  See Appendix~\ref{app:limits} for the values of the limits.  See Section~\ref{sec:analysis} for details.}
 \label{fig:potential_limits}
\end{figure}

\begin{figure}[t!]
 \centering
 \includegraphics[width=.49\textwidth]{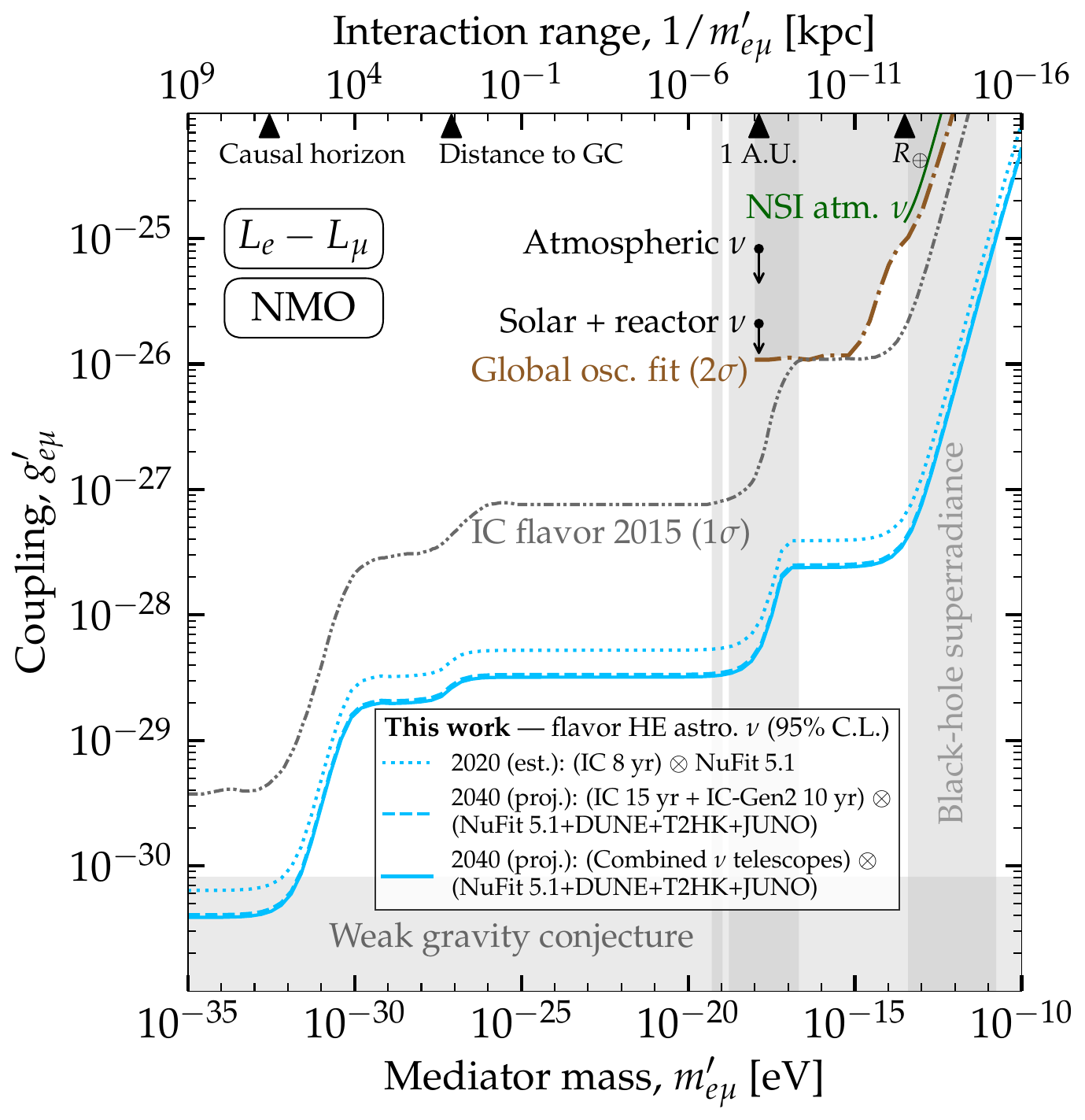}
 \includegraphics[width=.49\textwidth]{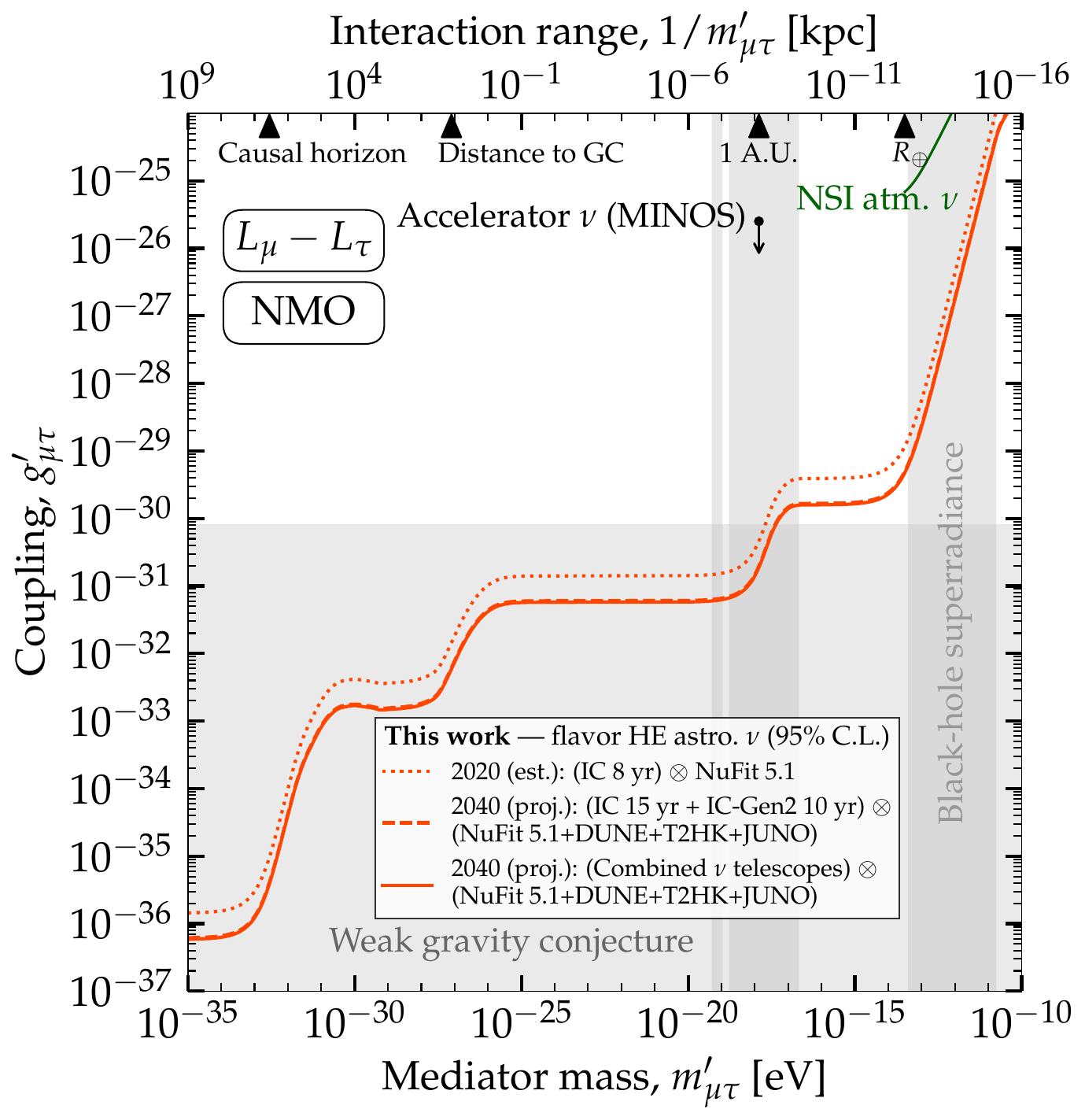}
 \caption{
 \textbf{\textit{Estimated present-day (2020) and projected (2040) upper limits (95\%~C.L.) on the coupling strength, $g_{\alpha\beta}^\prime$, of the new boson, $Z_{\alpha\beta}^\prime$, with mass $m_{\alpha\beta}^\prime$, that mediates flavor-dependent long-range neutrino interactions.}}  Same as \figu{limits_3models}, but now showing also the projected 2040 limits using either IceCube plus IceCube-Gen2, or those combined with Baikal-GVD, KM3NeT, P-ONE, and TAMBO.  {\it Left:} Limits on neutrino-electron interactions under the $L_e-L_\mu$ gauge symmetry.  {\it Right:} Limits on neutrino-neutron interactions under the $L_\mu-L_\tau$ gauge symmetry;  existing limits from accelerator neutrinos in MINOS (95\% C.L.) are from \Refe~\cite{Heeck:2010pg}.   See Section~\ref{sec:analysis} for details.}
 \label{fig:g_vs_m_120}
\end{figure}

Figure~\ref{fig:potential_limits} (also Table~\ref{tab:bounds}) shows the resulting upper limits on the long-range potentials. A limit has value of $V_{\alpha\beta}^\star$ such that the integral of the posterior, $\int_0^{V_{\alpha\beta}^\star} \mathcal{P}(V_{\alpha\beta}) dV_{\alpha\beta}$, equals a desired credible level (C.L.).  We show limits at the 95\%~C.L in \figu{potential_limits} (also in Figs.~\ref{fig:limits_3models}, \ref{fig:g_vs_m_120}, and \ref{fig:plots_for_et}).  Differences in the limits obtained under different symmetries are moderate because, under the assumption of $(1/3,2/3,0)_{\rm S}$, their predicted flavor compositions at Earth are only moderately different; see Figs.~\ref{fig:flav_ratio} and \ref{fig:flav_ratio_let}.

Figure~\ref{fig:potential_limits} reveals modest improvement in the limits between 2020 and 2040, of a factor of 2.5--3, and marginal improvement in 2040 between using only IceCube plus IceCube-Gen2 and using all future available telescopes.  Given the significant increase in the combined neutrino detection rate between 2020 and 2040 (see Fig.~1 in \Refe~\cite{Song:2020nfh}), and between using only IceCube plus IceCube-Gen2 and all telescopes, this is unexpected at first glance.  Yet, close inspection of the variation of the flavor composition at Earth with the long-range potential in \figu{flav_ratio} reveals the subtle reason.  For the canonical choice of flavor composition at the sources of $(1/3, 2/3, 0)_{\rm S}$, which we have adopted to obtain the limits on the long-range potential, the flavor composition at Earth approaches the center of the flavor triangle --- where the likelihood of flavor measurement is largest and from whence the limit is predominantly derived --- from the direction along which the gradients of the 2020 and 2040 likelihoods resemble each other the most.

Figure~\ref{fig:g_vs_m_120} shows how the upper limits on the potential translate into upper limits on the coupling, $g_{\alpha\beta}^\prime$, as a function of the mediator mass, $m_{\alpha\beta}^\prime$.  In \figu{g_vs_m_120}, each limit is an isocontour of potential in the $g_{\alpha \beta}^\prime$--$\,m_{\alpha \beta}^\prime$ plane; we use \equ{pot_total} to translate the limits on $V_{\alpha\beta}$ in \figu{potential_limits} into the limits in \figu{g_vs_m_120}.  Thus, the step-like transitions in \figu{g_vs_m_120} have the same origin as those of the potential itself (\figu{potential}): the limits on $g_{\alpha\beta}^\prime$ are stronger for smaller mediator masses, where the potential is due to a larger number of electrons or neutrons.  Figure~\ref{fig:limits_3models} spotlights our 2020 projections combining all neutrino telescopes.

For $L_e-L_\mu$ (and also for $L_e-L_\tau$, see Appendix~\ref{app:results_le_ltau}), \figu{g_vs_m_120} reveals that already our 2020 estimate is better, by about one order of magnitude, than the proof-of-principle sensitivity based on 2015 IceCube flavor measurements from \Refe~\cite{Bustamante:2018mzu}. (However, a fair comparison is not possible due to differences in their statistical significance and methods.). For $L_\mu-L_\tau$, we estimate and forecast limits based on the flavor composition  for the first time.  The limits for $g_{\mu\tau}^\prime$ shown in \figu{g_vs_m_120} were obtained by fixing the mixing strength to $(\xi-\sin \theta_W \chi)=5 \times 10^{-24}$ (see Section~\ref{subsec:models}).  Because $V_{\mu \tau} \propto g_{\mu\tau}^\prime (\xi-\sin \theta_W \chi)$ (see Eqs.~(\ref{equ:pot_general}) and (\ref{equ:Gab})), given an upper limit on $V_{\mu \tau}$, a smaller value of the mixing strength, which is plausible, would entail a weaker limit on $g_{\mu\tau}^\prime$. (Alternatively, the limits for $L_\mu-L_\tau$ in \figu{g_vs_m_120} can be interpreted as being on the product of the coupling times the mixing strength, without any assumption on the value of the latter.)

Figure~\ref{fig:g_vs_m_120}, and also Figs.~\ref{fig:limits_3models} and \ref{fig:plots_for_et}, show that, for $L_e-L_\mu$ and $L_e-L_\tau$, our limits, both estimated for 2020 and forecast for 2040, improve on existing ones obtained from atmospheric neutrinos~\cite{Joshipura:2003jh}, solar and reactor neutrinos~\cite{Bandyopadhyay:2006uh}, reinterpretations~\cite{Wise:2018rnb} of bounds on the coefficients of non-standard neutrino interactions (NSI)~\cite{Super-Kamiokande:2011dam,Ohlsson:2012kf,Gonzalez-Garcia:2013usa,Coloma:2020gfv} (see Appendix~\ref{app:reinterp_nsi_limits}), and the recent global oscillation fit from \Refe~\cite{Coloma:2020gfv}.  For $L_\mu-L_\tau$, light mediators in the mass range that we consider are largely unconstrained, except towards high masses, from accelerator neutrinos~\cite{Heeck:2010pg}
and NSI.

\textbf{\textit{In summary, our results show that, already today, the estimated flavor sensitivity of IceCube and the uncertainty in the mixing parameters have the potential to significantly improve the constraints on flavor-dependent long-range neutrino interactions.}}

\section{Future improvements}
\label{sec:future}

The goal of our analysis is to estimate the reach of flavor composition of high-energy astrophysical neutrinos to constrain flavor-dependent long-range neutrino interactions rather than providing detailed results based on real flavor measurements.  In doing so, we incorporated assumptions in our analysis towards facilitating our estimates.  Below, we list them and point out with envisioned improvements.
\begin{description}
 \item[Alternative flavor composition choices.] 
  To produce our limits, we limited ourselves only to the case where the flavor composition at Earth is the canonical expectation of $(1/3,1/3,1/3)_\oplus$, from neutrino production via the full pion decay production, $(1/3,2/3,0)_{\rm S}$.  This was due to the unavailability of the projected flavor sensitivity of neutrino telescopes to alternative choices of flavor composition (Section~\ref{sec:he_nu_analysis_choices}), but our methods are general and applicable also in those alternatives.  We encourage experimental collaborations to make publicly available the flavor sensitivity of their experiments under assumptions other than the canonical one.
 \item[Flavor-measurement capabilities of upcoming detectors.]  
  Our forecasts for upcoming neutrino telescopes --- Baikal-GVD, IceCube-Gen2, KM3NeT, P-ONE, TAMBO --- assume that their capabilities to measure the flavor composition will be similar to those of IceCube.  This is a necessary assumption given the absence of their realistic capabilities at the time of writing.  However, it is foreseeable already that differences in their geometries --- most notably, in the spacing between photomultiplier strings --- will imply differences between their detection capabilities.  Further, we have not considered the possibility of proposed improvements in flavor-tagging techniques, like those provided by muon and neutron echoes~\cite{Li:2016kra}.  As the specific flavor-measurement capabilities of different upcoming experiments become better defined, future versions of our analysis will be able to incorporate them.
 \item[Using the energy dependence of the flavor composition.]
  Our results are based on the comparison of the energy-averaged flavor composition at Earth {\it vs.}~the flavor composition measured over an energy interval, assuming that it is constant within it (Section~\ref{sec:he_nu_analysis_choices}).  However, long-range interactions modify the flavor composition in an energy-dependent way, so there may be additional probing power to be reaped from measuring the flavor composition in multiple energy bins.  Present-day event rates are likely insufficient to do this, but IceCube-Gen2 may be able to~\cite{IceCube-Gen2:2020qha}.  Further, doing this could allow us to distinguish between the variation of the flavor composition with energy that stems from long-range interactions from the one that stems from the neutrino production mechanism changing with energy, \eg, from full pion decay chain at low energies to muon-damped at high energies~\cite{Kashti:2005qa, Kachelriess:2006fi, Kachelriess:2007tr, Hummer:2010ai, Mehta:2011qb, Winter:2013cla, Bustamante:2015waa, Bustamante:2020bxp}.
 \item[Astrophysical uncertainties.] 
  To produce our results, we assumed that the energy spectrum of the high-energy astrophysical neutrinos is a power law $\propto E^{-\gamma}$, with fixed $\gamma = 2.5$, compatible with IceCube analyses that combine HESE events and through-going tracks~\cite{IceCube:2015gsk}.  However, the value of $\gamma$ is only uncertainly known and varies somewhat depending on the data set used to measure it~\cite{IceCube:2020wum, IceCube:2021uhz}.  Further, the IceCube observations admit a description with differently shaped energy spectra; \eg, \Refes~\cite{Palladino:2018evm, Capanema:2020rjj, IceCube:2020wum, Ambrosone:2020evo, IceCube:2021uhz, Fiorillo:2022rft}.  Finally, because the diffuse flux of neutrinos is the addition of contributions from sources located at different distances, it depends on how the number density of sources evolves with redshift, which we have neglected in our analysis.  Incorporating uncertainties in the shape of the neutrino spectrum and the redshift distribution of sources would likely weaken the bounds that we report, though a full analysis is beyond the scope of this paper.
 \item[Computing neutrino propagation.]
  When computing the long-range matter potential (Section~\ref{subsec:LRI}), we do so at the location where the neutrinos are detected, at IceCube, rather than propagating neutrinos from their astrophysical sources to Earth and computing the changing potential at every point along their trajectory.  This simplified treatment allows us to first put limits on the long-range potential and then to translate those into limits on $g_{\alpha\beta}^\prime$.  Our method overestimates the influence of faraway electrons and neutrons, but mainly for limits at large mediator masses. Reference~\cite{Coloma:2020gfv} is an example of how to compute long-range interactions during propagation, though only for neutrino propagation inside the Earth and the Sun.
 \item[Screening due to relic neutrinos.]
  If the background of relic neutrinos contains $\nu_e$ and $\bar{\nu}_e$ in equal proportions, it may partially screen the long-range potential due to cosmological electrons~\cite{Dolgov:1995hc, Blinnikov:1995kp, Joshipura:2003jh}.  The Debye length~\cite{Joshipura:2003jh}, \ie, the distance at which the screening becomes important, is a factor-of-10 shorter than the long-range interaction range, so screening would affect limits at mediator masses below about $10^{-30}$~eV.
\end{description}


\section{Summary and outlook}
\label{sec:conc}

Neutrinos may conceivably undergo interactions with matter beyond those contained in the Standard Model.  If so, they must necessarily be weak up to neutrino energies of a few hundred GeV in order to have gone undetected so far.  Yet, at higher energies, accessible via recently detected astrophysical neutrinos, they may be significant or even prominent.  If discovered, they would represent compelling evidence of new physics.

We have studied the sensitivity to new flavor-dependent interactions between neutrinos and electrons and neutrinos and neutrons, today and in the future.  They result from gauging the lepton-number symmetries $L_e-L_\mu$, $L_e-L_\tau$, or $L_\mu-L_\tau$ that are already global symmetries of the Standard Model.  The interaction is mediated by a new neutral gauge boson that, if ultra-light (\ie, lighter than $10^{-10}$~eV), subtends a prodigious interaction range, from km to Gpc.  As a result, neutrinos may be affected by the aggregated matter potential sourced by large collections of nearby and faraway matter --- the Earth, Moon, Sun, Milky Way, and the cosmological distribution of matter.  The interactions may modify neutrino oscillations, especially at high energies, and, if significant, may suppress them.  

Thus, we have estimated and forecast constraints on long-rage interactions using the highest-energy neutrinos known: the TeV--PeV astrophysical neutrinos discovered by IceCube~\cite{Aartsen:2013bka, Aartsen:2013jdh, Aartsen:2014gkd, Aartsen:2015rwa, Aartsen:2016xlq, Ahlers:2018fkn, IceCube:2020wum, Ackermann:2022rqc}.  We employ the flavor composition of their diffuse flux, \ie, the proportion of $\nu_e$, $\nu_\mu$, and $\nu_\tau$ in it: since the flavor composition reflects the effect of neutrino oscillations en route to Earth, it may reveal the presence of flavor-dependent long-range interactions. 

We combine the estimated flavor-measuring capabilities of IceCube with present-day uncertainties on the standard neutrino mixing parameters that drive neutrino oscillations~\cite{Esteban:2020cvm, NuFIT}.  Our work improves and extends \Refe~\cite{Bustamante:2018mzu}, chiefly by adopting superior statistical methods, and by considering the $L_\mu-L_\tau$ symmetry.   Our analysis choices are conservative and realistic (Section~\ref{sec:he_nu_analysis_choices}).   Because we rely on estimates of the flavor sensitivity, our goal is to showcase the capacity of neutrino telescopes, rather than provide high-precision results, and to motivate further work.

Presently, there is large uncertainty in the measurement of the flavor composition at IceCube and moderate-to-large uncertainties on the neutrino mixing parameters.  At first glance, this should reduce sensitivity to long-range interactions.  Surprisingly, we find that \textbf{\textit{using the estimated current flavor sensitivity of IceCube and current mixing parameter uncertainties, high-energy astrophysical neutrinos could tightly constrain long-range interactions, surpassing existing limits (see \figu{limits_3models}).}}

In the coming two decades, the above limitations will likely be overcome by next-generation neutrino oscillation experiments --- DUNE~\cite{Abi:2020wmh}, Hyper-Kamiokande~\cite{Abe:2018uyc}, and JUNO~\cite{An:2015jdp} --- and by high-energy neutrino telescopes --- IceCube-Gen2~\cite{IceCube-Gen2:2020qha}, Baikal-GVD \cite{Avrorin:2019vfc}, KM3NeT~\cite{Adrian-Martinez:2016fdl}, P-ONE~\cite{P-ONE:2020ljt}, and TAMBO~\cite{Romero-Wolf:2020pzh}.  By 2040, nominally, we anticipate modest improvements from repeating our analysis unchanged but possible substantial gains by upgrading it to leverage higher event rates and potential advancements in  flavor composition measurement (Section~\ref{sec:future}).

The discovery of a new fundamental interaction would mark groundbreaking progress.  Our findings show that high-energy astrophysical neutrinos, already today, have the potential to probe this more rigorously than present-day constraints.



\acknowledgments 
We thank Pilar Coloma, Francis Halzen, Masoom Singh, and Yu-Dai Tsai for their helpful discussions and crucial inputs.
S.K.A., S.D., and A.N.~acknowledge the support from the Department of Atomic Energy (DAE), Govt.~of India, under the Project Identification no.~RIO 4001.~S.K.A.~is supported by the Young Scientist Project [INSA/SP/YSP/144/201 7/1578] from the Indian National Science Academy (INSA). S.K.A.~acknowledges the financial support from the Swarnajayanti Fellowship (sanction order no.~DST/SJF/PSA-05/2019-20) provided by the Department of Science and Technology (DST), Govt.~of India.~S.K.A.~and A.N.~receive the financial support from the Research Grant (sanction order no.~SB/SJF/2020-21/21) provided by the Science and Engineering 
Research Board (SERB), Govt. of India, under the Swarnajayanti Fellowship project. S.K.A.~would like to thank the United States-India Educational Foundation (USIEF) for providing the financial support through the Fulbright-Nehru Academic and Professional Excellence Fellowship (Award no.~2710/F-N APE/2021). M.B.~is supported by the {\sc Villum Fonden} under project no.~29388.
The numerical simulations are carried out using the ``SAMKHYA:~High-Performance Computing Facility'' at the Institute of Physics, Bhubaneswar, India.


\begin{appendix}


\section{Interaction potential under the $L_\mu-L_\tau$ symmetry}
\label{app:potential_lmu_ltau}

\setcounter{figure}{0} 
\setcounter{equation}{0}
\renewcommand\thefigure{A\arabic{figure}}
\renewcommand\theHfigure{A\arabic{figure}}

Under the $L_e-L_\mu$ and $L_e-L_\tau$ symmetries, neutrinos couple directly to electrons via $Z_{e\beta}^\prime$ $(\beta=\mu,\tau)$, \ie, via the term $\mathcal{L}_{Z^\prime}$ of the Lagrangian, \equ{lag_zprime} in the main text; see Fig.~\ref{fig:feyn}.  Under the $L_\mu-L_\tau$ symmetry, such direct interaction does not occur due to the absence of naturally occurring muons and tauons.  But, after breaking the $L_\mu-L_\tau$ symmetry, there are terms in the interaction Lagrangian that mix the field strength tensors and the two massive bosons, $Z$ and $Z_{\mu\tau}^\prime$, which facilitate neutrino interactions with matter; see Fig.~\ref{fig:feyn}. 

For ordinary, electrically neutral matter, this interaction is
\begin{equation}
 \label{equ:Vmutau}
 \mathcal{L}_{\rm int}
 =
 -g'(\xi-\sin\theta_W\chi)\frac{e}{\sin\theta_W \cos\theta_W}
 J'_\rho J_3^\rho \;,
\end{equation}
where $J^\prime_\rho = \bar{\nu}_\mu \gamma_\rho P_L\nu_\mu-\bar{\nu}_\tau \gamma_\rho P_L\nu_\tau$, $J_3^\rho = -\frac{1}{2}\bar{e}\gamma^\rho P_L e+\frac{1}{2}\bar{u}\gamma^\rho P_L u-\frac{1}{2}\bar{d}\gamma^\rho P_L d$, and the term $(\xi-\sin\theta_W\chi)$ accounts for $Z$--$Z^\prime$ mixing (Fig.~\ref{fig:feyn}). 

The time component of $J_3^\rho$, relevant for static matter, is
\begin{equation}
 J_3^0
 =
 -\frac{1}{4}\left[\bar{e}\gamma^0(1-\gamma^5)e-\bar{u}\gamma^0(1-\gamma^5)u+\bar{d}\gamma^0(1-\gamma^5)d\right] \;,
\end{equation}
which, for an unpolarized medium, becomes
\begin{equation}
 J_3^0
 =
 -\frac{1}{4}(n_e-n_u+n_d)
 =
 -\frac{1}{4}(n_e-n_p+n_n) \;,
\end{equation}
where $n_f$ ($f = e, u, d)$ is the number density of electrons, up quarks, and down quarks.  For electrically neutral medium, $n_p = n_e$, so, $J_3^0 = -n_n/4$.  Replacing this in \equ{Vmutau} yields the potential due to neutrons under the $L_\mu-L_\tau$ symmetry,
\begin{equation}
 V_{\mu\tau}
 = 
 g_{\mu\tau}^\prime
 (\xi-\sin\theta_W\chi)\frac{e}{4 \sin\theta_W \cos\theta_W} n_n \;.
\end{equation}


\section{Evolution of the modified mixing angles with long-range potential}
\label{app:evol_mix_angles}

\setcounter{figure}{0} 
\setcounter{equation}{0}
\renewcommand\thefigure{B\arabic{figure}}
\renewcommand\theHfigure{B\arabic{figure}}

\begin{figure}[t!]
 \centering
 \includegraphics[width=\textwidth]{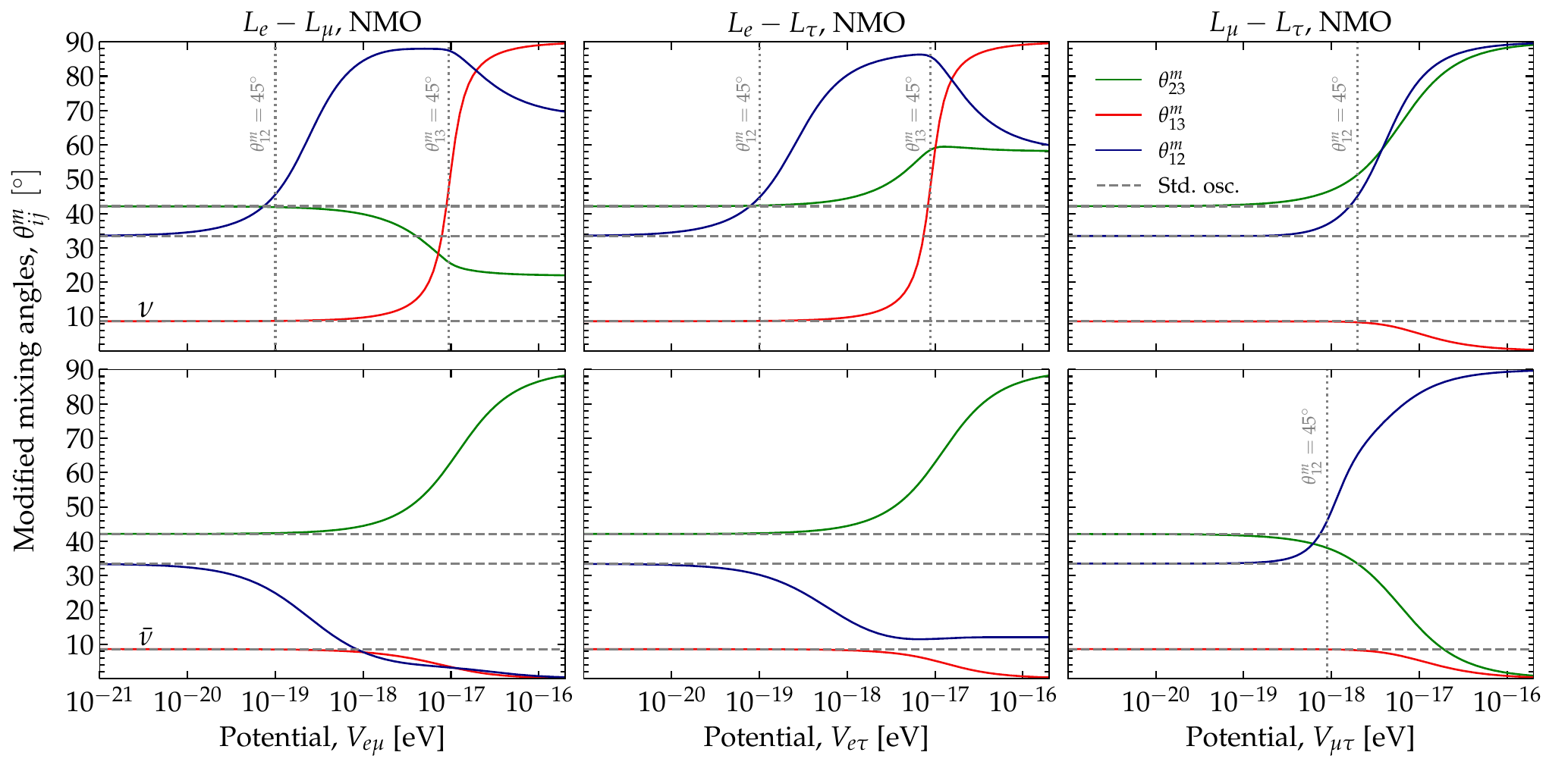}
\caption{\textbf{\textit{Modified mixing angles as functions of the long-range matter potential induced by the $U(1)$ gauge symmetries $L_e-L_\mu$ (left column), $L_e-L_\tau$ (middle column), and $L_\mu-L_\tau$ (right column).}}  Results are for neutrinos and antineutrinos at a fixed energy of 100~TeV, assuming normal mass ordering (NMO).   Vertical lines mark the values of the potential for which $\theta_{12}^m$ and $\theta_{13}^m$ become resonant.  Values of the mixing angles under standard oscillations (``Std.~osc.''), \ie, for $V_{\alpha\beta} = 0$, are shown for comparison. The standard mixing parameters are fixed at their present-day best-fit values from NuFit 5.1~\cite{Esteban:2020cvm, NuFIT}.  See Section~\ref{subsec:transition-prob} for details.}
 \label{fig:mix_angle_lemmt_nmo}
\end{figure}

Figure~\ref{fig:mix_angle_lemmt_nmo} shows the evolution of the modified mixing angles with the long-range potential.  In the main text, the average oscillation probability, \equ{prob_avg}, depends on the unitarity matrix $\mathbf{U}^m$ that diagonalizes the Hamiltonian, \equ{hamiltonian_tot}, which includes contribution from the long-range matter potential.  The matrix $\mathbf{U}^m$ is parametrized as the PMNS matrix, but in terms of three new mixing angles, $\theta_{12}^m$, $\theta_{23}^m$ and $\theta_{13}^m$, and one CP-violating phase, $\dcp^m$, that depend on the long-range potential.  Reference~\cite{Agarwalla:2021zfr} showed analytically that $\delta^m$ remains unchanged from its vacuum value, $\dcp$.

\section{Reinterpretation of Super-Kamiokande NSI limits}
\label{app:reinterp_nsi_limits}

\setcounter{table}{0} 
\setcounter{figure}{0} 
\setcounter{equation}{0}
\renewcommand\thefigure{C\arabic{figure}}
\renewcommand\theHfigure{C\arabic{figure}}

Using the atmospheric neutrino data, the Super-Kamiokande experiment obtained bounds on the neutrino non-standard interaction (NSI)
parameters involving only down quarks in the medium, \ie, $\vert \varepsilon^d_{\mu\mu} - \varepsilon^d_{\tau\tau} \vert < 0.049$
at 90\% C.L.~\cite{Super-Kamiokande:2011dam}. In our convention, we consider the NSIs due to electrons in the medium, \ie, $\vert \varepsilon_{\mu\mu} -\varepsilon_{\tau\tau} \vert = 3 \times \vert \varepsilon^d_{\mu\mu} - \varepsilon^d_{\tau\tau} \vert < 0.147$,
since $N_d \approx 3 N_e$ inside the Earth. Reference~\cite{Wise:2018rnb} reinterpreted these bounds on the NSI parameters
as limits on long-range neutrino interactions under the $L_e-L_\mu$ and $L_e-L_\tau$ symmetries.  Below, we revisit explicitly
these limits and extend them to the $L_\mu-L_\tau$ symmetry. We obtain limits on the potentials $V_{e\mu}$, $V_{e\tau}$, and $V_{\mu\tau}$;
in Figs.~\ref{fig:limits_3models}, \ref{fig:g_vs_m_120}, \ref{fig:plots_for_et}, and \ref{fig:g_vs_m_120_IMO}, we show them
translated into limits on the coupling strength, via the definition of the potential, \equ{pot_total}.

The NSI matter potential is
\begin{eqnarray}
 \mathbf{V}_{\rm NSI}
 =
 V_{\rm CC}
 \begin{pmatrix}
  \varepsilon_{ee} & \varepsilon_{e \mu} & \varepsilon_{e \tau}\\
  \varepsilon_{e \mu}^\ast & \varepsilon_{\mu \mu} & \varepsilon_{\mu \tau}\\
  \varepsilon_{e \tau}^\ast & \varepsilon_{\mu \tau}^\ast & \varepsilon_{\tau \tau}
 \end{pmatrix} \;,
\end{eqnarray}
where $V_{\rm CC}$ is the charged-current SM potential from electrons, as defined for \equ{pot_lri_matrix} in the main text.
The parameters $\varepsilon_{\alpha\beta}$ represent the strength of NSI between neutrinos of flavor $\alpha$ and $\beta$
interacting with electrons, $u$ quarks, and $d$ quarks, normalized to the electron abundance.
See \Refes~\cite{Ohlsson:2012kf, Miranda:2015dra, Farzan:2017xzy} for reviews on NSI. Identifying $\mathbf{V}_{\rm NSI}$
with the long-range matter potentials $\mathbf{V}_{e\mu}$, $\mathbf{V}_{e\tau}$, and $\mathbf{V}_{\mu\tau}$ in
\equ{pot_lri_matrix}, yields $V_{e\mu} = V_{\rm CC} \varepsilon_{ee}$ (with $\varepsilon_{ee}
= -\varepsilon_{\mu\mu}$ and $\varepsilon_{\tau\tau} = 0$), $V_{e\tau} = V_{\rm CC} \varepsilon_{ee}$ (with $\varepsilon_{ee}
= -\varepsilon_{\tau\tau}$  and $\varepsilon_{\mu\mu} = 0$), and $V_{\mu\tau} = V_{\rm CC} \varepsilon_{\mu\mu}$ (with $\varepsilon_{\mu\mu}
= -\varepsilon_{\tau\tau}$  and $\varepsilon_{ee} = 0$).
Therefore, the NSI bounds $\vert \varepsilon_{\mu\mu} - \varepsilon_{\tau\tau} \vert < 0.147$ \cite{Super-Kamiokande:2011dam}
translate into $V_{e\mu} < 0.147 V_{\rm CC}$, $V_{e\tau} < 0.147 V_{\rm CC}$, and $V_{\mu\tau} < 0.147 V_{\rm CC}/2$.

\section{Additional results under the $L_e-L_\tau$ symmetry}
\label{app:results_le_ltau}

\renewcommand\thefigure{D\arabic{figure}}
\renewcommand\theHfigure{D\arabic{figure}}
\renewcommand\thetable{D\arabic{table}}
\renewcommand\theHtable{D\arabic{table}}
\setcounter{figure}{0} 
\setcounter{equation}{0}
\setcounter{table}{0}

In the main text, we show mainly results for the $L_e-L_\mu$ and $L_\mu-L_\tau$ symmetries. Below we show results for the $L_e-L_\tau$ symmetry, under the normal mass ordering.

\begin{figure}[t!]
 \centering
 \includegraphics[width=\textwidth]{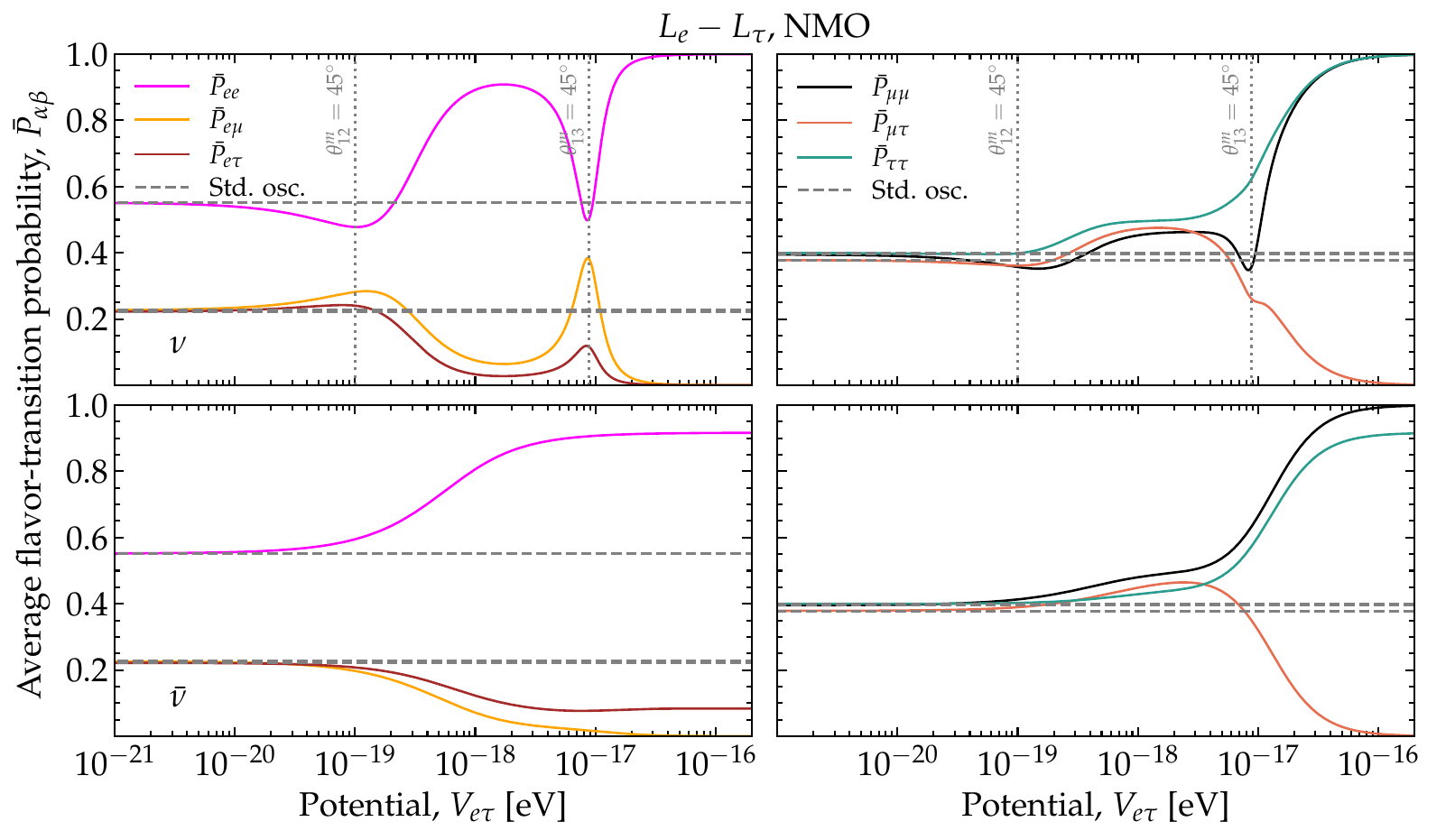}
 \caption{
 \textit{\textbf{Average flavor-transition probabilities, \equ{prob_avg}, as functions of the new matter potential induced by the $U(1)$ gauge symmetry $L_e-L_\tau$.}}  Same as \figu{prob_lemmt_nmo}, but for the $L_e-L_\tau$ symmetry.  See Section~\ref{subsec:transition-prob} and Appendix~\ref{app:results_le_ltau} for details.
 \label{fig:prob_let_nmo}}
\end{figure}

Figure~\ref{fig:prob_let_nmo} shows the oscillation probabilities as functions of the long-range matter potential, $V_{e\tau}$.  Their features are similar to those for the $L_e-L_\mu$ symmetry in \figu{prob_lemmt_nmo}, except swapping $\bar{P}_{e\mu}$ and $\bar{P}_{e\tau}$.

\begin{figure}[b!]
 \centering
 \includegraphics[scale=0.3]{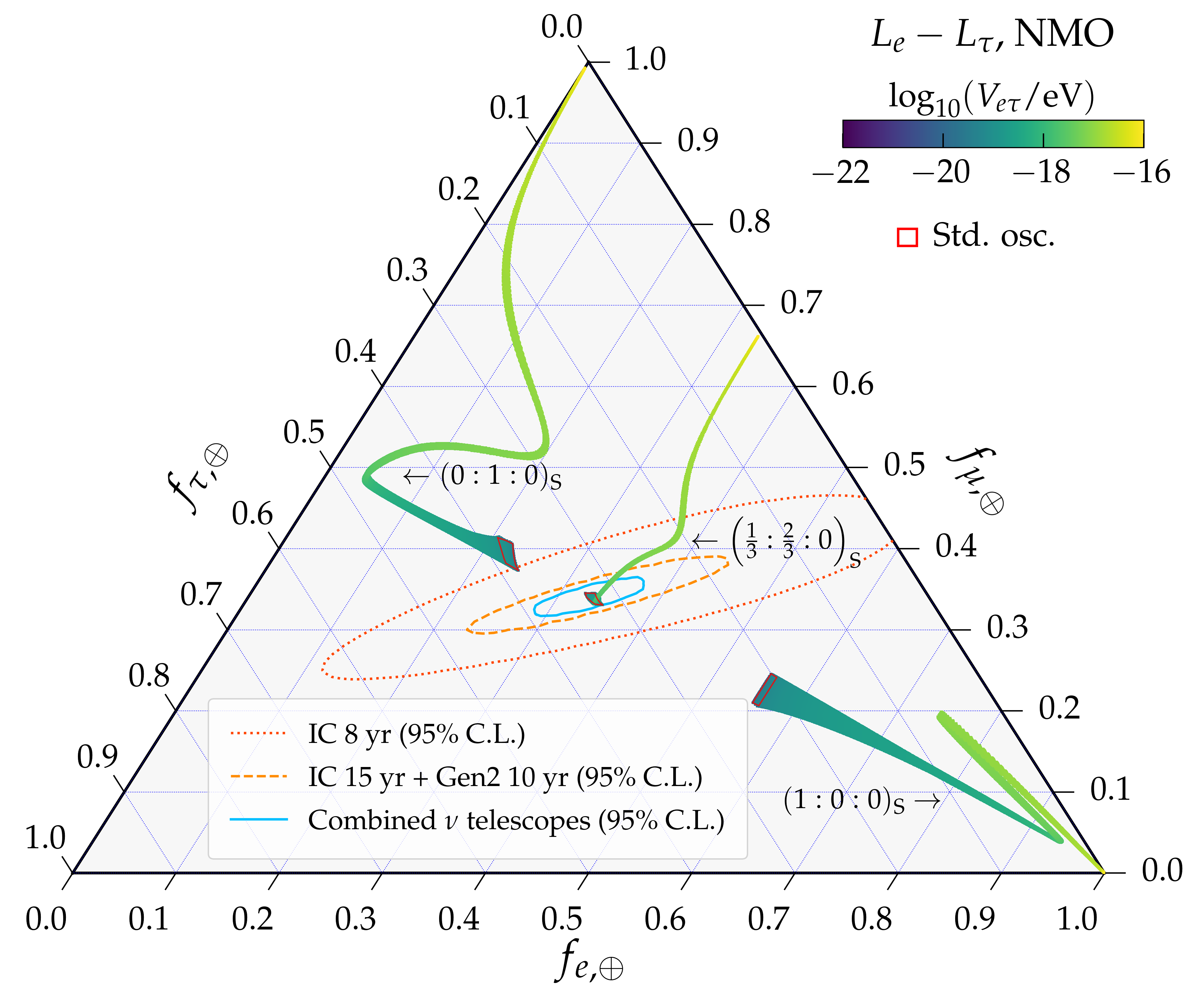}
 \caption{\textbf{\textit{Flavor composition of high-energy astrophysical neutrinos at Earth, $f_{\alpha, \oplus}$, as a function of the long-range matter potential $V_{e\tau}$ under $L_e-L_\tau$.}}  Same as \figu{flav_ratio}, but for the $L_e-L_\tau$ symmetry.  See Section~\ref{sec:he_nu_flavor_ratios} and Appendix~\ref{app:results_le_ltau} for details.}
 \label{fig:flav_ratio_let}
\end{figure}

Figure~\ref{fig:flav_ratio_let} shows the corresponding neutrino flavor composition at Earth.  Their features track those of the oscillation probabilities in \figu{prob_let_nmo}.  

\begin{figure}[t!]
 \centering
 \includegraphics[width=.49\textwidth]{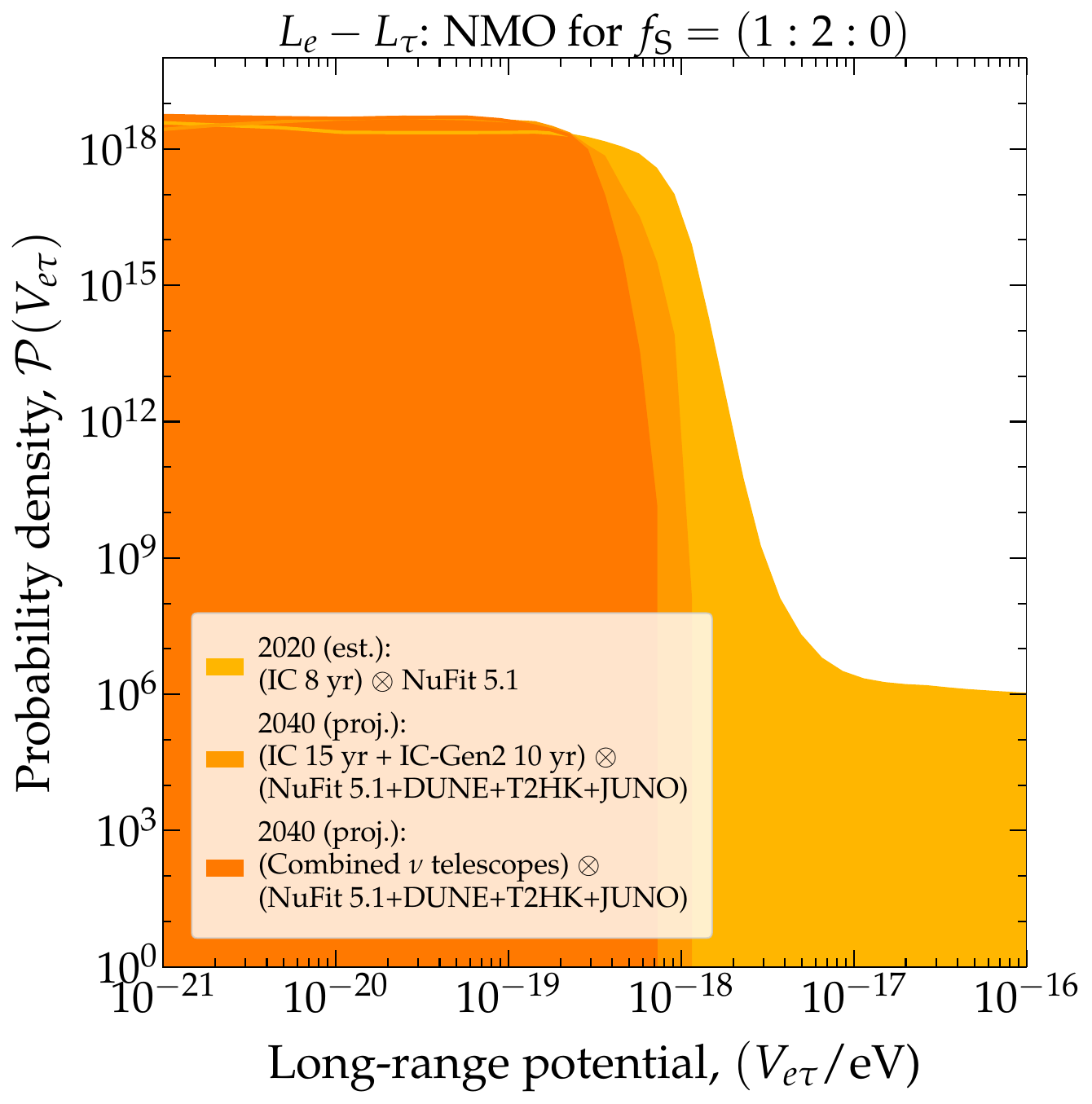}
 \includegraphics[width=.49\textwidth]{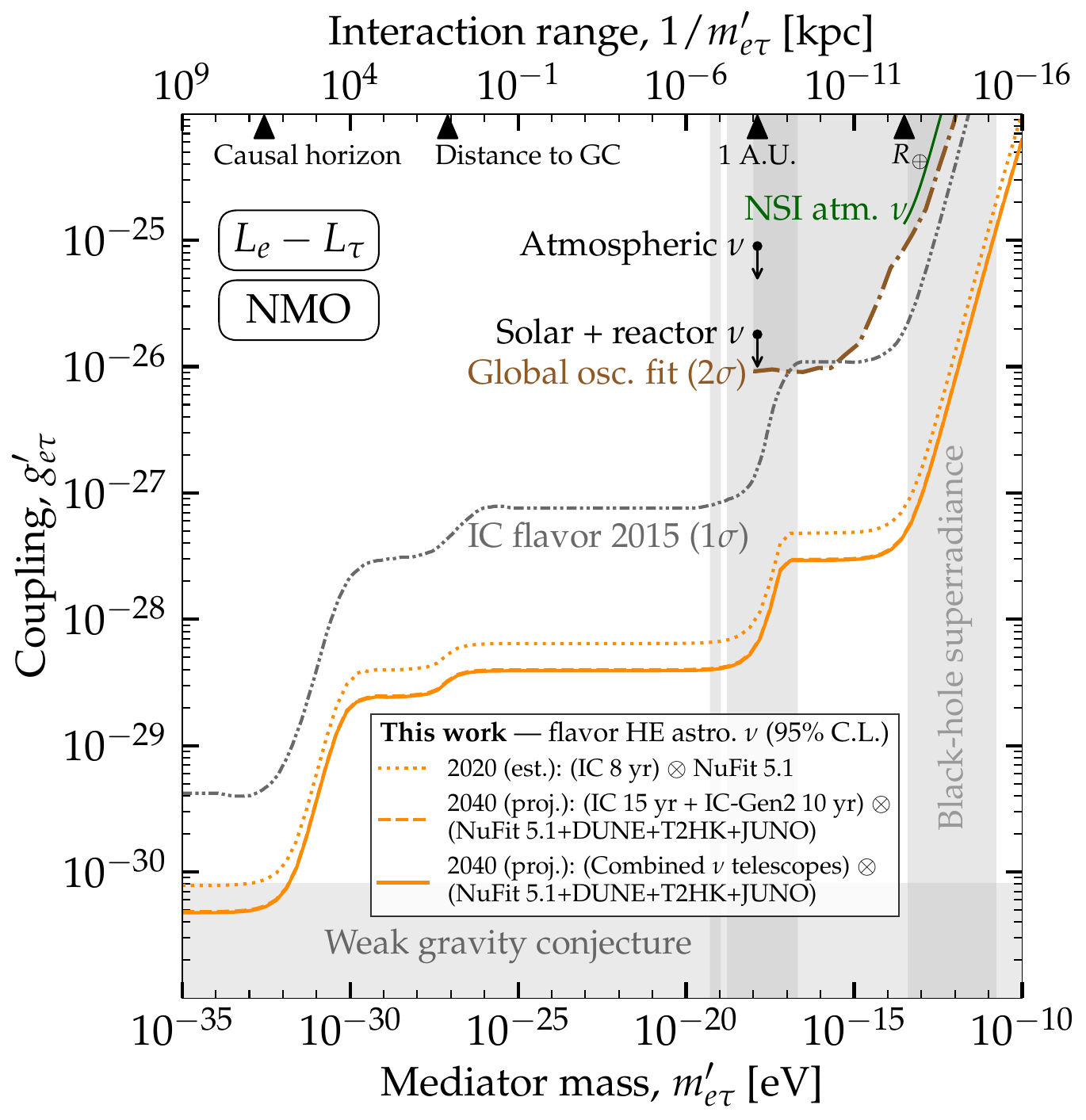}
 \caption{\textbf{\textit{Constraints on the long-range matter potential $V_{e\tau}$, under the $L_e-L_\tau$ symmetry.}}  {\it Left:}  Posterior probability density of $V_{e\tau}$.  Same as \figu{posterior}, but for $V_{e\tau}$.  {\it Right:}  Estimated present-day (2020) and projected (2040) upper limits (95\%~C.L.) on the coupling strength, $g_{e\tau}^\prime$, of the new boson, $Z_{e\tau}^\prime$, with mass $m_{e\tau}^\prime$.  Same as \figu{g_vs_m_120}, but for $V_{e\tau}$.  See Section~\ref{sec:results} and Appendix~\ref{app:results_le_ltau} for details.}
 \label{fig:plots_for_et}
\end{figure}

Figure~\ref{fig:plots_for_et} shows the posterior on $V_{e\tau}$ and limits on the coupling, $g_{e\tau}^\prime$, obtained following the procedure described in Section~\ref{sec:stat_analysis}. The limits are similar to those for $V_{e\mu}$ under the $L_e-L_\mu$ symmetry.

\section{Results assuming inverted neutrino mass ordering}
\label{appendix_e}

\renewcommand\thefigure{E\arabic{figure}}
\renewcommand\theHfigure{E\arabic{figure}}
\setcounter{table}{0} 
\setcounter{figure}{0} 
\setcounter{equation}{0}

In the main text, we show mainly results obtained assuming normal neutrino mass ordering.  Below, we show results assuming inverted ordering.  There are differences in the oscillation probabilities: under inverted ordering, resonant features are prominent for antineutrinos rather than for neutrinos.  However, because we average the flavor composition at Earth between neutrinos and antineutrinos, the limits on the long-range matter potentials that we obtain (\figu{potential_limits}) are largely insensitive to the mass ordering.

\begin{figure}[t!]
 \centering
 \includegraphics[width=\textwidth]{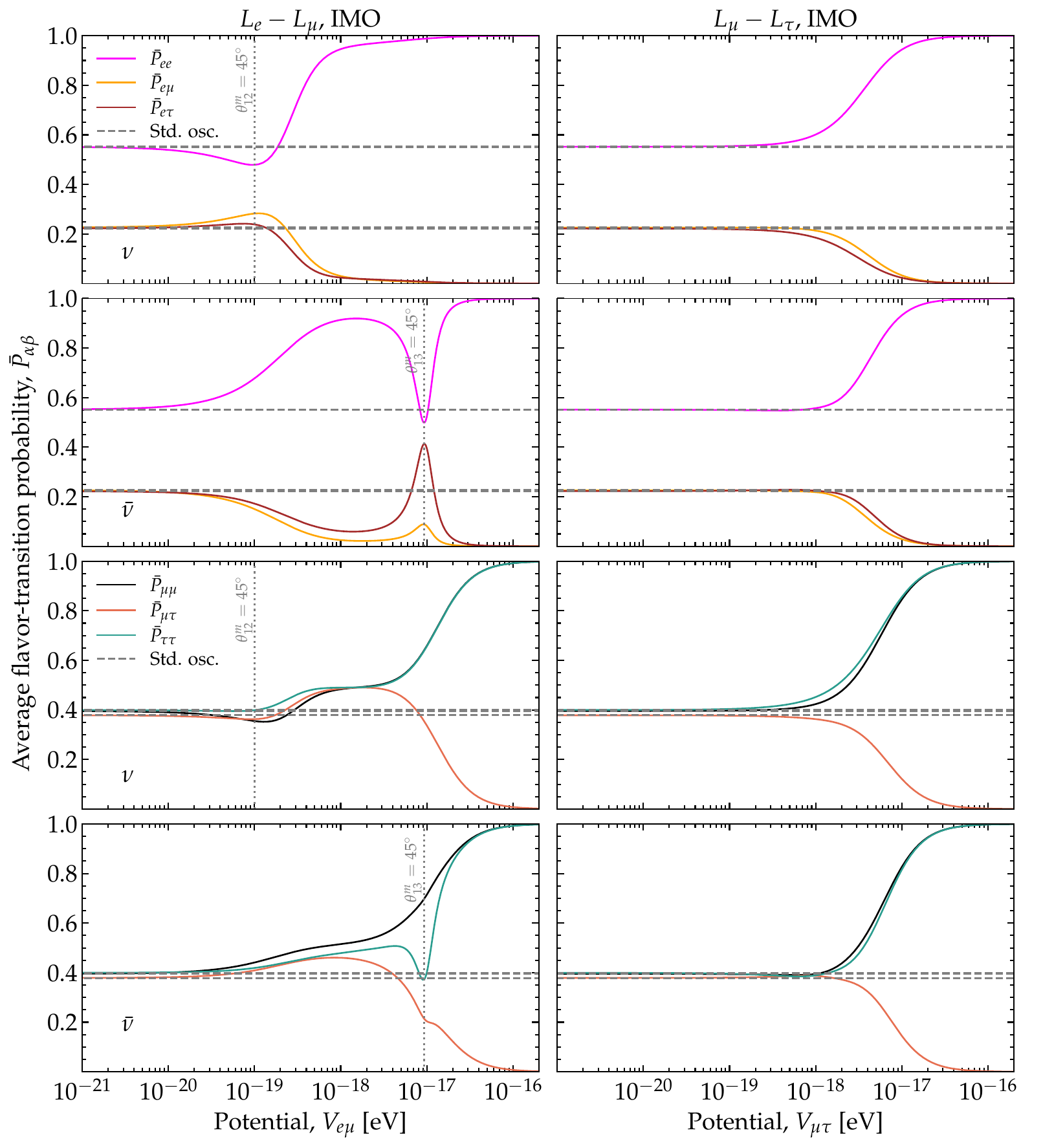}
 \caption{\textit{\textbf{Average flavor-transition probabilities, \equ{prob_avg}, as functions of the new matter potential induced by the $U(1)$ gauge symmetries $L_e-L_\mu$ (left column) and $L_\mu-L_\tau$ (right column).}}  Same as Fig.~\ref{fig:prob_lemmt_nmo}, but assuming inverted neutrino mass ordering (IMO).  See \figu{prob_let_imo} for the probabilities under $L_e-L_\tau$.  See Section~\ref{subsec:transition-prob} for details.}
 \label{fig:prob_lemmt_imo}
\end{figure}

\begin{figure}[t!]
 \centering
 \includegraphics[width=\textwidth]{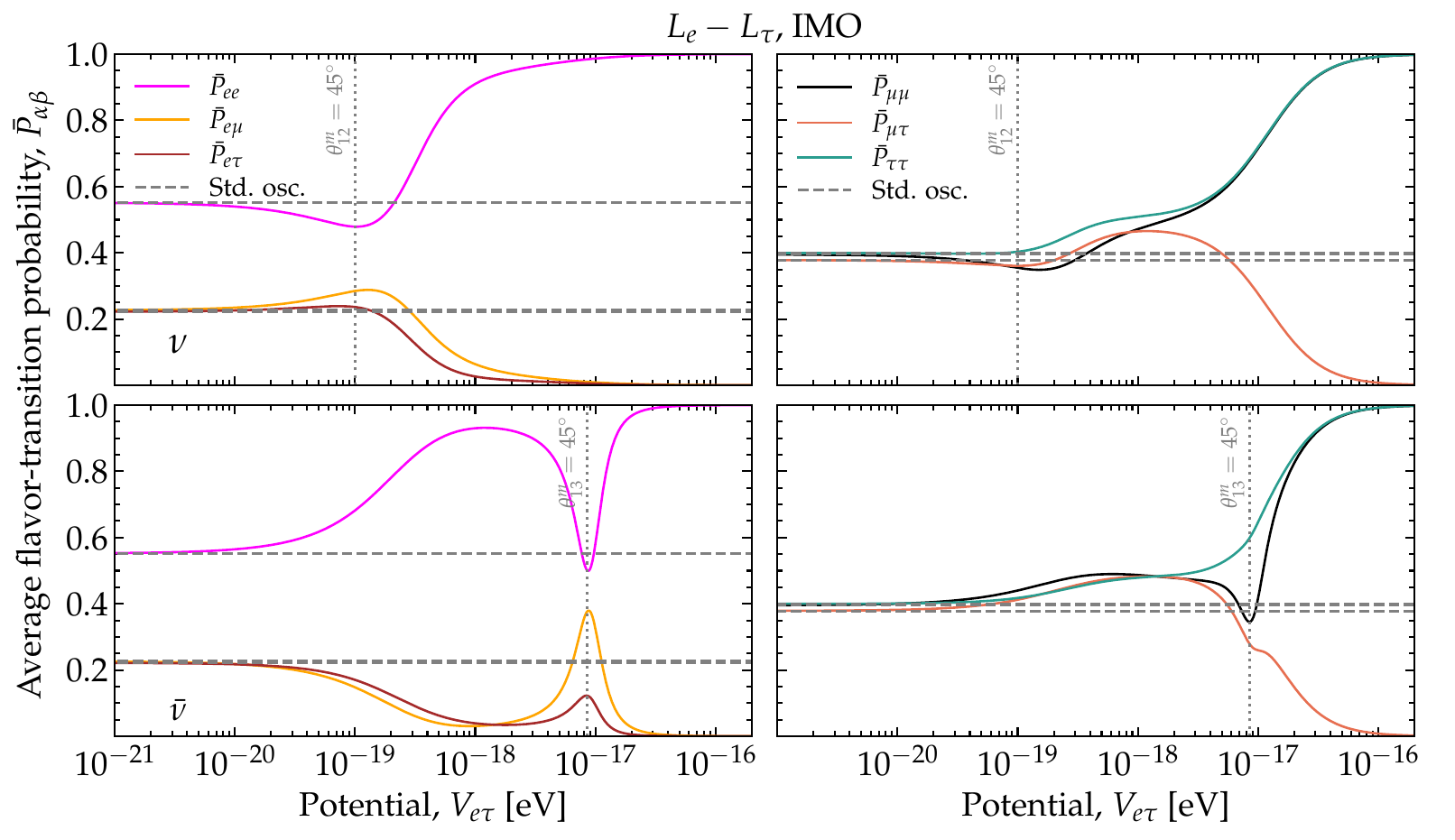}
 \caption{\textit{\textbf{Average flavor-transition probabilities, \equ{prob_avg}, as functions of the new matter potential induced by the $U(1)$ gauge symmetry $L_e-L_\tau$.}}  Same as \figu{prob_let_nmo}, but for inverted mass ordering (IMO).  See \figu{prob_lemmt_imo} for the probabilities under the other two symmetries.  See Section~\ref{subsec:transition-prob} for details.}
 \label{fig:prob_let_imo}
\end{figure}

Figures~\ref{fig:prob_lemmt_imo} and~\ref{fig:prob_let_imo} show the oscillation probabilities as functions of the long-range matter potentials.  

\begin{figure}[t!]
 \centering
 \includegraphics[width=.49\textwidth]{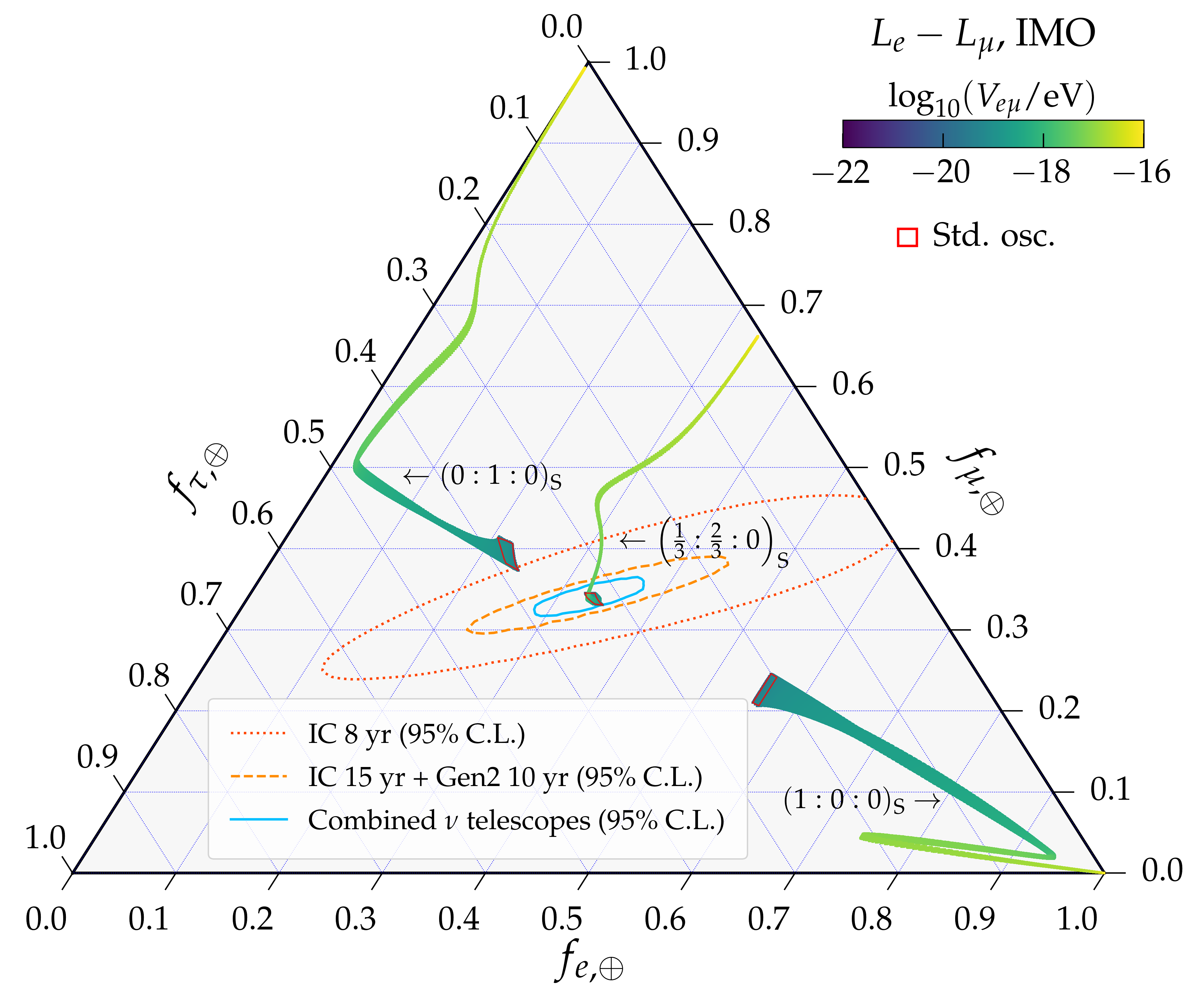}
 \includegraphics[width=.49\textwidth]{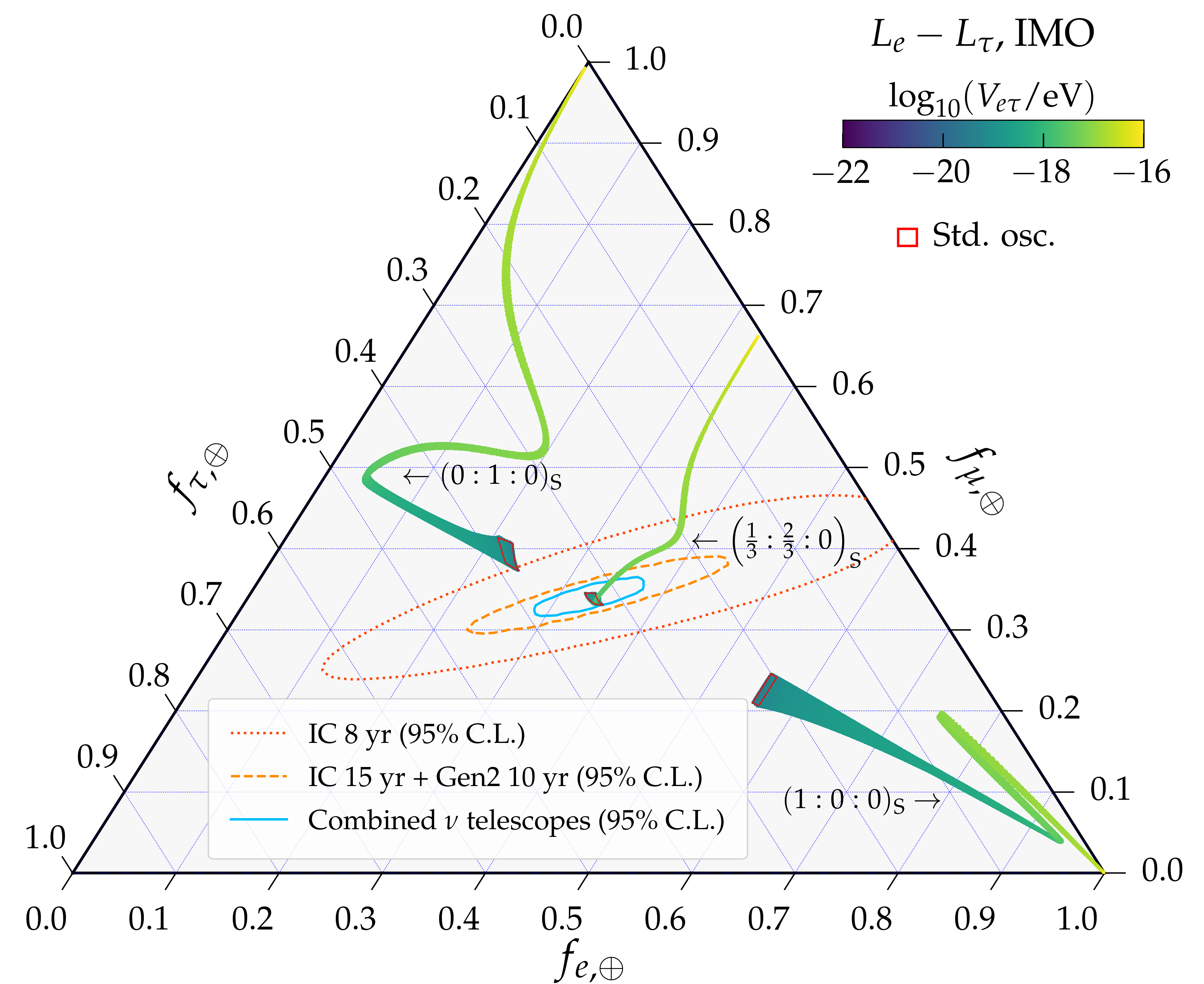}    
 \includegraphics[width=.49\textwidth]{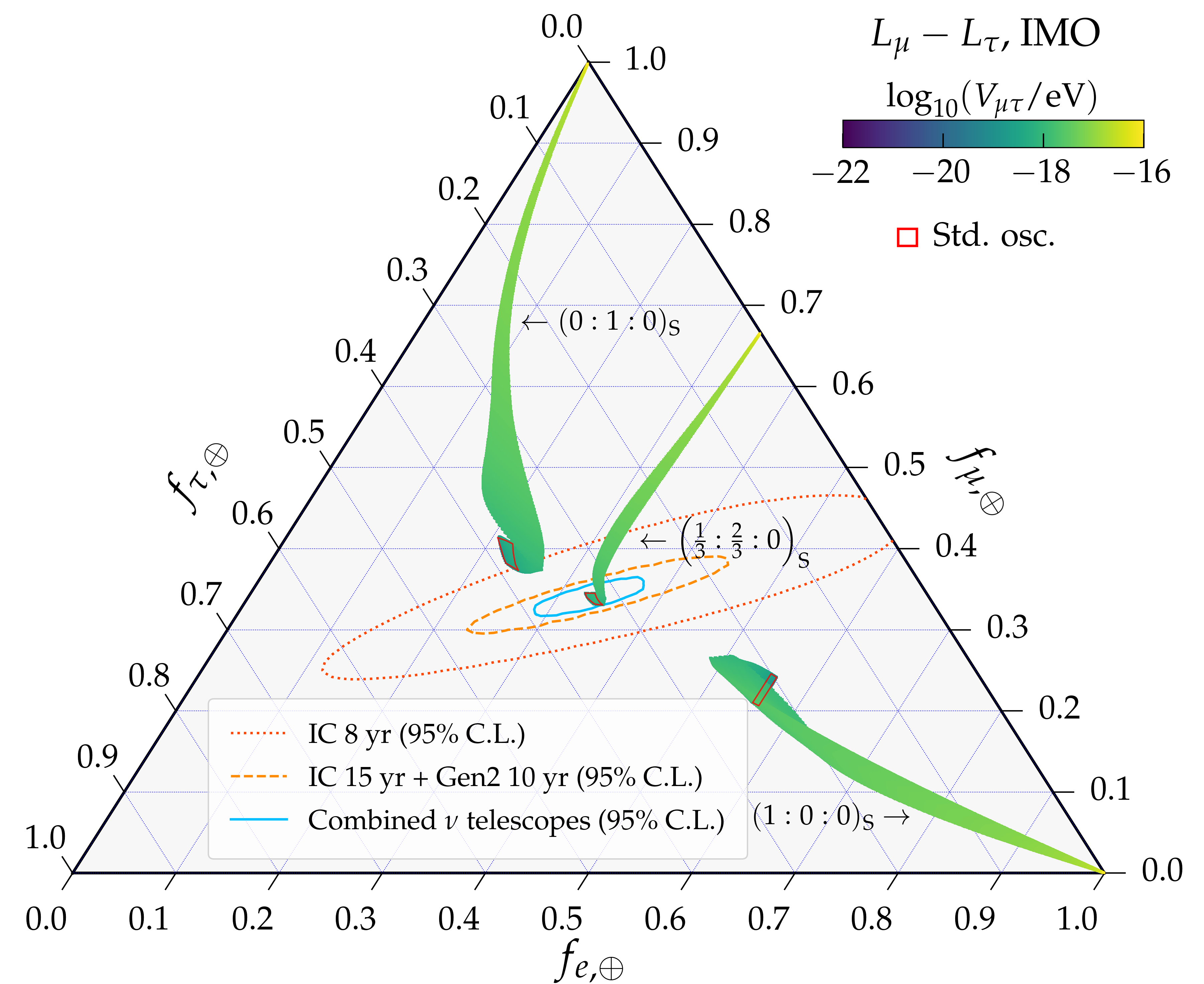}
 \caption{\textbf{\textit{Flavor composition of high-energy astrophysical neutrinos at Earth, $f_{\alpha, \oplus}$, as a function of the long-range matter potential $V_{e\mu}$, under $L_e-L_\mu$ (top left), $V_{e\tau}$, under $L_e-L_\tau$ (top right), and $V_{\mu\tau}$, under $L_\mu-L_\tau$ (bottom).}}  Same as Figs.~\ref{fig:flav_ratio} and \ref{fig:flav_ratio_let}, but for inverted mass ordering (IMO).  See Section~\ref{sec:he_nu_flavor_ratios} for details.}
 \label{fig:flavor_IMO}
\end{figure}

Figure~\ref{fig:flavor_IMO} shows the corresponding neutrino flavor composition at Earth.

\begin{figure}[t!]
 \centering
 \includegraphics[width=.49\textwidth]{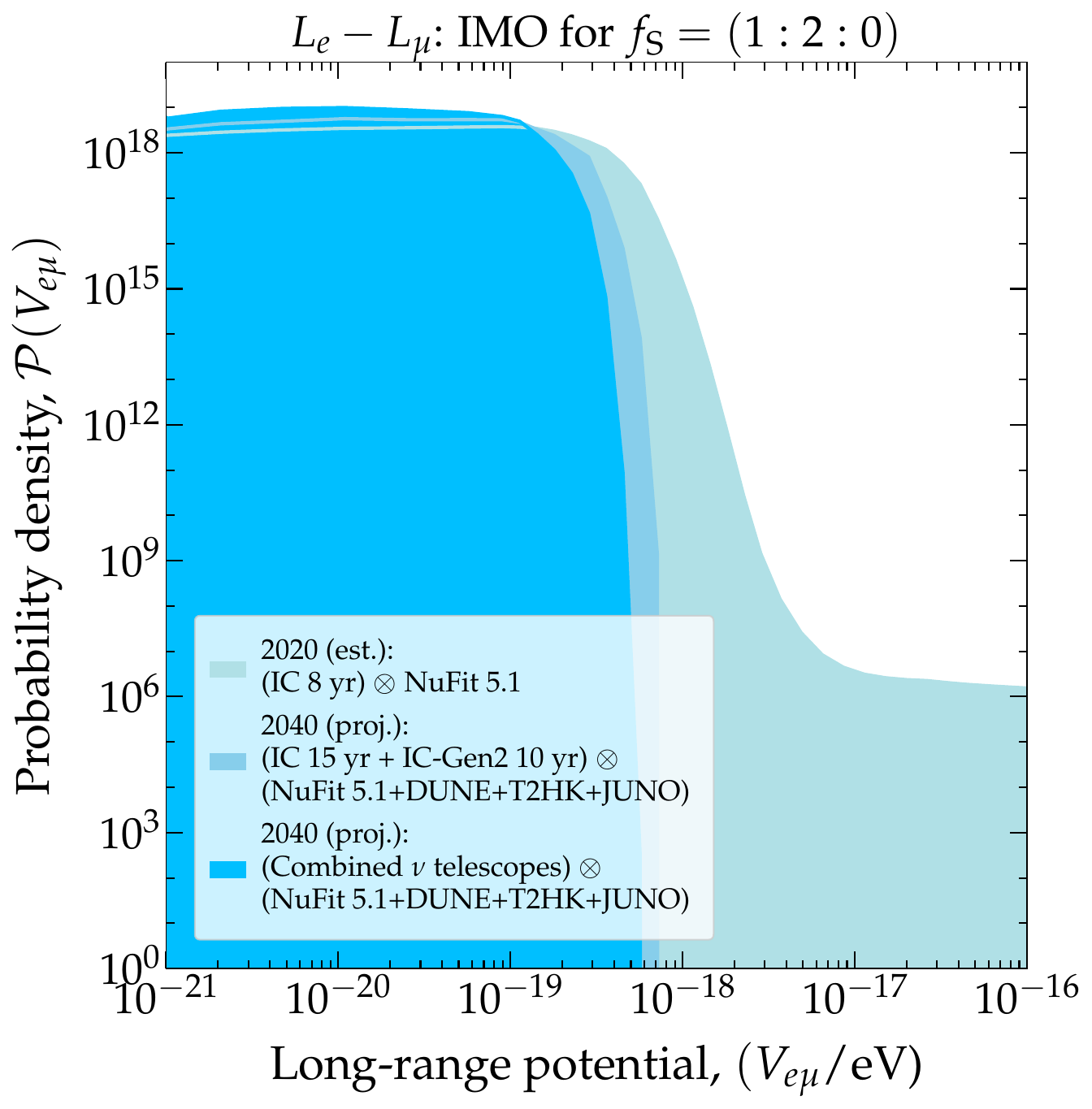}
 \includegraphics[width=.49\textwidth]{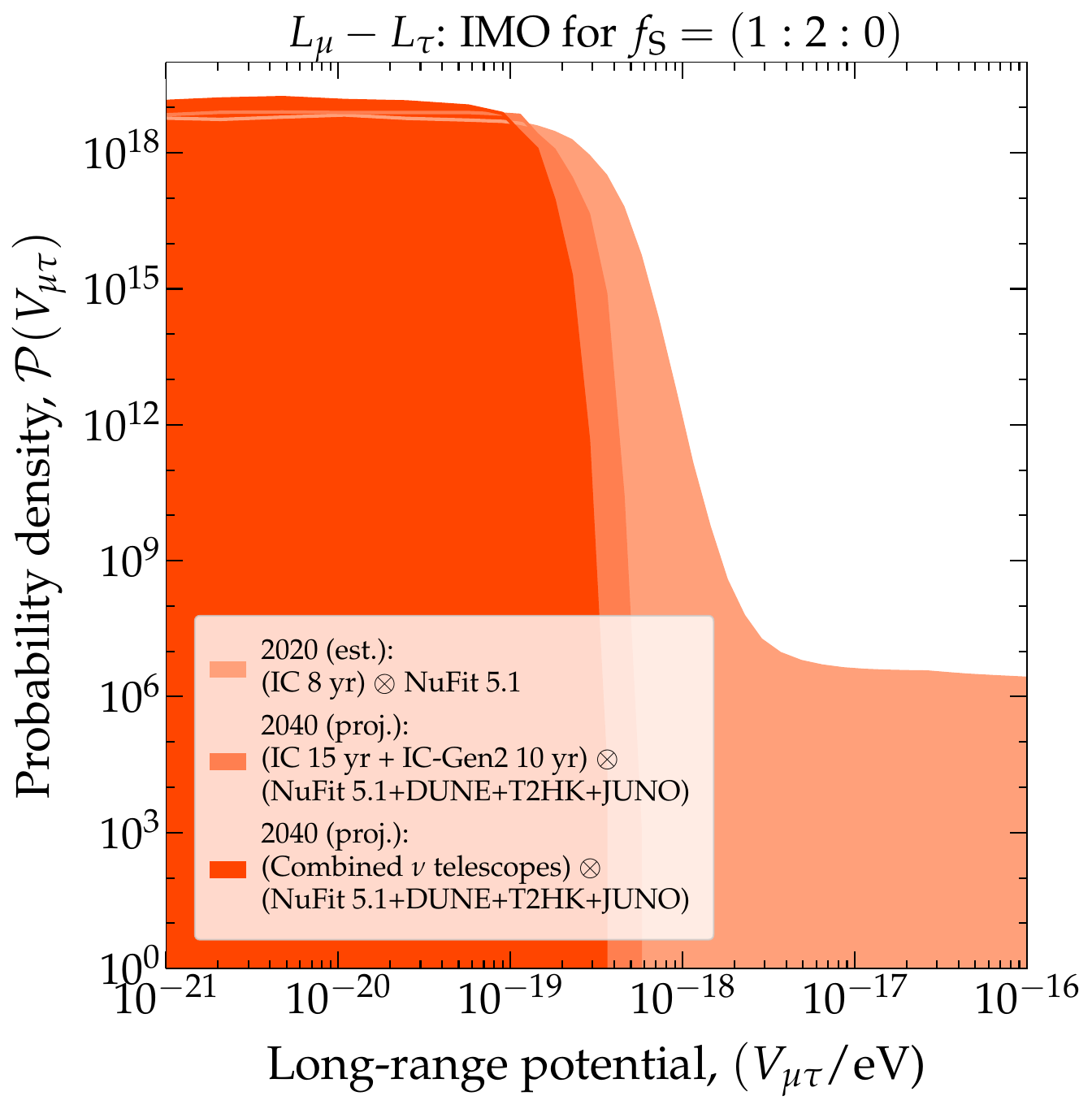}    
 \includegraphics[width=.49\textwidth]{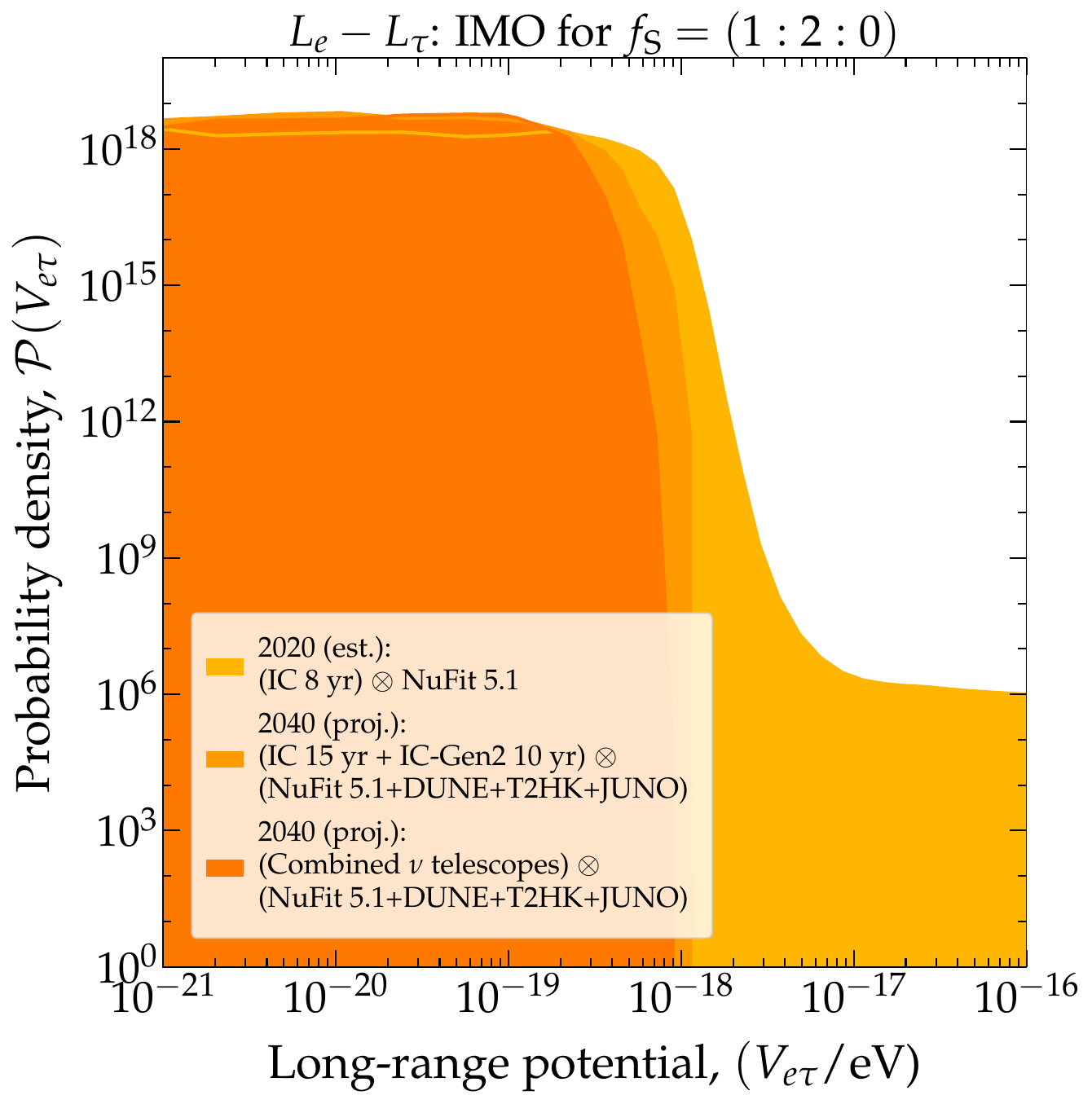}
 \caption{\textbf{\textit{Posterior probability density of the long-range matter potentials $V_{e \mu}$, under $L_e-L_\mu$ (top left), $V_{e\tau}$, under $L_e-L_\tau$ (top right), and $V_{\mu \tau}$, under $L_\mu-L_\tau$ (bottom).}}  Same as Figs.~\ref{fig:posterior} and \ref{fig:plots_for_et}, but for inverted neutrino mass ordering (IMO).  See Section~\ref{sec:stat_analysis} for details.}
 \label{fig:posterior_IMO}
\end{figure}

Figure~\ref{fig:posterior_IMO} shows the posterior probability distribution of the long-range potentials, obtained following the procedure described in Section~\ref{sec:stat_analysis}.

\begin{figure}[t!]
 \centering
 \includegraphics[width=.49\textwidth]{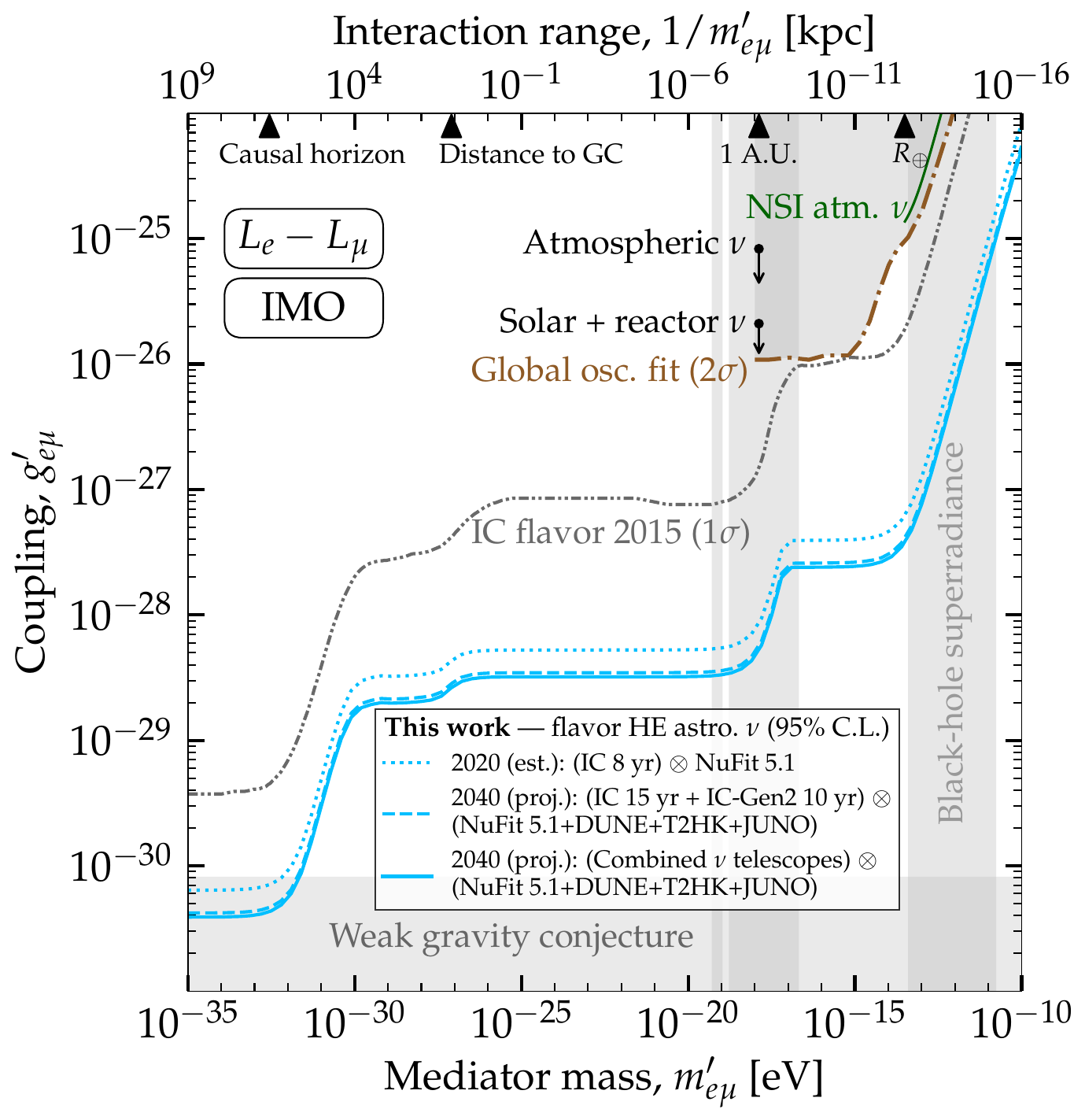}
 \includegraphics[width=.49\textwidth]{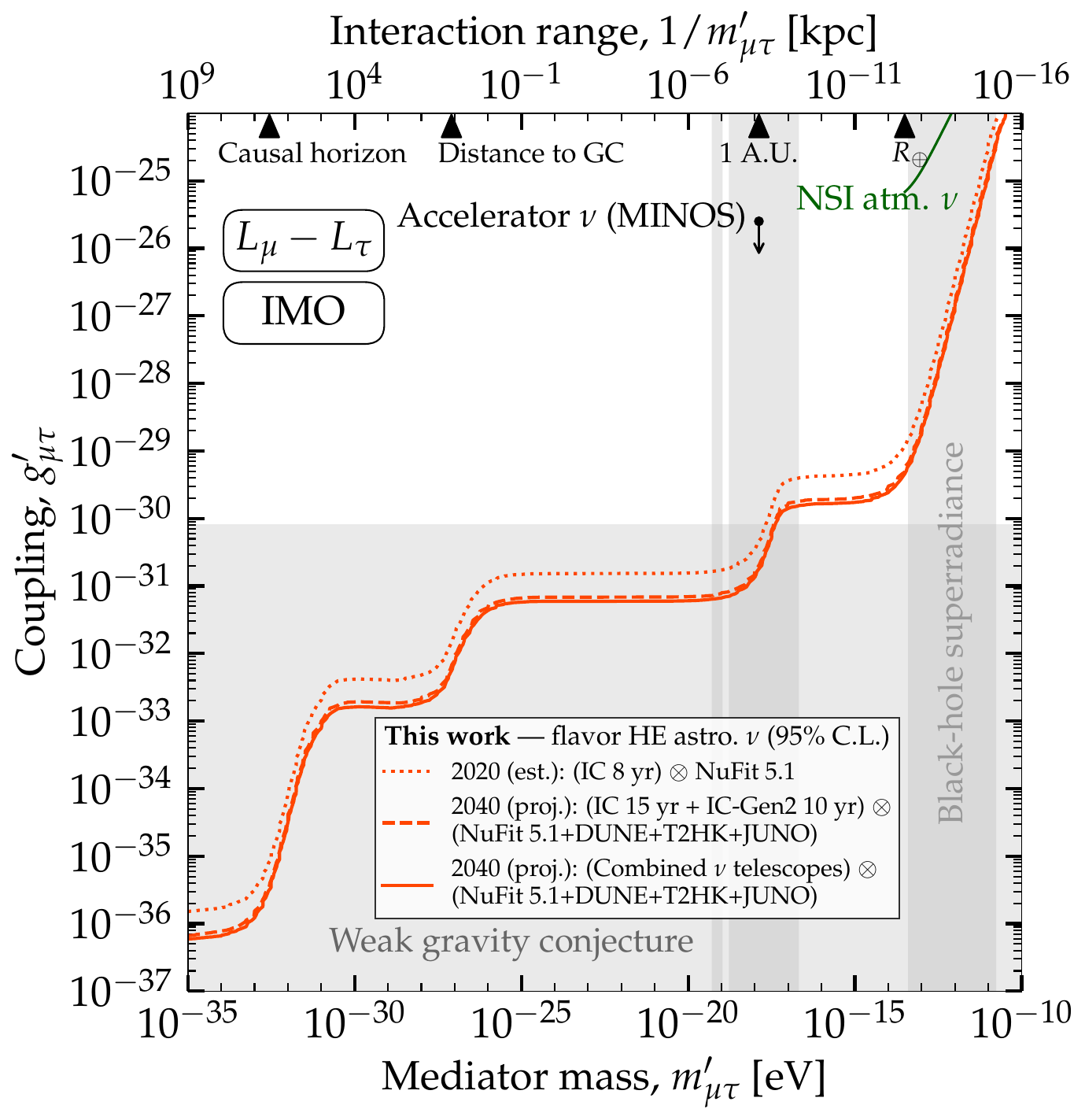}    
 \includegraphics[width=.49\textwidth]{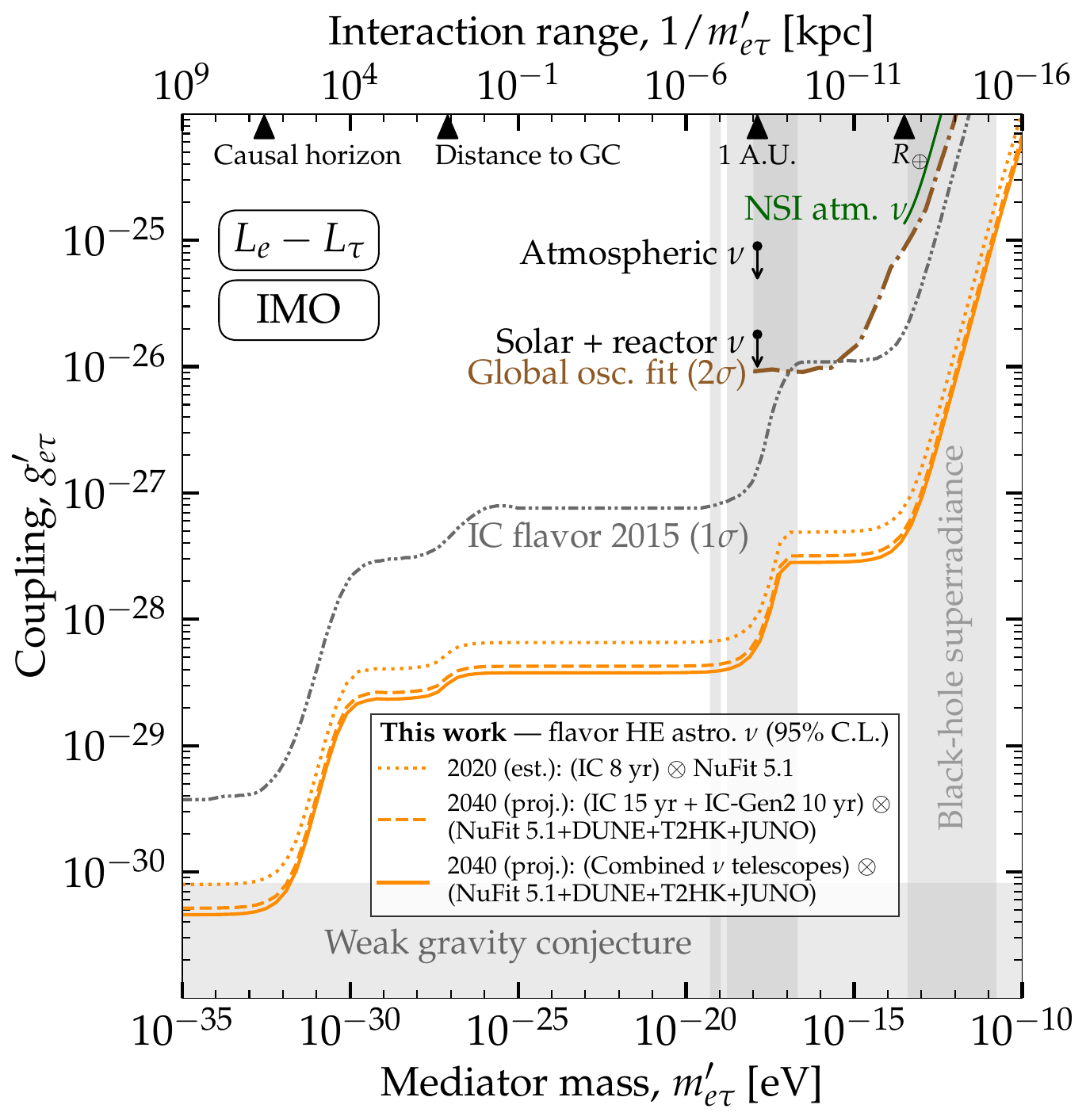}
 \caption{
 \textbf{\textit{Estimated present-day (2020) and projected (2040) upper limits (95\%~C.L.) on the coupling strength, $g_{\alpha\beta}^\prime$, of the new boson, $Z_{\alpha\beta}^\prime$, with mass $m_{\alpha\beta}^\prime$, that mediates flavor-dependent long-range neutrino interactions.}}  Same as Figs.~\ref{fig:g_vs_m_120} and \ref{fig:plots_for_et}, but for inverted neutrino mass ordering (IMO).  See Section~\ref{sec:results} for details.}
 \label{fig:g_vs_m_120_IMO}
\end{figure}

Figure~\ref{fig:g_vs_m_120_IMO} shows the corresponding upper limits on the couplings, $g_{\alpha\beta}^\prime$.

\section{Table of limits on the long-range potential}
\label{app:limits}

\renewcommand\thefigure{F\arabic{figure}}
\renewcommand\theHfigure{F\arabic{figure}}
\renewcommand\thetable{F\arabic{table}}
\renewcommand\theHtable{F\arabic{table}}
\setcounter{table}{0} 
\setcounter{figure}{0} 
\setcounter{equation}{0}

Table~\ref{tab:bounds} shows the upper limits on the long-range matter potentials obtained in Section~\ref{sec:results}.  Figure \ref{fig:potential_limits} in the main text shows a graphical representation of them and Figs.~\ref{fig:limits_3models}, \ref{fig:g_vs_m_120}, \ref{fig:plots_for_et}, and \ref{fig:g_vs_m_120_IMO} shows them translated into limits on the coupling, $g_{\alpha\beta}^\prime$.

\begin{table}[b!]
 \centering
 \begin{tabular}{ | c | *{7}{>{\centering\arraybackslash}p{1.0cm} |}}
  \hline
  \multirow{3}{*}{Observation epoch} &
  \multicolumn{6}{c|}{Upper limit (95\%~C.L.) on potential [$10^{-19}$~eV]} \\
  &
  \multicolumn{2}{c|}{$V_{e\mu}$} &
  \multicolumn{2}{c|}{$V_{e\tau}$} &
  \multicolumn{2}{c|}{$V_{\mu\tau}$} \\
  & NMO & IMO & NMO & IMO & NMO & IMO \\
  \hline
  2020 (est.): IC 8 yr 
  & 3.11 & 2.95 & 4.41 & 4.58 & 1.79 & 1.87 \\
  2040 (proj.): IC 15 yr + Gen2 10 yr 
  & 1.11 & 1.28 & 1.69 & 1.93 & 0.731 & 0.837 \\
  2040 (proj.): Combined $\nu$ telescopes
  & 1.08 & 1.10 & 1.63 & 1.52 & 0.702 & 0.727 \\
  \hline
  \end{tabular}
  \caption{\textbf{\textit{Estimated present-day (2020) and projected (2040) upper limits (95\%~C.L.) on the long-range matter potentials $V_{e\mu}$, $V_{e\tau}$, and $V_{\mu\tau}$.}}  Results are for normal (NMO) and inverted neutrino mass ordering (IMO).  See Section~\ref{sec:results} for details.}
 \label{tab:bounds}
\end{table}

\end{appendix}


\clearpage
\bibliographystyle{JHEP}
\bibliography{lrf_ref}


\end{document}